\documentclass[a4paper, 11pt]{article}
\usepackage[utf8]{inputenc}

\RequirePackage{amsthm,amsmath,amsfonts,amssymb,subcaption,algorithm,algpseudocode}
\RequirePackage[authoryear]{natbib}
\RequirePackage[colorlinks,citecolor=blue,urlcolor=blue,backref=page,backref=page]{hyperref}
\RequirePackage{graphicx}
\RequirePackage{nicefrac}
\RequirePackage{mathtools}
\mathtoolsset{showonlyrefs=true}
\RequirePackage{authblk}
\RequirePackage[left=3.5cm,right=3.5cm,top=3.5cm,bottom=4.8cm]{geometry}

\title{A Latent Position Co-Clustering Model for Multiplex Networks}

\author[1, 2]{C.J. Clarke\thanks{courtney.clarke@ucdconnect.ie}}
\author[1]{Michael Fop\thanks{michael.fop@ucd.ie}}
\affil[1]{School of Mathematics and Statistics, University College Dublin, Ireland}
\affil[2]{Taighde Éireann Centre for Research Training in Foundations of Data Science}

\begin{document}

\maketitle

\begin{abstract}
Multiplex networks are increasingly common across diverse domains, motivating the development of clustering methods that uncover patterns at multiple levels. Existing approaches typically focus on clustering either entire networks or nodes within a single network. We address the lack of a unified latent space framework for simultaneous network- and node-level clustering by proposing a latent position co-clustering model (LaPCoM), based on a hierarchical mixture-of-mixtures formulation. LaPCoM enables co-clustering of networks and their constituent nodes, providing joint dimension reduction and two-level cluster detection. At the network level, it identifies global homogeneity in topological patterns by grouping networks that share similar latent representations. At the node level, it captures local connectivity and community patterns. The model adopts a Bayesian nonparametric framework using a mixture of finite mixtures, which places priors on the number of clusters at both levels and incorporates sparse priors to encourage parsimonious clustering. Inference is performed via Markov chain Monte Carlo with automatic selection of the number of clusters. LaPCoM accommodates both binary and count-valued multiplex data. Simulation studies and comparisons with existing methods demonstrate accurate recovery of latent structure and clusters. Applications to real-world social multiplexes reveal interpretable network-level clusters aligned with context-specific patterns, and node-level clusters reflecting social patterns and roles.
\end{abstract}

\clearpage
\section{Introduction} \label{Sec_Introduction}
Networks are widely used to model pairwise relationships among individuals, with applications spanning the social, biological, and physical sciences. A multiplex is a collection of networks (or views) that share a common set of nodes but have different types of relationships between the nodes in each view. Natural examples of multiplex networks arise in a variety of domains, including social networks \citep{DAngelo:2019:Methods:Eurovision, Ren:2023:Methods:DPERGM}, longitudinal social networks such as co-voting networks \citep{Yin:2022:Methods:ERGM}, brain connectivity networks \citep{Taya:2016:Example:Brains, Lunagomez:2021:Methods:Distances, Yin:2022:Methods:ERGM}, and systems biology, particularly in the study of cancer gene interactions \citep{Lunagomez:2021:Methods:Distances}.

Clustering and community detection are common tasks in the analysis of network data. Traditional approaches to network and multiplex clustering focus on either clustering entire networks or clustering nodes within a single network. However, in some applications, such as multiplex social networks capturing different types of relationships among the same individuals \citep[e.g., ][]{Magnani:2013:Aarhus:Data,DAngelo:2023:PP:ILPCM}, it may be desirable to perform {\em co-clustering}, inspired by classical data analysis methods that simultaneously cluster rows (observations) and columns (variables) in a data matrix \citep[e.g.,][]{Biernacki:2023:CoClustering, Battaglia:2024:CoClustering}. We extend this idea to multiplex networks by proposing a model to jointly cluster both the networks and their constituent nodes. This approach enables multilevel insights into the underlying structure of multiplex data. Indeed, clustering at both levels can yield a deeper understanding of complex systems by revealing patterns across and within networks.

With a focus on modelling relations between nodes, a common model for analysing network data is the latent position model \citep[LPM,][]{Hoff:2002:Methods:LPM}, which embeds nodes in a low-dimensional latent space and posits that the probability of an edge between two nodes depends on their relative positions. The LPM has become a foundational tool in network analysis and has inspired numerous extensions. An important extension of the LPM is the latent position cluster model \citep[LPM,][]{Handcock:2007:Methods:LPCM}, which incorporates a Gaussian mixture model to induce clustering of the nodes in a network. Several extensions of the LPM and LPCM have been proposed to analyse multiplex networks. For instance, \cite{GolliniMurphy:2016:Methods:LSJM} introduced the latent space joint model, where a single latent space is assumed to have generated the multiplex. However, inference is carried out in a variational framework where a collection of variational latent spaces, one for each network, are averaged to estimate the generating latent space. Similarly, \cite{DAngelo:2019:Methods:Eurovision} assumed a common latent space shared across networks, which may overlook network-specific variation, while \cite{DAngelo:2023:PP:ILPCM} extended this approach to clustering the nodes of a multiplex by adapting the LPCM within a Bayesian nonparametric (BNP) framework. \cite{MST:2017:Methods:Village} proposed a multivariate Bernoulli model, using a conditional LPM to model dyadic dependence within each layer. \cite{Sweet:2013:Methods:HLSM} introduced a hierarchical framework that extends single-network models, including the LPM, to multiplex data. However, this requires estimating a separate latent space for each network, which becomes computationally intensive as the number of networks increases. 

Several network-clustering methods have been developed, where entire networks are grouped based on structural characteristics. These approaches typically use distance- or summary-statistic-based representations of networks, followed by standard clustering algorithms. However, they often rely on additional information, such as nodal class labels \citep{Sweet:2019:Methods:Ensembles}, node attributes \citep{Brandes:2011:Methods:Ensembles}, or network-level labels \citep{Lunagomez:2021:Methods:Distances}. Building on the stochastic block model \citep[SBM,][]{Lee:2019:SBM}, recent work has proposed co-clustering frameworks that simultaneously cluster both nodes and networks. These include methods based on spectral clustering \citep{Pensky:2024:Methods:DIMPLE}, sparse subspace techniques \citep{Noroozi:2024:Methods:Sparse}, tensor decompositions \citep{Jing:2021:Methods:TWIST}, minimisation algorithms \citep{Fan:2022:Methods:ALMA}, and hierarchical or multilayer SBM extensions \citep{Chabert-Liddel:2024:Methods:SBM:Ecology, Stanley:2024:Methods:sMLSBM, Josephs:2024:Methods:NSBM}.

In addition to latent space and SBM approaches, alternative approaches based on mixture models offer a flexible framework for capturing heterogeneity in multiplex networks by representing observed networks as draws from a mixture of latent structures. Within this class, both finite and nonparametric mixtures have been explored. For example, mixtures of exponential random graph models (ERGMs) have been applied using Dirichlet process mixtures (DPMs) \citep{Ren:2023:Methods:DPERGM} and finite mixtures \citep{Yin:2022:Methods:ERGM}. \cite{Rebafka:2024:Methods:Competitor} proposed a finite mixture of SBMs for networks with varying node sets, using hierarchical clustering to infer the number of components. \cite{Mantziou:2024:Methods:Competitor} introduced a mixture of measurement error models with latent representative networks, extending to node-level clustering via block structures. This was further developed by \cite{Barile:2024:Methods:Mantziou} into a BNP approach using a location-scale DPM. Other approaches include mixtures of network-clustering probability models, such as generalised linear (mixed) models \citep{SignorelliWit:2020:Methods:Competitor}, and latent space mixtures with component-specific embeddings and nonparametric priors \citep{Durante:2017:Methods:PopNet}.

In this paper, we propose a {\em latent position co-clustering model} ({\em LaPCoM}) for multiplex network data. Existing methods largely focus on node-clustering or network-clustering only, often require a fixed number of clusters, and offer limited interpretability. The main contribution of {\em LaPCoM} is to provide a co-clustering framework that simultaneously identifies network-level clusters and node-level clusters within each network-level cluster, leveraging the interpretability of latent space models. Each network-level cluster is represented by a distinct latent space, providing a visual and low dimensional summary of the shared structural features of those networks. Within each latent space, a second mixture model, based on the LPCM, is used to uncover communities of nodes. Importantly, {\em LaPCoM} employs a BNP approach via a mixture of finite mixtures (MFM) model \citep{FS:2021:Methods:MFM_TS} at both levels, enabling automatic inference of the number of clusters from the data and principled uncertainty quantification. Informed by foundational work in latent position modelling \citep{Hoff:2002:Methods:LPM} and its extensions to clustering and multiplex settings \citep{Handcock:2007:Methods:LPCM, Durante:2017:Methods:PopNet, DAngelo:2019:Methods:Eurovision,DAngelo:2023:PP:ILPCM}, {\em LaPCoM} unifies these ideas into a cohesive framework for simultaneous co-clustering and low-dimensional representation of multiplex networks. 

The paper is structured as follows: Section~\ref{Sec_LaPCoM} describes the LPM, the LPCM and the proposed {\em LaPCoM}; Section~\ref{Sec_Inference} outlines the inferential procedure; Section~\ref{Sec_SimulationStudies} explores the performance of {\em LaPCoM} on simulated data; Section~\ref{Sec_Applications} applies the proposed {\em LaPCoM} to a selection of illustrative datasets; and Section~\ref{Sec_Discussion} concludes with a discussion and potential avenues for future research. 

The \textsf{R} \citep{R} code to implement {\em LaPCoM} is freely available at \url{https://github.com/cjclarke258/LaPCoM}.

\section{Latent Position Co-Clustering Model} \label{Sec_LaPCoM}

\subsection{The Latent Position Model} \label{SubSec_LPM}
A \emph{network} is a data structure which records relations among a set of $N$ nodes, describing how these nodes are connected to one another. A network is typically represented by an $N \times N$ square adjacency matrix $\boldsymbol{Y}$ with element $y_{ij}$ equal to 1 if nodes $i$ and $j$ are connected, and 0 otherwise. This particular representation is specific to binary networks; however, other types of networks exist, such as weighted networks, where the ties between the nodes of the network have weights assigned to them. For example, one can consider count networks in which the elements of the adjacency matrix can take non-negative integer values. 

The latent position model \citep[LPM,][]{Hoff:2002:Methods:LPM} models the probability of an edge between nodes $i$ and $j$ based on their unobserved positions $\boldsymbol{z}_i$ and $\boldsymbol{z}_j$ in a $p$-dimensional latent space. Edges are assumed independent given the latent positions, with nodes closer in the latent space more likely to connect. Self-loops are excluded ($y_{ii} = 0 \ \forall \ i$). The model assumes that $y_{ij} \sim {\cal P}(\lambda_{ij})$, where ${\cal P}$ is an appropriate distribution, with $\lambda_{ij} = \mathbb{E}[y_{ij}]$ and $f(\lambda_{ij}) = \alpha - \|\boldsymbol{z}_i - \boldsymbol{z}_j\|_2^2$, where $f$ is an appropriate link function. For example, the logit link function and the Bernoulli distribution are used when modelling binary networks \citep{Hoff:2002:Methods:LPM}, while a log link and the Poisson distribution may be used when the edge weights are counts \citep{Gwee:2025:LSPM}. Additionally, $\alpha$ is an intercept controlling overall connectivity, and the squared Euclidean distance penalises long-range connections \citep{GolliniMurphy:2016:Methods:LSJM}. 

The latent position cluster model \citep[LPCM,][]{Handcock:2007:Methods:LPCM} extends the LPM by assuming that the latent positions arise from a finite mixture of $K$ multivariate normals: $\boldsymbol{z}_i \sim \sum_{k=1}^K \pi_k \mathcal{MVN}_p(\boldsymbol{\mu}_k, \sigma_k^2 \boldsymbol{\mathbb{I}}_p)$, where $\pi_k$ is the cluster probability, $\boldsymbol{\mu}_k$ the cluster mean, and $\sigma_k^2 \boldsymbol{\mathbb{I}}_p$ a spherical covariance matrix reflecting rotational invariance. This enables community detection while preserving latent space interpretability. Extensions to multiplex networks include \cite{GolliniMurphy:2016:Methods:LSJM}, \cite{DAngelo:2019:Methods:Eurovision,DAngelo:2023:PP:ILPCM}, and \cite{Durante:2017:Methods:PopNet}.

\subsection{A Latent Position Co-Clustering Model for
Multiplex Networks ({\em LaPCoM})} \label{SubSec_LaPCoM}
{\em LaPCoM} is a mixture-of-mixtures model that clusters networks (views) within a multiplex while simultaneously clustering the nodes within each network-level group. The model consists of a two-level mixture formulation: a top-level mixture captures structural similarities across networks, while a lower-level mixture clusters nodes within each network-level group based on their connectivity patterns. This hierarchical formulation enables joint modelling of global (network-level) structure and local (node-level) variation. Global homogeneity reflects shared topological features across views, where nodes exhibit similar connectivity profiles along certain relational dimensions. Local heterogeneity captures within-network features, such as community structure or node-specific roles.

In the following, we consider generic valued multiplex networks, represented by a collection of adjacency matrices $\boldsymbol{Y}^{(1)}, \ldots, \boldsymbol{Y}^{(m)}, \ldots, \boldsymbol{Y}^{(M)}$, where each entry $y_{ij}^{(m)}$ represents the relationship between nodes $i$ and $j$ in view $m$, $m = 1, \ldots, M$. This relationship can be binary or weighted, depending on the nature of the data. We model each network using an appropriate distribution ${\cal P}$ with associated link function $f$, consistent with the general latent position formulation introduced above.

Our model assumes that each network $\boldsymbol{Y}^{(m)}$ is generated from a network-level finite mixture distribution:
\begin{equation}
    \boldsymbol{Y}^{(m)}  \sim \sum_{g = 1}^G \tau_g \prod_{\underset{i \neq j}{i,j}} {\cal P}(\lambda_{g,ij}), \quad m = 1, \dots, M,
\end{equation}
where $G$ is the number of components in the network-level mixture, $\boldsymbol{\tau}$ are the mixing proportions such that $\tau_g \geq 0$ for all $g$ and $\sum_{g = 1}^G \tau_g = 1$. The parameter $\lambda_{g,ij}$ denotes the expected value $\mathbb{E}[y^{(m)}_{ij}]$, and has the associated link function $f(\lambda_{g,ij}) = \alpha - \|\boldsymbol{z}_{g,i} - \boldsymbol{z}_{g,j}\|_2^2$. For example, if the multiplex consists of count-valued adjacency matrices, then $\lambda_{g,ij}$ represents the rate parameter corresponding to component $g$, ${\cal P}$ is the Poisson distribution, and the log link function implies $\log \lambda_{g,ij} = \alpha - \|\boldsymbol{z}_{g,i} - \boldsymbol{z}_{g,j}\|_2^2$. The latent positions $\boldsymbol{z}_{g,i}$ capture the connectivity patterns of the nodes across all network views arising from component $g$. 

To induce clustering of the nodes in addition to clustering of the networks, {\em LaPCoM} assumes that each component-specific latent space has positions originating from a node-level finite mixture distribution:
\begin{equation} \label{Equation:NodeMixture}
    \boldsymbol{z}_{g,i} \sim \sum_{k = 1}^{K_g} \pi_{gk} \mathcal{MVN}_2(\boldsymbol{\mu}_{gk}, \boldsymbol{\Sigma}_{gk}), \quad i=1,\dots,N,
\end{equation}
\noindent where $K_g$ is the number of components in the node-level mixture within the $g$\textsuperscript{th} network-level component, $\boldsymbol{\pi}_g$ represents the mixing proportions, defined such that $\pi_{gk} \geq 0$ for all $k$ and $\sum_{k = 1}^{K_g} \pi_{gk} = 1$, and $\boldsymbol{\Sigma}_{gk} = \mathrm{diag}(\sigma_{gk,1}^2, \sigma_{gk,2}^2)$. Note that, for visualisation purposes and ease of interpretability, we will focus on latent spaces of dimension $p = 2$. {\em LaPCoM} incorporates the following prior distributions for the mixture components:
\begin{eqnarray*} \label{Equation:LPM_Priors}
    \boldsymbol{\mu}_{gk} & \sim & \mathcal{MVN}_2(\boldsymbol{0}, \boldsymbol{\mathbb{I}}_2), \quad k = 1, \dots, K_g, \ \mathrm{and} \ g = 1, \dots, G,\\
    \sigma_{gk,q}^2 & \sim & \mathcal{IG}(u_{\sigma^2}, v_{\sigma^2}), \quad q=1, 2, \ k = 1, \dots, K_g, \ \mathrm{and} \ g = 1, \dots, G,
\end{eqnarray*}
\noindent
where $\mathcal{IG}$ denotes the Inverse-Gamma distribution. 

At both network and node level, {\em LaPCoM} employs a mixture of finite mixtures (MFMs) framework. Unlike standard finite mixtures that fix the number of components $K$, MFMs adopt a Bayesian nonparametric (BNP) approach by placing a prior on $K$, allowing for uncertainty while retaining a finite mixture structure \citep{MillerHarrison:2015:MFM}. A key development in this framework is the extension of \cite{FS:2021:Methods:MFM_TS}, who introduced a dynamic MFM, where the concentration parameter of the symmetric Dirichlet prior on the mixture weights depends on $K$. This leads to a richer, generalised class of BNP models. Crucially, they highlight the important distinction between the number of components $K$ and the number of clusters (active components), $K_+$. The distribution of $K_+$ is influenced by both the prior on $K$ and the Dirichlet hyperparameter, allowing for greater modelling flexibility. In contrast, Dirichlet process mixtures (DPMs) assume an infinite number of components and focus on estimating the number of clusters $K_+$. While widely used, DPMs lack the finite tractability of MFMs, which can approximate DPMs asymptotically when the prior on $K$ is diffuse. Notably, \cite{FS:2021:Methods:MFM_TS} employ a three-parameter translated beta-negative-binomial (BNB) prior distribution on $K$, which provides flexible control over both the expected number of components and the tail behaviour of the induced prior on $K_+$. A detailed justification for the use of this MFM framework over alternatives, such as overfitted finite mixtures \citep{MW:2016:Methods:SFM:Original}, is provided in Section~10 of the Supplementary Material.

Following \citet{FS:2021:Methods:MFM_TS}, {\em LaPCoM} places a translated BNB prior on the number of components at both network and node level:
\begin{equation}
    G - 1 \sim \mathcal{BNB}(a_G, b_G, c_G) \qquad K_g - 1 \sim \mathcal{BNB}(a_K, b_K, c_K) \quad \ g = 1, \dots, G,
\end{equation}
where $\mathcal{BNB}$ denotes the beta-negative-binomial (BNB) distribution. Additionally, we employ a dynamic MFM formulation in which the hyperparameter of the Dirichlet prior on the mixture weights depends on the number of components, inducing a shrinkage that encourages sparsity in the mixture weights, leading to the emptying of superfluous components. Specifically:
\begin{equation} \label{Equation:SFM_Priors}
    \boldsymbol{\tau} \sim \mathcal{D} \Big(\frac{e}{G}\Big), \qquad
    \boldsymbol{\pi}_{g} \sim \mathcal{D} \Big(\frac{w_g}{K_g}\Big), \quad g = 1, \dots, G,
\end{equation}
\noindent where $\mathcal{D}$ denotes the Dirichlet distribution.

To implement both network-level and node-level clustering within our model, we introduce a set of latent allocation variables. At the network level, we define the $M \times G$ binary matrix $\boldsymbol{C}$, such that $C_g^{(m)} = 1$ if network $m$ is allocated to component $g$. Within each network-level cluster $g$, we define the $N \times K_g$ binary matrix $\boldsymbol{S}_g$, such that $S_{gk}^{(i)} = 1$ if node $i$ is allocated to component $k$. 

The remaining priors of the model are assumed to be as follows:
\begin{eqnarray*} \label{Equation:OtherPriors}
    e & \sim & \mathcal{F}(l_G, r_G),\\
    \boldsymbol{C}^{(m)} & \sim & \mathcal{MN}(\boldsymbol{\tau}) \quad \forall \ m = 1, \dots, M,\\
    w_g & \sim & \mathcal{F}(l_K, r_K) \quad \forall \ g = 1, \dots, G,\\
    \boldsymbol{S}_g^{(i)} & \sim & \mathcal{MN}(\boldsymbol{\pi}_{g}) \quad \forall \ g = 1, \dots, G \ \mathrm{and} \ \forall \ i = 1, \dots, N,\\
    \alpha & \sim & \mathcal{N}(m_{\alpha}, s_{\alpha}^2),
\end{eqnarray*}
\noindent where, $\mathcal{F}$ denotes the Fisher-Snedecor distribution, $\mathcal{MN}$ denotes the multinomial distribution and $\mathcal{N}$ denotes the univariate normal distribution.

\section{Inference} \label{Sec_Inference}

\subsection{MCMC Algorithm} \label{SubSec_MCMCAlgorithm}
With hyperparameters $\boldsymbol{H} = \{a_G, b_G, c_G, l_G, r_G, m_{\alpha}, s_{\alpha}, a_K, b_K, c_K, l_K, r_K, u_{\sigma^2}, v_{\sigma^2}\}$, the joint posterior distribution of {\em LaPCoM} is as follows:
\begin{eqnarray*}
    & & \mathbb{P} \bigg(G, e, \boldsymbol{\tau}, \boldsymbol{C}, \alpha, \Big\{K_g, w_g, \boldsymbol{\pi}_g, \boldsymbol{S}_g, \{\boldsymbol{\mu}_{gk}, \boldsymbol{\Sigma}_{gk}\}_{k=1}^{K_g}\Big\}_{g = 1}^G, \{\boldsymbol{Z}_g\}_{g = 1}^G \mid \boldsymbol{\mathcal{Y}} \bigg)\\
    & & \propto \mathbb{P}\big(\boldsymbol{\mathcal{Y}} \mid \boldsymbol{C}, \alpha, \{\boldsymbol{Z}_g\}_{g=1}^G\big) \times \mathbb{P}(\boldsymbol{C} \mid \boldsymbol{\tau}) \times \mathbb{P}(\boldsymbol{\tau} \mid G, e) \times \mathbb{P}(G \mid \boldsymbol{H}) \times \mathbb{P}(e \mid \boldsymbol{H}) \times \mathbb{P}(\alpha \mid \boldsymbol{H})\\
    & & \quad \times \prod_{g = 1}^G \mathbb{P}\Big(\boldsymbol{Z}_g \mid \boldsymbol{S}_g, \{\boldsymbol{\mu}_{gk}, \boldsymbol{\Sigma}_{gk}\}_{k=1}^{K_g}\Big) \times \prod_{g = 1}^G \prod_{k = 1}^{K_g} \mathbb{P}(\boldsymbol{\mu}_{gk} \mid \boldsymbol{H}) \times \prod_{g = 1}^G \prod_{k = 1}^{K_g} \mathbb{P}(\boldsymbol{\sigma}_{gk}^2 \mid \boldsymbol{H})\\
    & & \quad \times \prod_{g = 1}^G \mathbb{P}(\boldsymbol{S}_g \mid \boldsymbol{\pi}_{g}) \times \prod_{g = 1}^G \mathbb{P}(\boldsymbol{\pi}_{gk} \mid K_g, w_g) \times \prod_{g = 1}^G \mathbb{P}(K_g \mid \boldsymbol{H}) \times \prod_{g = 1}^G \mathbb{P}(w_g \mid \boldsymbol{H}).
\end{eqnarray*}

Inference is carried out using a Metropolis-within-Gibbs MCMC algorithm. Additionally, the algorithm incorporates the telescoping sampling procedure of \citet{FS:2021:Methods:MFM_TS} to adaptively sample mixture components and cluster allocations. A brief description of the MCMC algorithm is described below. The Supplementary Material includes a complete list of notation (Section~1), details on our choice of hyperparameters (Section~2), detailed derivations of all full conditionals (Section~3), a pseudocode description of the MCMC algorithm (Section~4), and details of the algorithm initialisation (Section~5).

The MCMC algorithm iterates through the following steps, for $t = 1, \dots, T$, where $t$ denotes the current iteration and $T$ is the total number of iterations:

\begin{enumerate}

    \item Sample $C_g^{(m)[t + 1]}$ for each network $m = 1, \dots, M$ from a multinomial distribution such that 
    $Pr(C^{(m)[t + 1]} = g \mid \dots) \propto \tau_g^{[t]} \prod_{i \neq j} p(y^{(m)}_{ij}; \lambda_{g,ij}^{[t]})$, where $p$ is the pdf of distribution ${\cal P}$ and $f(\lambda^{[t]}_{g,ij}) = \alpha^{[t]} - \|\boldsymbol{z}^{[t]}_{g,i} - \boldsymbol{z}^{[t]}_{g,j}\|_2^2$.

    \item Compute $M_g^{[t+1]}$ for each $g = 1, \dots, G^{[t]}$, the number of networks allocated to each network-level component; determine $G_+^{[t+1]}$, the number of active (non-empty) components; relabel the network-level mixture parameters so that the first $G_+^{[t+1]}$ components are non-empty.

    \item For $g = 1, \dots, G_+^{[t + 1]}$:

    \begin{enumerate}

        \item Propose a block update $\hat{\boldsymbol{Z}}_g$, conditioned on $\boldsymbol{S}_g^{[t]}$, with proposal distribution $\hat{\boldsymbol{z}}_{g,i} \sim \mathcal{MVN}\Big(\boldsymbol{z}_{g,i}^{[t]}, \, \delta_Z^2 \boldsymbol{\Sigma}_{gk}^{[t]}\Big), \ i = 1, \dots, N$, where $\delta_Z$ is a scaling factor. Accept $\hat{\boldsymbol{Z}}_g$ as $\boldsymbol{Z}_g^{[t+1]}$ with the appropriate probability; otherwise, set $\boldsymbol{Z}_g^{[t+1]} = \boldsymbol{Z}_g^{[t]}$.

        \item For each $i = 1, \dots, N$, sample $S_{gk}^{(i)[t + 1]}$ from a multinomial distribution with \small{$Pr(S_g^{(i)[t + 1]} \mid \dots) \propto \pi_{gk}^{[t]} \Big|\boldsymbol{\Sigma}_{gk}^{[t]}\Big|^{-\frac{1}{2}} \exp\Big[-\frac{1}{2} \Big(\boldsymbol{z}_{g,i}^{[t+1]} - \boldsymbol{\mu}_{gk}^{[t]}\Big)^T \boldsymbol{\Sigma}_{gk}^{[t]^{-1}}\Big(\boldsymbol{z}_{g,i}^{[t+1]} - \boldsymbol{\mu}_{gk}^{[t]}\Big)\Big]$}.

        \item Calculate $N_{gk}^{[t+1]}$ for each $k = 1, \dots, K_g^{[t]}$, the number of nodes allocated to each node-level component; determine $K_{g+}^{[t+1]}$, the number of active components; and relabel the node-level mixture parameters so that the first $K_{g+}^{[t+1]}$ components are non-empty.

        \item For each $k = 1, \dots, K_{g+}^{[t+1]}$, sample $\boldsymbol{\mu}_{gk}^{[t+1]}$ from $\mathcal{MVN}(\boldsymbol{\mu}_{gk}^*, \boldsymbol{\Sigma}_{gk}^*)$ and ${\sigma}_{gk,q}^{2\,[t+1]}$ for $q = 1, 2$ from $\mathcal{IG}(u_{\sigma^2}^*, v_{\sigma^2}^*)$, where $\boldsymbol{\Sigma}_{gk}^*$, $\boldsymbol{\mu}_{gk}^*$, $u_{\sigma^2}^*$, and $v_{\sigma^2}^*$ are updated based on the current model parameters, as derived in Section~3 of the Supplementary Material.

        \item Sample $K_g^{[t+1]}$ from a multinomial distribution with probabilities $\mathbb{P}\big(k^* \mid \boldsymbol{S}_g^{[t+1]}, w_g^{[t]}\big) \propto \mathbb{P}(k^*) \frac{(w_g^{[t]})^{K_{g+}^{[t+1]}} k^*!}{(k^*)^{K_{g+}^{[t+1]}} \big(k^* - K_{g+}^{[t+1]}\big)!} \prod_{k=1}^{K_{g+}^{[t+1]}} \frac{\Gamma\left(N_{gk}^{[t+1]} + \frac{w_g^{[t]}}{k^*}\right)}{\Gamma\left(1 + \frac{w_g^{[t]}}{k^*}\right)}$, where $\mathbb{P}(k^*)$ is the prior distribution, $k^*$ ranges from $K_{g+}^{[t]}$ to $K_{\max}$, and $K_{\max}$ is the maximum number of node-level components considered; $\Gamma(\cdot)$ is the gamma function.

        \item Propose $\log(\hat{w}_g)$ from the proposal distribution $\mathcal{N}\big(\log(w_g^{[t]}), s_w^2\big)$. Accept $\hat{w}_g$ as $w_g^{[t+1]}$ with the appropriate probability; otherwise, set $w_g^{[t+1]} = w_g^{[t]}$.

        \item If $K_g^{[t+1]} > K_{g+}^{[t+1]}$, add $K_g^{[t+1]} - K_{g+}^{[t+1]}$ empty components by sampling $\boldsymbol{\mu}_{gk}^{[t+1]}$ and $\boldsymbol{\sigma}_{gk}^{2\,[t+1]}$ from their respective priors, for $k = K_{g+}^{[t+1]} + 1, \dots, K_g^{[t+1]}$.

        \item Sample $\boldsymbol{\pi}_g^{[t+1]}$ from $\mathcal{D}\big(\boldsymbol{\psi}_g^{[t+1]}\big)$, where $\psi_{gk}^{[t+1]} = \frac{w_g^{[t+1]}}{K_g^{[t+1]}} + N_{gk}^{[t+1]}, \ k = 1, \dots, K_g^{[t+1]}$.

    \end{enumerate}

    \item Propose $\hat{\alpha}$ using the proposal distribution $\mathcal{N}(\alpha^{[t]}, \delta_{\alpha}^2 s_{\alpha}^{*2})$, where $\delta_{\alpha}$ is a scaling factor. Accept $\hat{\alpha}$ as $\alpha^{[t+1]}$ with the appropriate probability; otherwise, set $\alpha^{[t+1]} = \alpha^{[t]}$.

    \item Sample $G^{[t+1]}$ from a multinomial distribution with probabilities $\mathbb{P}\big(g^* \mid \boldsymbol{C}^{[t+1]}, e^{[t]}\big) \propto \mathbb{P}(g^*) \frac{(e^{[t]})^{G_+^{[t+1]}} g^*!}{(g^*)^{G_+^{[t+1]}} \big(g^* - G_+^{[t+1]}\big)!} \prod_{g=1}^{G_+^{[t+1]}} \frac{\Gamma \left(M_g^{[t+1]} + \frac{e^{[t]}}{g^*}\right)}{\Gamma \left(1 + \frac{e^{[t]}}{g^*}\right)}$, where $\mathbb{P}(g^*)$ is the prior distribution, and $g^* = G_+^{[t]}, G_+^{[t]} + 1, \dots, G_{\max}$, with $G_{\max}$ the maximum number of network-level components considered.

    \item Propose $\log(\hat{e})$ from the proposal distribution $\mathcal{N}\big(\log(e^{[t]}), s_e^2\big)$. Accept $\hat{e}$ as $e^{[t+1]}$ with the appropriate probability; otherwise, set $e^{[t+1]} = e^{[t]}$.

    \item If $G^{[t+1]} > G_+^{[t+1]}$, add $G^{[t+1]} - G_+^{[t+1]}$ empty components by sampling the corresponding hyperparameters from their priors, for $g = G_+^{[t+1]} + 1, \dots, G^{[t+1]}$, assuming no clustering structure in these additional latent spaces.

    \item Sample $\boldsymbol{\tau}^{[t + 1]}$ from $\mathcal{D}\Big(\zeta_1^{[t + 1]}, \dots, \zeta_{G^{[t + 1]}}^{[t + 1]}\Big)$, where $\zeta_g^{[t + 1]} = \frac{e^{[t + 1]}}{G^{[t + 1]}} + M_g^{[t + 1]}$.
    
\end{enumerate}

\subsection{Hyperparameter Choices} \label{SubSec_HyperparameterChoices}
Table~\ref{tab:HyperChoices} summarises our hyperparameter choices; we provide some details below. We use non-informative priors for the intercept $\alpha$ and node-level cluster means $\boldsymbol{\mu}_{gk}$. Cluster variances $\sigma_{gk,q}^2$ follow an $\mathcal{IG}(u_{\sigma^2}, v_{\sigma^2})$ prior, with parameters chosen to keep clusters tight, yielding expected variances around 0.2 for small to moderate networks and 0.1 for larger ones. We adopt a $\mathcal{BNB}(8,18,10)$ prior on the number of mixture components $G$ and $K_g$, that favours a moderate number of clusters but allows a heavy tail for flexibility and avoids excessive shrinkage. Initial values $G_0$ and $K_0$ are set to 2, balancing flexibility and shrinkage. Coupled with a Dirichlet shrinkage prior on mixing proportions, this enables adaptive growth of the number of components. Upper bounds $G_{\max}$ and $K_{\max}$ are set lower than \cite{FSMW:2019:Methods:SFMvsDPM} recommend, to reduce computational cost, with $K_{\max}$ tailored to network size and $G_{\max}$ chosen mainly for computational efficiency. Finally, node-level Dirichlet concentration parameters have an $\mathcal{F}(6,3)$ prior as in \cite{FS:2021:Methods:MFM_TS}, reflecting moderate homogeneity. Preliminary tests validate these settings as appropriate for the data’s complexity and structure. Full details of these choices can be found in Section~2 of the Supplementary Material.

\begin{table*}[!b]
\centering
\caption{Hyperparameter choices.}
\label{tab:HyperChoices}
\centering
\resizebox{\linewidth}{!}{ 
\begin{tabular}{@{}llllll@{}}
    \hline
    Parameter & Prior Distribution & Hyperparameters & Recommended Values \\
    \hline
    $\alpha$ & $\alpha \sim \mathcal{N}(m_{\alpha},s_{\alpha})$ & $m_{\alpha},s_{\alpha}$ & $m_{\alpha} = 0, s_{\alpha} = 1$ \\ 
    $\boldsymbol{\mu}_{gk}$ & $\boldsymbol{\mu}_{gk} \sim \mathcal{MVN}_2(\boldsymbol{\mu}_0,\boldsymbol{\Sigma}_{\mu})$ & $\boldsymbol{\mu}_0,\boldsymbol{\Sigma}_{\mu}$ & $\boldsymbol{\mu}_0 = \boldsymbol{0}, \boldsymbol{\Sigma}_{\mu} = \boldsymbol{\mathbb{I}}_2$ \\
    $\sigma_{gk,q}^2$ & $\sigma_{gk,q}^2 \sim \mathcal{IG}(u_{\sigma^2},v_{\sigma^2})$ & $u_{\sigma^2},v_{\sigma^2}$ & $u_{\sigma^2} = \begin{cases} 11 & \text{ if } N < 60 \\ 21 & \text{ if } N \geq 60 \end{cases}$, $v_{\sigma^2} = 2$ \\
    $G$ & $G - 1 \sim \mathcal{BNB}(a_G, b_G, c_G)$ & $a_G, b_G, c_G$ & $a_G = 8, b_G = 18, c_G = 10$ \\
    $G_+$ & Induced by $\mathbb{P}(G)$ & - & - \\
    $G_0$ & - & - & $G_0 = 2$ \\
    $G_{\text{max}}$ & - & - & $G_{\text{max}} = \begin{cases} 5 & \text{ if } M < 60 \\ 10 & \text{ if } M \geq 60 \end{cases}$ \\
    $e$ & $e \sim \mathcal{F}(l_G, r_G)$ & $l_G, r_G$ & $l_G = 6, r_G = 3$ \\
    $K_g$ & $K_g - 1 \sim \mathcal{BNB}(a_K, b_K, c_K)$ & $a_K, b_K, c_K$ & $a_K = 8, b_K = 18, c_K = 10$ \\
    $K_{g+}$ & Induced by $\mathbb{P}(K_g)$ & - & - \\
    $K_0$ & - & - & $K_0 = 2$ \\
    $K_{\text{max}}$ & - & - & $K_{\text{max}} = \begin{cases} \frac{N}{5} + 2 & \text{ if } N < 60 \\ \frac{N}{10} + 2 & \text{ if } N \geq 60 \end{cases}$ \\
    $w_g$ & $w_g \sim \mathcal{F}(l_K, r_K)$ & $l_K,r_K$ & $l_K = 6, r_K = 3$ \\
    \hline
\end{tabular}
}
\end{table*}

\subsection{Post-Processing} \label{SubSec_PostProcessing}
To ensure valid inference, we address the identifiability issue due to label switching, which arises because the mixture model likelihood is invariant to component label permutations. Following \cite{DAngelo:2023:PP:ILPCM}, we combine strategies from \cite{FS:2011:PP:LabelSwitching} and \cite{WadeGhahramani:2018:PP:LSP}. This section describe this process for the network-level mixture; given the co-clustering nature of {\em LaPCoM}, the same procedure is applied analogously at the node level. As a first step, we apply the method of \cite{WadeGhahramani:2018:PP:LSP} to the posterior samples of the clustering vector $\boldsymbol{C}$, using the variation of information and posterior similarity matrix to obtain an optimal partition, $\hat{\boldsymbol{C}}$. This is implemented via the \verb|mcclust.ext| package in \textsf{R} \citep{Wade:2015:mcclust.ext}. From $\hat{\boldsymbol{C}}$, we obtain a posterior point estimate of the number of network-level clusters, $\hat{G}_+$, as the number of unique cluster labels present in the optimal partition.

Next, we cluster the vectorised latent spaces $\boldsymbol{Z}_g$ (at the network level) into $\hat{G}_+$ groups using $K$-means, following \cite{FS:2011:PP:LabelSwitching}. This provides a classification sequence for each MCMC iteration, assigning consistent labels to components. A permutation check ensures each sequence is a valid relabelling of ${1, \dots, \hat{G}_+}$; only iterations satisfying this condition are retained. Valid iterations are then used to relabel all component-specific parameters and the cluster allocation vector $\boldsymbol{C}$ for coherence. This relabelling procedure is applied at the node-level using the cluster means $\boldsymbol{\mu}_{gk}$ as the component-specific parameter.

To ensure label consistency across parameters, we align the optimal clustering vector $\hat{\boldsymbol{C}}$ with its relabelled counterpart $\hat{\boldsymbol{C}}_{\text{relabelled}}$ by comparing each $\boldsymbol{C}^{[t]}$ and $\boldsymbol{C}^{[t]}_{\text{relabelled}}$ via cross-tabulation. The most frequent label-matching pattern across iterations is used to permute $\hat{\boldsymbol{C}}$, yielding the final aligned clustering $\hat{\boldsymbol{C}}^*$. This step ensures coherence between $\hat{\boldsymbol{C}}$ and the labels of cluster-specific parameters such as $\{\boldsymbol{Z}_g\}_{g=1}^{\hat{G}_+}$, avoiding mismatches that can arise from applying independent post-processing procedures.

The posterior mode of the number of clusters from \cite{WadeGhahramani:2018:PP:LSP} may differ from that implied by the relabelled parameters using \cite{FS:2011:PP:LabelSwitching}, as the two methods target different aspects of the posterior and operate independently. We adopt the mode from the former as our primary estimate and use it as a reference in the \cite{FS:2011:PP:LabelSwitching} relabelling step. Although reversing this order is possible, it can yield different results, highlighting the impact of methodological choices on posterior summaries. When the posterior is multimodal, we select the smaller mode to favour parsimony.

As in standard LPMs, the use of Euclidean distances leads to identifiability issues due to invariance under rotation, reflection, and translation. Following \cite{Hoff:2002:Methods:LPM}, we apply a Procrustes transformation to each sampled latent space to resolve this. In {\em LaPCoM}, this is done ``offline'': after MCMC sampling and label correction, each cluster-specific latent space is aligned to the first retained sample for that cluster. The same transformation (scaling, rotation, and translation) is applied to the corresponding cluster means $\boldsymbol{\mu}_g$, with only scaling applied to the variances $\boldsymbol{\Sigma}_g$, since they are diagonal matrices.

\section{Simulation Studies} \label{Sec_SimulationStudies}
To assess the performance of {\em LaPCoM}, we conduct two simulation studies: Section~\ref{SubSec_SS1} assesses clustering performance across various scenarios, and Section~\ref{SubSec_SS2} compares {\em LaPCoM} with four relevant models. In all settings, we align estimated and true cluster labels using the \verb|matchClasses| function from the \textsf{R} package \verb|e1071| \citep{Meyer:2023:Extra:R:e1071}, which maximises the number of correctly matched cases.

\subsection{Study 1: Clustering Performance} \label{SubSec_SS1}
This simulation study evaluates the clustering performance of {\em LaPCoM} across eight scenarios, in which different multiplex networks are generated under varying combinations of the number of nodes, number of networks, and number of clusters at both levels. The full details of these scenarios are provided in Table~3 in Section~6 of the Supplementary Material. For each scenario, 10 multiplexes are generated using the specified parameters. Two models are fitted: {\em LaPCoM} and a simplified variant, \emph{mono-LaPCM}, which clusters the networks only, omitting the node-level mixture as in~\eqref{Equation:NodeMixture} and instead assumes $\boldsymbol{z}_{g,i} \sim \mathcal{MVN}_2(\boldsymbol{0}, \boldsymbol{\mathbb{I}}_2)$. This comparison enables an assessment of the contribution of the node-level mixture, allowing us to evaluate whether incorporating the additional clustering structure in the latent positions improves model performance. Scenario-specific scaling factors ensure suitable acceptance rates. MCMC chains run for 300,000 iterations, with an additional 90,000 as burn-in and thinning every 300\textsuperscript{th} sample to yield 1,000 posterior draws. Network-level allocations, $\boldsymbol{C}$, are initialised via $K$-means.

Posterior summaries of the number of network-level clusters, $G_+$, inferred from the 10 simulated multiplexes across eight scenarios show that both {\em LaPCoM} and \emph{mono-LaPCM} accurately recovered $\hat{G}_+ = 2$ in 100\% of draws for all but scenario B. In scenario B, \emph{mono-LaPCM} assigned 80\% posterior mass to $\hat{G}_+ = 2$ and 20\% to $\hat{G}_+ = 1$, while {\em LaPCoM} split mass between $\hat{G}_+ = 2$ (80\%) and $\hat{G}_+ = 4$ (20\%). Both models consistently recovered the latent space, each with a mean Procrustes correlation (PC) of 0.99 across all scenarios. PC was calculated using the \verb|protest| function from the \verb|vegan| package \citep{Oksanen:2024:R:vegan}. the interquartile range (IQR) was 0.01, indicating low variability in performance. Clustering performance, assessed by the adjusted Rand index (ARI; \cite{Hubert:1985:Extra:ARI}), was perfect (mean = 1, standard deviation (SD) = 0) in all but scenario B, where \emph{mono-LaPCM} achieved a mean ARI of 0.80 (SD = 0.42) and {\em LaPCoM} 0.97 (SD = 0.09). These results demonstrate comparable performance overall, with improved clustering performance for {\em LaPCoM} in some settings due to its additional flexibility given by the node-level mixture.

\begin{figure}[!t]
    \centering
    \includegraphics[width = \textwidth]{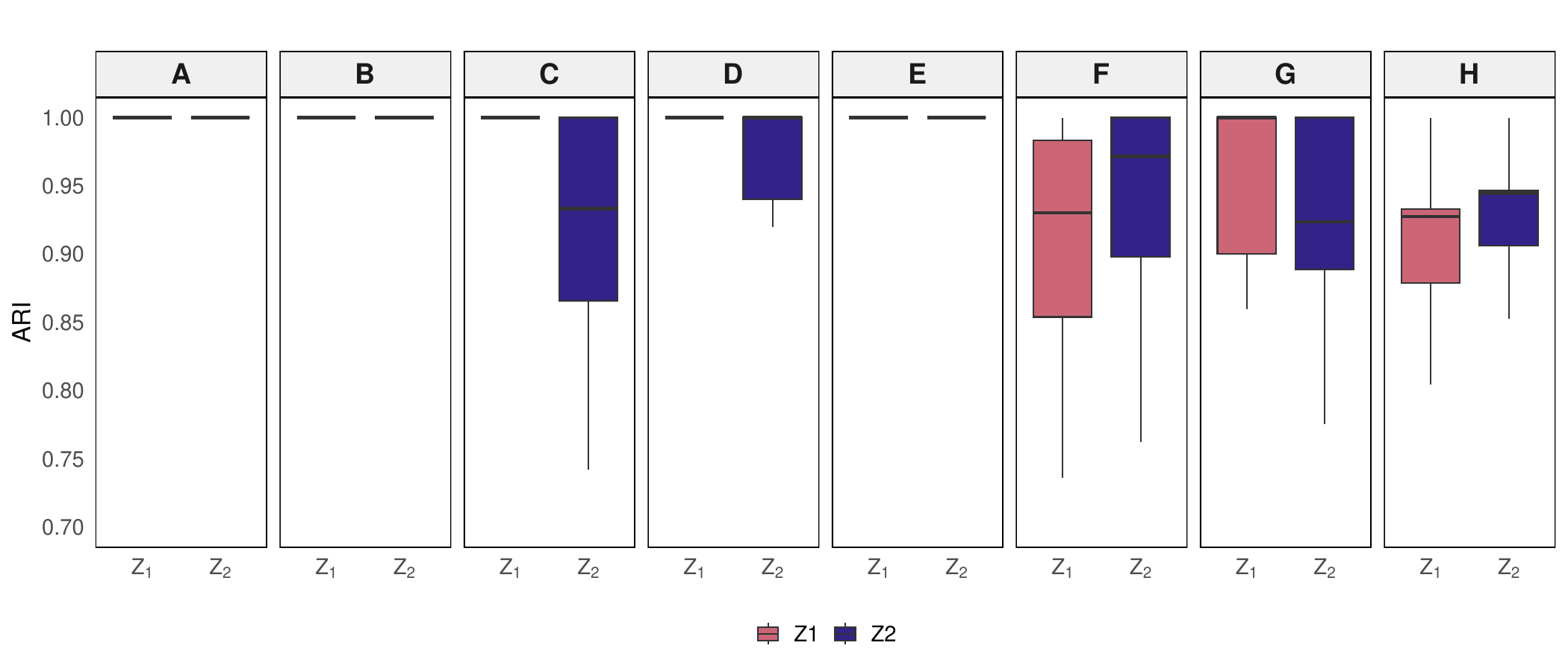}
    \caption{Boxplots comparing the Adjusted Rand Index (ARI) across 10 simulations for each scenario, between the estimated clustering solution to the true clustering solution, illustrating the clustering performance of {\em LaPCoM}.}
    \label{fig:SS1_Results_Node_Clustering}
\end{figure}

Figure~\ref{fig:SS1_Results_Node_Clustering} shows node-level clustering performance, assessed via ARI, across 10 simulations per scenario using {\em LaPCoM} (not applicable to \emph{mono-LaPCM}). Median ARIs exceed 0.92 with IQRs below 0.14. Scenarios A and B, each with one node-level cluster per $G^* = 2$ latent spaces, achieved perfect median ARIs of 1.00 (IQR = 0.00). In scenarios C and D ($K_g = \{1, 2\}$), clustering improved in D (more nodes), with median ARI for the second latent space increasing from 0.93 (IQR = 0.13) to 1.00 (IQR = 0.06), while the first latent space remained perfect. More complex scenarios G and H ($K_g = \{2, 3\}$) showed a slight drop in median ARI for the first latent space (1.00 to 0.93) and a minor increase for the second (0.92 to 0.95), with reduced IQRs in H indicating greater stability. These results confirm that {\em LaPCoM} achieves good node-level clustering performance across diverse multiplex settings.

\subsection{Study 2: Comparison with Other Methods} \label{SubSec_SS2}
This simulation study compares the clustering performance of {\em LaPCoM} against four competing models across five scenarios, in which different multiplex networks are generated under varying combinations of the number of nodes, number of networks, and number of clusters at both levels. For each scenario, 10 multiplexes are generated, using the parameters provided in Table~4 in Section~6 of the Supplementary Material.

While many relevant methods exist (see Section~\ref{Sec_Introduction}), we select one representative from each major category for comparison: a mixture of LPMs \citep[\emph{PopNet},][]{Durante:2017:Methods:PopNet}, a finite mixture of SBMs \citep{Rebafka:2024:Methods:Competitor} via the \verb|graphclust| \textsf{R} package \citep{Rebafka:2023:Methods:Competitor:R:graphclust}, a mixture of measurement error models \citep{Mantziou:2024:Methods:Competitor}, and a mixture of generalised linear (mixed) models \citep{SignorelliWit:2020:Methods:Competitor}. Code for all models is publicly available. Each model represents a distinct latent structure approach.

As in the first simulation study, {\em LaPCoM} adjusts scaling factors for the intercept and latent positions in each scenario to maintain suitable acceptance rates. \cite{Mantziou:2024:Methods:Competitor}’s model uses $C_{\text{max}}=10$ (maximum number of network-level clusters) and $B=2$ (SBM blocks), with other hyperparameters set as recommended or slightly adapted per scenario. Unlike {\em LaPCoM}, it employs a $\mathcal{G}(1,400)$ prior on the Dirichlet concentration parameter to enforce cluster shrinkage. The other three competing models use default hyperparameters. {\em LaPCoM}, \emph{PopNet}, and \cite{Mantziou:2024:Methods:Competitor}'s model run for 100,000 iterations including 30,000 burn-in, and thinning every 100\textsuperscript{th} iteration for 700 samples. For {\em LaPCoM}, network-level allocations, $\boldsymbol{C}$, are initialised via $K$-means clustering.

Table~\ref{tab:SS2_Results_G+} summarises a simulation comparing {\em LaPCoM} with four competing models over five scenarios (10 replicates each). For each simulation, the posterior mode of the number of clusters, $\hat{G}_+$, with 80\% credible intervals is reported. The \citeauthor{Mantziou:2024:Methods:Competitor}, \citeauthor{SignorelliWit:2020:Methods:Competitor}, and \emph{graphclust} models provided poor performance, generally overestimating or underestimating the number of clusters. \emph{PopNet} reliably recovered the true $\hat{G}_+$ with narrow intervals. {\em LaPCoM} matched this accuracy, consistently identifying the correct $\hat{G}_+$, though with slightly wider intervals due to its added flexibility.

\begin{table*}[!b]
\centering
\caption{Posterior mode of $G_+$, denoted $\hat{G}_+$, with 80\% credible intervals in brackets, for {\em LaPCoM} and the four competing models under the five simulation scenarios. The true generating number of network-level clusters is denoted by $G^*$.}
\label{tab:SS2_Results_G+}
\centering
\resizebox{\linewidth}{!}{
\begin{tabular}{@{}cccccccc@{}}
    \hline
    Scenario & $G^*$ & \citeauthor{Mantziou:2024:Methods:Competitor} & \citeauthor{SignorelliWit:2020:Methods:Competitor} & \texttt{graphclust} & \emph{PopNet} & {\em LaPCoM} \\
    \hline
    I     & 2 & 10 (10, 10)  &  2 (2, 3) &    1  (1, 3)   &  2 (2, 2) &  2 (2, 3) \\ 
    II    & 2 & 10 (9, 10)   &  2 (2, 7) &    3  (3, 5)   &  2 (2, 2) &  2 (2, 2) \\ 
    III   & 3 & 10 (10, 10)  &  2 (2, 2) &    5  (4, 6)   &  3 (3, 3) &  3 (3, 3) \\ 
    IV    & 4 & 10 (10, 10)  &  2 (2, 2) &    8  (6, 9)   &  4 (4, 4) &  4 (3, 4) \\ 
    V     & 4 & 10 (10, 10)  &  2 (2, 2) &    10 (5, 11)  &  4 (4, 4) &  4 (3, 4) \\
    \hline
\end{tabular}
}
\end{table*}

Table~\ref{tab:SS2_Results_Network_Clustering} reports mean ARI values concerning network clustering performance (SDs in brackets). \emph{PopNet} achieved perfect ARIs in all five scenarios, followed by {\em LaPCoM}, which consistently attained mean ARIs above 0.86 with SDs $\leq 0.17$. Both models substantially outperformed the other three across all scenarios, highlighting good network clustering performance. These results highlight {\em LaPCoM}’s performance and suitability for clustering a multiplex.

\begin{table*}[!b]
\centering
\caption{Mean network clustering ARI (standard deviations in brackets) for the proposed model and the four competing models across five distinct scenarios.}
\label{tab:SS2_Results_Network_Clustering}
\centering
\resizebox{\linewidth}{!}{
\begin{tabular}{@{}ccccccc@{}}
    \hline
    Scenario & \citeauthor{Mantziou:2024:Methods:Competitor} & \citeauthor{SignorelliWit:2020:Methods:Competitor} & \texttt{graphclust} & \emph{PopNet} & {\em LaPCoM} \\
    \hline
    I        & 0.16 (0.10)           & 0.67 (0.26)              & 0.05 (0.09)           & 1.00 (0.00)    & 0.95 (0.12) \\ 
    II       & 0.39 (0.14)           & 0.76 (0.31)              & 0.59 (0.30)           & 1.00 (0.00)    & 1.00 (0.00) \\ 
    III      & 0.52 (0.14)           & 0.58 (0.08)              & 0.44 (0.20)           & 1.00 (0.00)    & 0.98 (0.07) \\ 
    IV       & 0.57 (0.04)           & 0.38 (0.06)              & 0.48 (0.14)           & 1.00 (0.00)    & 0.89 (0.17) \\ 
    V        & 0.62 (0.12)           & 0.40 (0.07)              & 0.47 (0.18)           & 1.00 (0.00)    & 0.86 (0.16) \\ 
    \hline
\end{tabular}
}
\end{table*}

Of the five models compared, only {\em LaPCoM}, \verb|graphclust| \citep{Rebafka:2023:Methods:Competitor:R:graphclust}, and \citeauthor{Mantziou:2024:Methods:Competitor}'s model support nodal clustering. {\em LaPCoM} models networks within a cluster as dependent, sharing a cluster-specific latent space. In a similar manner, \citeauthor{Mantziou:2024:Methods:Competitor} model networks as noisy observations of a latent representative network. In contrast, \citeauthor{Rebafka:2024:Methods:Competitor} treat networks as independent realisations of a cluster-specific SBM with network-specific latent variables. Thus, {\em LaPCoM} and \citeauthor{Mantziou:2024:Methods:Competitor} are comparable with a shared focus on interdependence, while the assumptions of \verb|graphclust| differ substantially.

\begin{figure}[!t]
    \centering
    \begin{subfigure}[!t]{\linewidth}
        \centering
        \includegraphics[width=\linewidth]{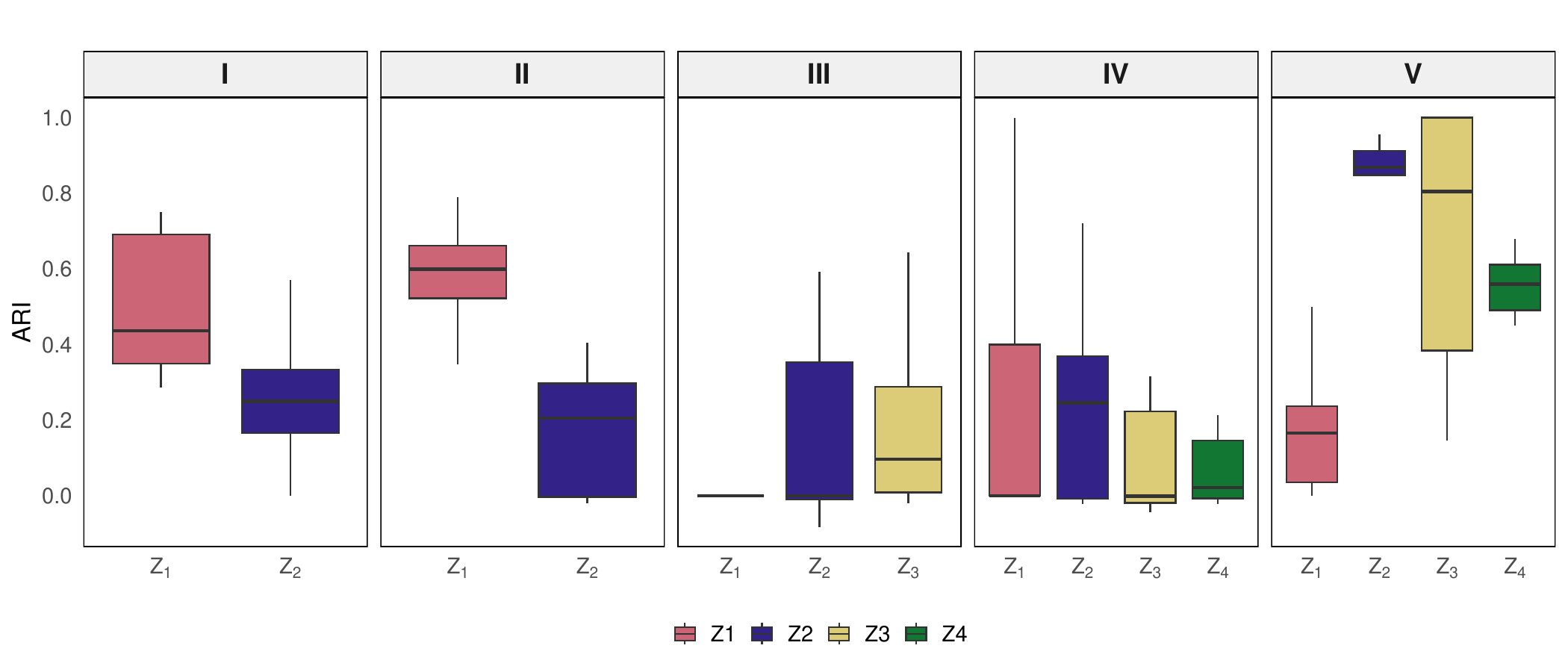}
        \caption{\citeauthor{Mantziou:2024:Methods:Competitor}'s model.}
        \label{fig:SS2_Results_Node_Clustering_Mantziou}
    \end{subfigure}
        
    \begin{subfigure}[!t]{\linewidth}
        \centering
        \includegraphics[width=\linewidth]{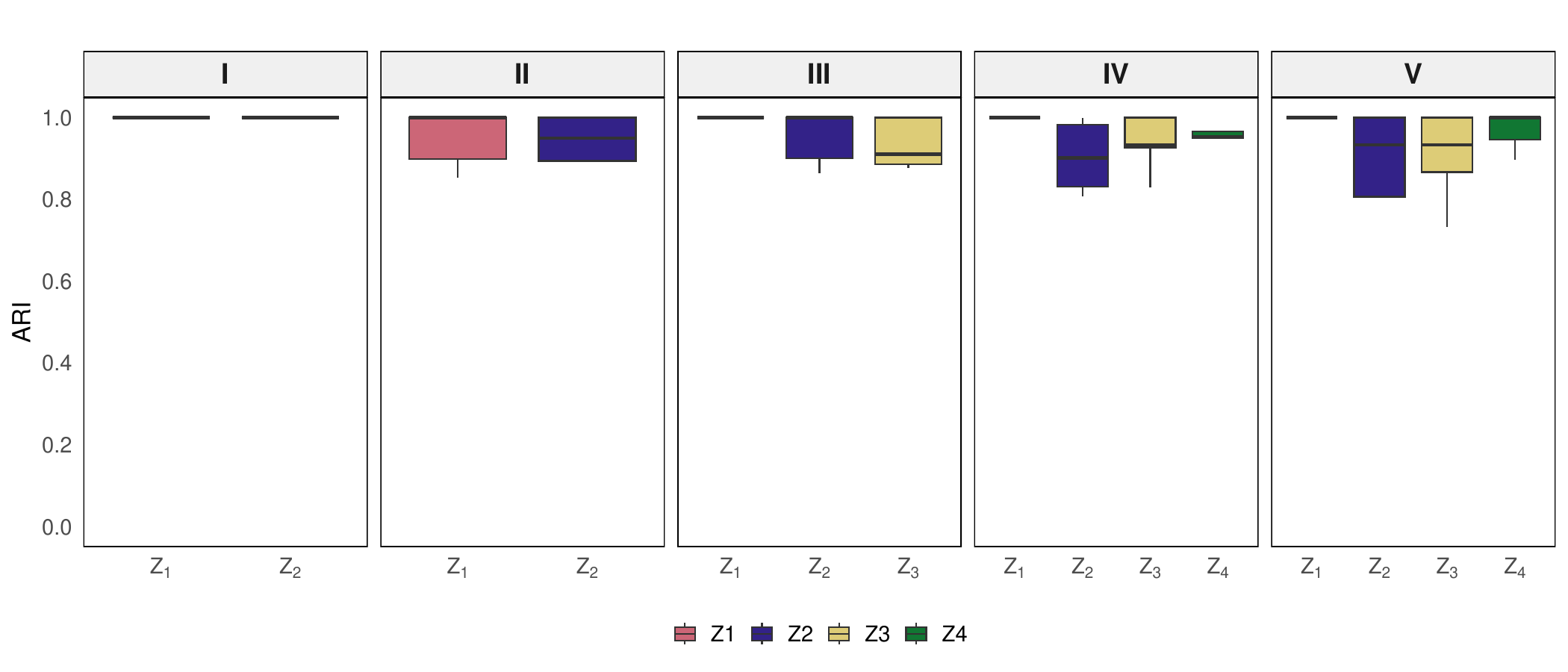}
        \caption{{\em LaPCoM}.}
        \label{fig:SS2_Results_Node_Clustering_LaPCoM}
    \end{subfigure}
    
    \caption{Comparison of nodal clustering performance measured by Adjusted Rand Index (ARI) across 10 simulations for each scenario. Results for (a) \citeauthor{Mantziou:2024:Methods:Competitor}'s model and (b) {\em LaPCoM}.}
    \label{fig:SS2_Results_Node_Clustering_Comparison}
\end{figure}

Figure~\ref{fig:SS2_Results_Node_Clustering_Comparison} compares node clustering performance via ARI across 10 simulations per scenario for \citeauthor{Mantziou:2024:Methods:Competitor}'s method (\ref{fig:SS2_Results_Node_Clustering_Mantziou}) and {\em LaPCoM} (\ref{fig:SS2_Results_Node_Clustering_LaPCoM}). Latent spaces differ across scenarios and are not comparable between panels. For example, $\boldsymbol{Z}_1$ in Scenario I differs from $\boldsymbol{Z}_1$ in Scenario II. \citeauthor{Mantziou:2024:Methods:Competitor}'s method shows poor performance, with median ARIs near 0.00 and an average IQR of 0.25. In contrast, {\em LaPCoM} achieves median ARIs above 0.90 and an average IQR of 0.07, demonstrating superior node-level clustering performance.

Tables~\ref{tab:SS2_Results_G+} and \ref{tab:SS2_Results_Network_Clustering} show that only {\em LaPCoM} and \emph{PopNet} perform competitively. Our {\em LaPCoM} offers several advantages over \emph{PopNet}. The average runtime for {\em LaPCoM} ranged from 3 hours (Scenario I) to 15 hours (Scenario V), compared to 6 and 25 hours respectively for \emph{PopNet}, making {\em LaPCoM} substantially faster. Additionally, although we ran 100,000 MCMC iterations for convergence, fewer iterations may suffice in practice. Moreover, {\em LaPCoM} supports clustering for both binary and weighted multiplexes, while \emph{PopNet} handles only binary data. Additionally, {\em LaPCoM} offers node-level clustering, which \emph{PopNet} lacks. 

\section{Illustrative Applications} \label{Sec_Applications}
This section illustrates the application of {\em LaPCoM} to a variety of multiplex datasets arising in different contexts.

In these illustrative data applications, we run the MCMC algorithm for multiple chains. To reconcile cluster-specific parameters across chains we post-process all chains and identify the overall mode of the number of active clusters, $\hat{G}_+^*$. Only chains matching this mode ($\hat{G}_+^{\text{chain}} = \hat{G}_+^*$) are retained. Among these, we compute the log-posterior value for each chain by evaluating the log-posterior at that chain's parameter estimates. We then select the chain with the highest log-posterior value and use its parameter estimates for all subsequent analyses. Robustness is evaluated by computing the ARI between its cluster allocations and those from the other retained chains.

\subsection{Krackhardt Advice Networks} \label{SubSec_Krackhardt}
We analyse the advice network dataset of \cite{Krackhardt:1987:Data}, consisting of $M = 21$ directed binary networks among $N = 21$ employees at a United States high-tech firm. Each network reflects one employee’s perception of who seeks advice from whom, with $y_{ij}^{(m)} = 1$ indicating that employee $m$ believes employee $i$ seeks advice from employee $j$. While the networks are directed, {\em LaPCoM} does not explicitly model directionality through sender or receiver effects. Instead, it implements node clustering without distinguishing advice-givers from recipients, capturing the overall structure of advice ties rather than their directional nuances. As noted in \cite{SignorelliWit:2020:Methods:Competitor}, some employees may share similar perceptions of the advice network, suggesting clusters of networks with common structure. Our {\em LaPCoM} model captures this by supporting both network- and node-level clustering. 

Scaling factors for the intercept and latent positions are tuned to ensure suitable acceptance rates. To assess robustness, we run 40 MCMC chains: one using the initialisation described in Section~5 of the Supplementary Material, and 39 with random perturbations. Each chain runs for 300,000 iterations, discarding an additional 90,000 as burn-in and thinning every 300\textsuperscript{th} draw, resulting in 1,000 posterior samples per chain. Network-level allocations, $\boldsymbol{C}$, are initialised via $K$-means clustering.

Post-processing (Section~\ref{SubSec_PostProcessing}) revealed that the optimal number of clusters was $\hat{G}_+ = 3$, appearing in 34 of 40 chains. Analyses were restricted to these chains, and the chain with the highest log-posterior value was selected for inference. Its estimated clustering partition matched those from all other retained chains (ARI $= 1$), confirming stability. 

\begin{figure}[!t]
    \centering
    \includegraphics[width = \textwidth]{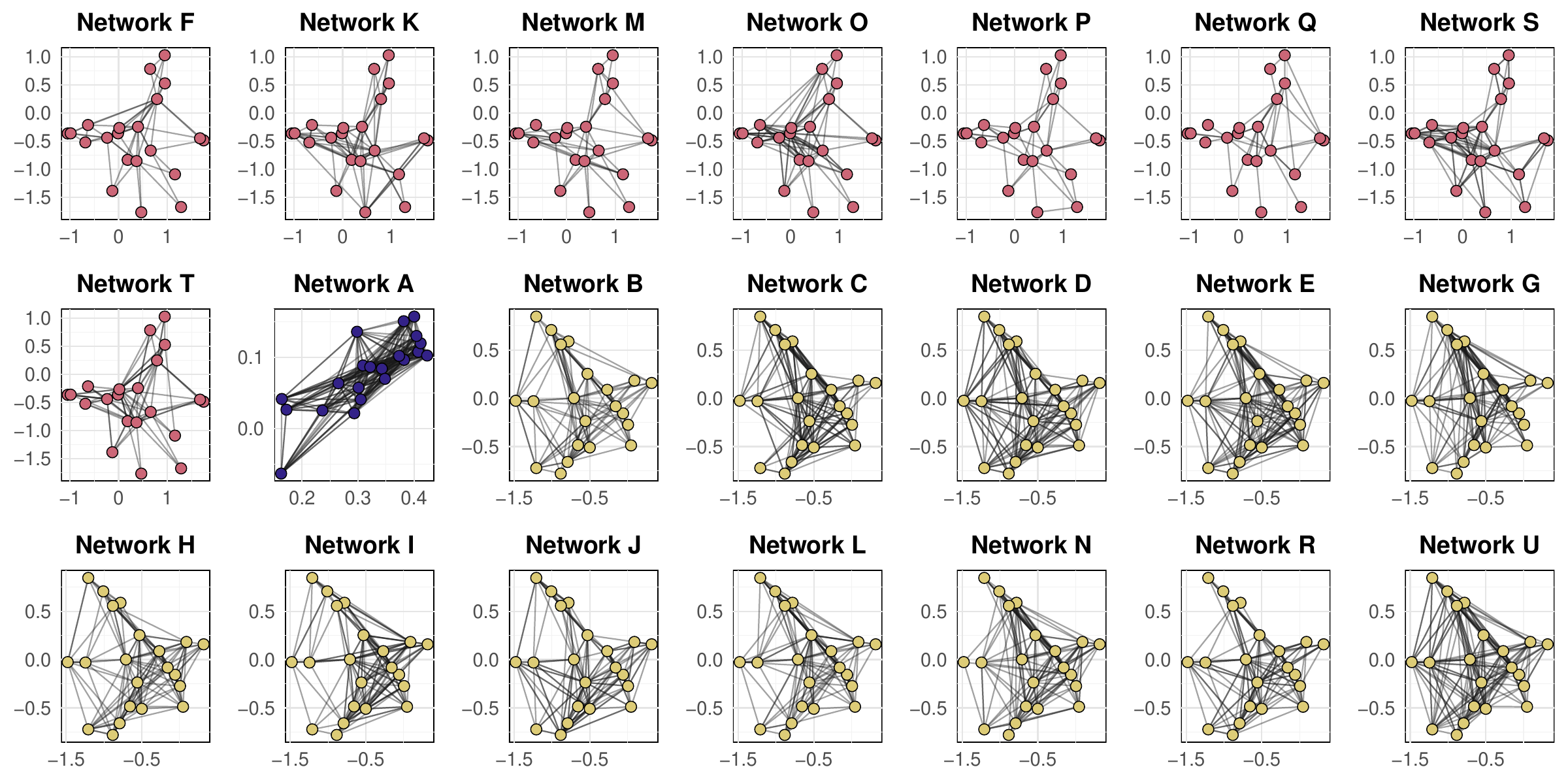}
    \caption{Latent space representation of the Krackhardt advice multiplex obtained from {\em LaPCoM}. The networks are reordered and coloured according to the clusters identified by {\em LaPCoM}.}
    \label{fig:Krack_Clustered_Networks_LS}
\end{figure}

Figure~\ref{fig:Krack_Clustered_Networks_LS} visualises employee positions in the estimated latent spaces by cluster. Edges are projected into the cluster-specific latent space, with proximate nodes indicating a higher likelihood of advice-seeking. Panels are ordered by cluster. The three clusters differed in perceived advice density: Cluster 1 (red, density 0.17) has the most diffuse configuration. Cluster 2 (blue, density 0.66) has a tightly packed latent space, indicating frequent advice-seeking and cluster 3 (yellow, density 0.38) shows moderate compactness. These spatial patterns align with the varying levels of advice-seeking across clusters.

It is worth noting that {\em LaPCoM} did not provide evidence of any node-level clustering within the $\hat{G}_+ = 3$ latent spaces. This outcome appears reasonable, as the networks projected onto their respective latent spaces in Figure~\ref{fig:Krack_Clustered_Networks_LS} shows no indication of distinct node-level groupings.

\subsubsection{Posterior Predictive Checks} \label{SubSubSec_KrackhardtPPCs}
Without a reference partition of the \cite{Krackhardt:1987:Data} dataset, we assess model fit using posterior predictive checks (PPCs), which compare observed and simulated multiplexes to evaluate adequacy of the obtained clustering. Using the final 500 samples of $\alpha$, $\boldsymbol{C}$, and $\{\boldsymbol{Z}_g\}_{g=1}^G$ from the selected chain, we generate 500 simulated multiplexes. Fit is evaluated using AUC of the precision-recall curve, $F_1$-score, density, \cite{Schieber:2017:Extra:Nature:Distance} network distance, and Hamming distance, reflecting the binary nature of the data.

For comparison, we applied the LPCM to each of the 21 networks using the \verb|latentnet| package in \textbf{R} \citep{Krivitsky:2024:Extra:R:latentnet:Manual, Krivitsky:2024:Extra:R:latentnet:Article}, fitting models with one to three node-level clusters. The \verb|ergmm| function was run with 5,000 retained samples, a burn-in of 1,500,000 iterations, and thinning every 50 iterations. The optimal number of clusters per network was selected via the Bayesian information criterion (BIC). Unlike {\em LaPCoM}, which shares three latent spaces across networks for parsimony, \verb|latentnet| fits a separate latent space for each network.

\begin{figure}[!t]
    \centering
    \includegraphics[width = \textwidth]{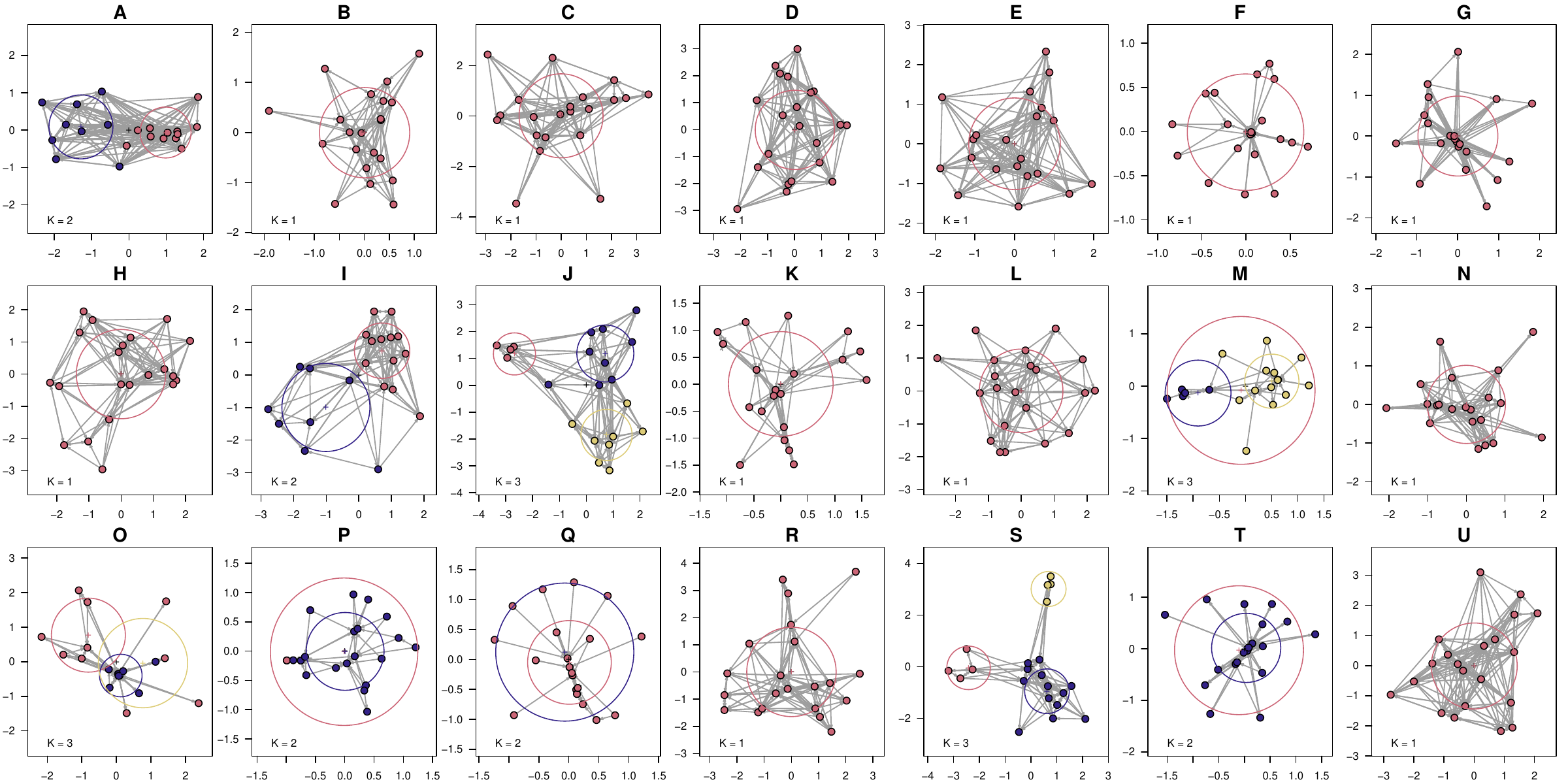}
    \caption{Latent space representation of the Krackhardt advice multiplex obtained from fitting a LPCM with \texttt{latentnet} to each network. The nodes of the networks are coloured according to the node-level clusters identified by \texttt{latentnet}.}
    \label{fig:Krack_PPC_latentnet_LS}
\end{figure}

The \verb|ergmm| function provided evidence for node-level clustering in 9 of the 21 networks: five were best fit by two clusters, four by three, and the remaining 12 by a single component. Figure~\ref{fig:Krack_PPC_latentnet_LS} shows the estimated latent spaces with nodes coloured by cluster using \verb|plot.ergmm|. Some partitions appear spurious. Unlike {\em LaPCoM}, which did not indicate evidence of node-level clustering, \verb|latentnet| fits separate latent spaces per network, allowing it to capture network-specific patterns. In contrast, {\em LaPCoM} uses shared cluster-specific latent spaces across networks, leading to more parsimonious representations.  

\begin{figure}[!t]
    \centering
    \includegraphics[width = \textwidth]{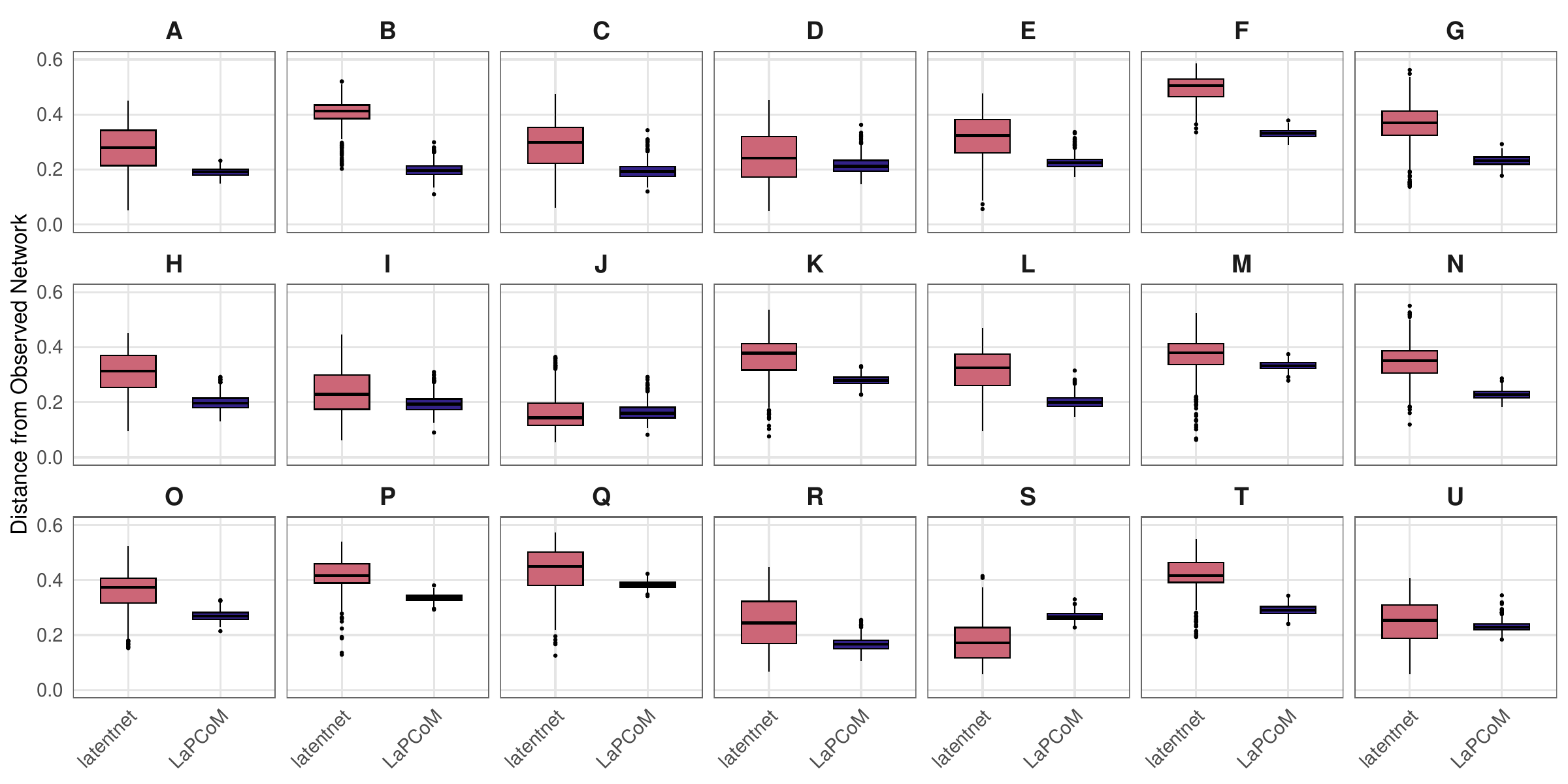}
    \caption{Boxplots showing the distribution of \citeauthor{Schieber:2017:Extra:Nature:Distance} dissimilarities across 500 multiplexes generated from the posterior predictive distribution, using both \texttt{latentnet} and {\em LaPCoM}.}
    \label{fig:Krack_PPC_Network_Distances}
\end{figure}

We report the \cite{Schieber:2017:Extra:Nature:Distance} dissimilarity metric, which ranges from 0 to 1, with lower values indicating better fit. Distances were computed for each of the $M$ networks across 500 simulated multiplexes. Figure~\ref{fig:Krack_PPC_Network_Distances} shows boxplots comparing \verb|latentnet| and {\em LaPCoM}. Posterior simulations for \verb|latentnet| generated networks with no edges in 13 instances. Excluding these cases, the median network distance was 0.33 for \verb|latentnet| and 0.24 for {\em LaPCoM}, with interquartile ranges of 0.10 and 0.03, respectively, indicating that {\em LaPCoM} achieves a better and more precise fit.

The other PPC metrics (AUC, $F_1$-score, density, and Hamming distance) are reported in Section~7 of the Supplementary Material and show similar results for both models, with a slight advantage for {\em LaPCoM}. It is important to note that \verb|latentnet| fits a separate latent space for each network, resulting in a total of 21 latent spaces. Consequently, some network-specific PPC metrics may favour \verb|latentnet| compared to {\em LaPCoM}, which models the entire multiplex using only three latent spaces. Nonetheless, {\em LaPCoM} achieves comparable, and in some cases better, results across PPC metrics while offering a more parsimonious representation of the multiplex.

\subsection{Aarhus Computer Science Department Social Networks} \label{SubSec_Aarhus}
This multiplex network comprises $M = 5$ undirected binary networks representing different types of interactions among $N = 61$ members of Aarhus University’s Department of Computer Science. Member roles include administrative staff, PhD students, postdocs, associate professors, and professors. Each network captures a distinct type of relationships: co-authorship, Facebook friendship, co-participation in leisure activities, having lunch together, and working together. See \cite{Magnani:2013:Aarhus:Data} for further details. Given that the data represent multiple interaction types among the same set of individuals, it is expected that some members will exhibit similar interaction patterns across networks, indicating the presence of network-level clusters. Simultaneously, individuals may form subgroups within each network based on social interactions and roles, suggesting the prescence of node-level clusters. Consequently, {\em LaPCoM} appears to be particularly well suited for modelling these data. 

Scaling factors for the intercept and latent positions were tuned to achieve good acceptance rates. The settings for the number of MCMC chains, iterations, burn-in, thinning, and initialisation of network-level cluster allocations followed those described in Section~\ref{SubSubSec_KrackhardtPPCs}.

The optimal number of network clusters, $\hat{G}_+$, was consistently estimated as 2 across all 40 chains. Two chains returned node-level cluster estimates of $K_{g+} = {6, 5}$; however, the estimate $\hat{K}_{2+} = 5$ was deemed spurious upon further inspection, as the resulting node partitions lacked interpretability. These chains were therefore excluded, leaving 38 for analysis. The chain with the highest log-posterior was selected for analysis of results. In this chain, the Facebook friendship network was assigned to its own cluster, while the remaining four networks were grouped together. Stability was confirmed with an ARI of 1 between the estimated clustering of the selected chain and those of the remaining chains.

\begin{figure}[!t]
    \centering
    \includegraphics[width = \textwidth]{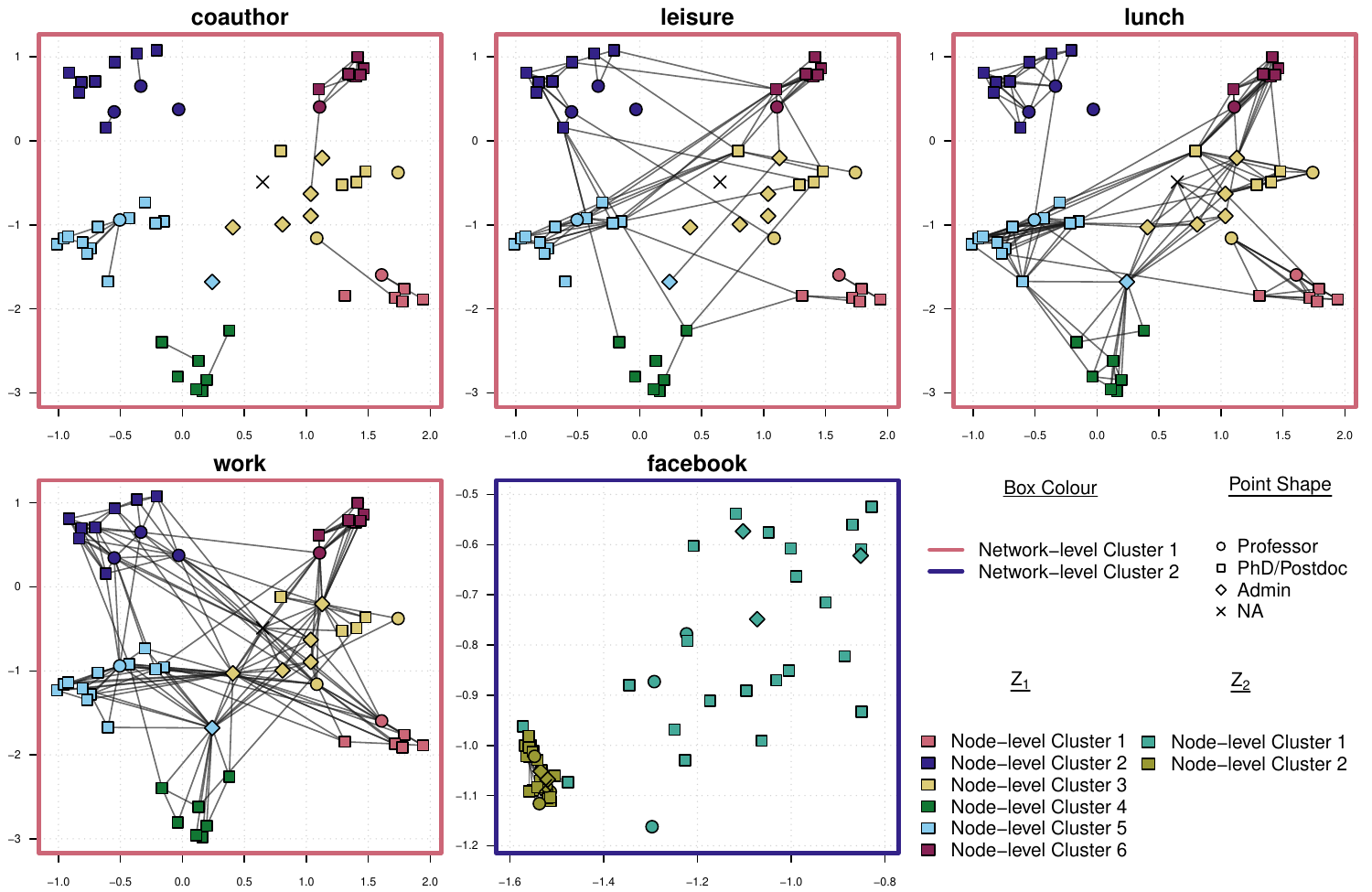}
    \caption{Latent space representation of the Aarhus multiplex from {\em LaPCoM}. Nodes coloured according to their node-level clusters; network-level cluster indicated by the outline colour of the plotting box. Shape of the points indicates the role of the department members.}
    \label{fig:Aarhus_Clustered_Networks_LS_Node_Clusters_Shape_Roles}
\end{figure}

Figure~\ref{fig:Aarhus_Clustered_Networks_LS_Node_Clusters_Shape_Roles} presents the posterior mean latent spaces for the Aarhus multiplex. Network-level clusters are indicated by the colour of the surrounding box, while node-level clusters are represented by the colour of the nodes. For these data, {\em LaPCoM} provided indication of node-level clustering. The posterior mode of the number of node-level clusters, $K_{2+}$, within the latent space of the second network-level cluster (denoted $\boldsymbol{Z}_2$, which corresponds solely to the Facebook network) was estimated as 2. Further examination showed that these two clusters corresponded to connected versus unconnected nodes. However, after applying post-processing (specifically the label-switching correction procedure of \cite{FS:2011:PP:LabelSwitching}) these clusters merged into a single cluster. This merging likely occurs because the unconnected nodes are difficult to position in the latent space, resulting in a high node-cluster variance that supports their merging. Despite this, we consider the $\hat{K}_{2+} = 2$ solution more interpretable, as the unconnected nodes likely represent individuals without Facebook accounts, thus justifying separate clusters. The posterior distribution of $K_{2+}$ exhibited considerable spread, reflecting uncertainty in this partition. In contrast, the latent space of the first network-level cluster, $\boldsymbol{Z}_1$, clearly exhibits six stable node-level clusters, each containing between six and fourteen nodes.

Node-level role data for department members are available, grouped here as administrative staff (6), professors (8, combining associate and full professors), PhD students and postdocs (43), and one unknown. Figure~\ref{fig:Aarhus_Clustered_Networks_LS_Node_Clusters_Shape_Roles} shows the latent spaces with node shapes representing roles. In the second cluster’s latent space, $\boldsymbol{Z}_2$ (Facebook network), no clear role-related patterns appear, suggesting no association between roles and node clusters. In contrast, the first cluster’s latent space, $\boldsymbol{Z}_1$, reveals role-related structure: five of the six node-level clusters include at least one professor alongside PhD students and postdoctoral researchers, reflecting a typical academic group structure with a principal investigator (PI) supervising early-career researchers. Leisure ties mostly form among PhDs and postdocs, consistent with their similar career stage and social activities. In the lunch network, one node-level cluster (green) appears relatively isolated, possibly indicating a separate lunch space, though overall, there is substantial inter-cluster mixing, as expected in a shared departmental environment. The work network shows strong intra- and inter-cluster ties, with professors collaborating among themselves and with supervisees, PhDs and postdocs interacting frequently, and administrative staff centrally positioned, highlighting their key support role.

\subsubsection{Posterior Predictive Checks} \label{SubSubSec_AarhusPPCs}
As in Section~\ref{SubSec_Krackhardt}, model fit is assessed using PPCs. We generate 500 replicate multiplexes from the posterior predictive distribution using the final 500 samples of $\alpha$, $\boldsymbol{C}$, and $\{\boldsymbol{Z}_g\}_{g=1}^G$ from the selected chain. Given the binary networks, we use the same PPC metrics as before: AUC of the precision-recall curve, $F_1$-score, density, network distance, and Hamming distance.

\begin{figure}[!t]
    \centering
    \includegraphics[width = \textwidth]{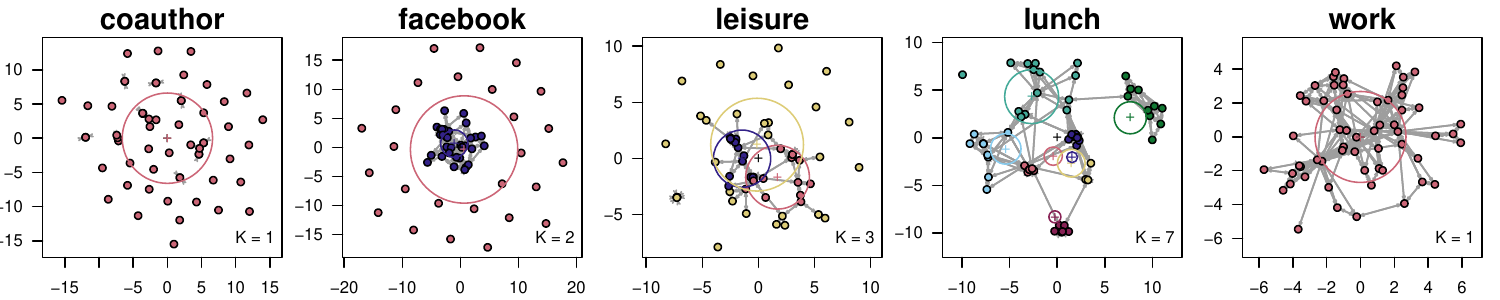}
    \caption{Latent space representation of the Aarhus multiplex obtained from fitting a LPCM with \texttt{latentnet} to each network; nodes coloured according to the node-level clusters.}
    \label{fig:Aarhus_PPC_latentnet_LS}
\end{figure}

To validate our findings, we also fitted LPCMs to each of the $M = 5$ networks individually using the \textsf{R} package \verb|latentnet| \citep{Krivitsky:2024:Extra:R:latentnet:Manual, Krivitsky:2024:Extra:R:latentnet:Article}. Models with 1--7 node-level clusters were estimated via the \verb|ergmm| function, using 5,000 retained samples, a burn-in of 1,500,000 iterations, and thinning every 50 iterations, as previously justified. The BIC guided model selection. We then computed the same five PPC metrics for comparison. The best-fitting models identified one, two, three, or seven clusters across the networks. Figure~\ref{fig:Aarhus_PPC_latentnet_LS} displays the resulting latent spaces, with node colours indicating cluster membership, as visualised using \verb|plot.ergmm|. Notably, the ``lunch'' and ``Facebook'' networks show node-level clusters that closely align with those from {\em LaPCoM}, reinforcing the validity of our model. In contrast, the ``co-authorship'', ``work'', and ``leisure'' networks show more pronounced differences. While both approaches aim to uncover latent clustering, \verb|latentnet| fits a separate latent space to each network. In comparison, {\em LaPCoM} uses only $\hat{G}_+ = 2$ latent spaces to model the association structure in the multiplex, offering a more parsimonious representation.

\begin{figure}[!t]
    \centering
    \includegraphics[width = \textwidth]{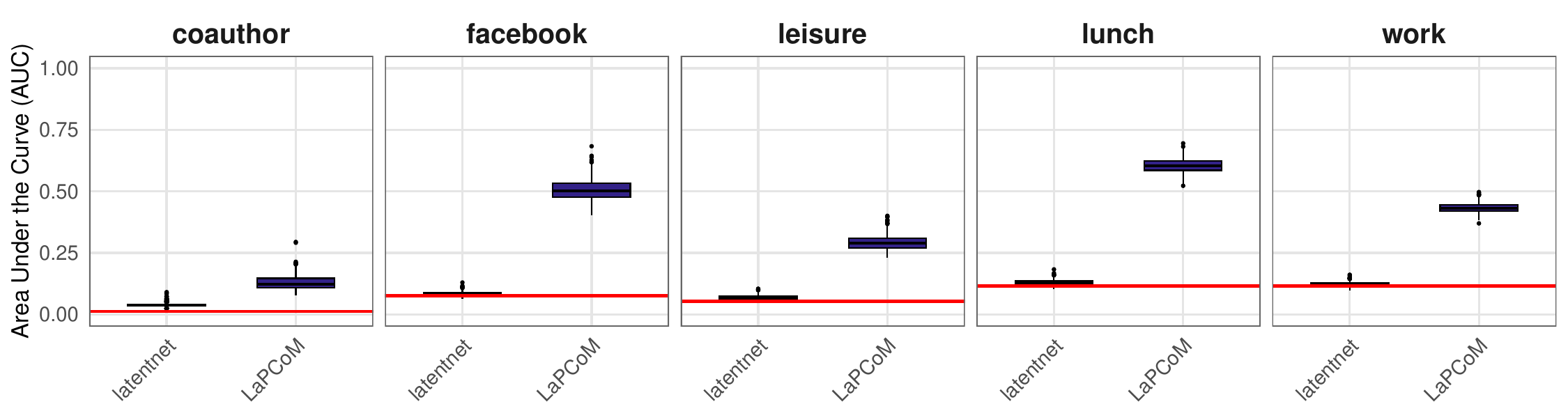}
    \caption{Boxplots showing the distribution of area under the curve (AUC) values across 500 multiplexes generated from the posterior predictive distribution, using both \texttt{latentnet} (pink) and {\em LaPCoM} (blue). Observed network density marked by a red horizontal line.}
    \label{fig:Aarhus_PPC_AUC}
\end{figure}

Figure~\ref{fig:Aarhus_PPC_AUC} shows boxplots of AUC values from 500 posterior predictive multiplexes for both \verb|latentnet| and {\em LaPCoM} at the network level. The observed network density is marked by a red horizontal line in each panel. AUC values above this line indicate good fit. For \verb|latentnet|, median AUCs generally align with the density, suggesting average fit. In contrast, {\em LaPCoM} consistently has all values, including the lower whisker, above the density line, indicating a better fit across networks.

The other four PPC metrics are reported in Section~8 of the Supplementary Material. Two show similar performance for both models, while the other two slightly favour {\em LaPCoM}. Overall, these results indicate that {\em LaPCoM} provides a more accurate and parsimonious fit than \verb|latentnet|.

\subsection{Primary School Face-to-face Interaction Networks} \label{SubSec_PSC}
This multiplex captures the temporal network of face-to-face interactions between students and teachers at a primary school in France. Originally collected for an epidemiological study \citep{Gemmetto:2014:App:PSC}, it has since been used in network analysis research \citep{Stehle:2011:App:PSC}. The dataset includes $N_S = 232$ students and $N_T = 10$ teachers over two school days (Thursday, October 1st, and Friday, October 2nd, 2009), spanning five grades divided into 10 classes. For analysis, we construct a multiplex in which each layer corresponds to a one hour snapshot of interactions. 
While {\em LaPCoM} does not directly model temporal dependence, the shared latent spaces at the network-level are capable of capturing similar features across the one-hour slots. We represent the data as a multiplex with $M = 16$ layers, each corresponding to an hour from 9 AM to 5 PM on each day. Nodes represent students and teachers; edges $y_{ij}^{(m)}$ count interactions between individuals $i$ and $j$ during hour $m$.

We apply our mixture of mixtures model, {\em LaPCoM}, to identify clusters of networks and nodes in this dataset. As before, scaling factors for the intercept and latent positions are tuned to ensure appropriate acceptance rates. Note that due to the large number of nodes in this application, we increased the value of $n_{\min}$ used to determine $K_{\max}$ to $n_{\min} = 25$. To assess robustness and stability, we run 20 independent MCMC chains. One chain is initialised as described in Section~5 of the Supplementary Material; the other 19 use the same method with added noise. Each chain runs for 250,000 iterations, discarding an additional 75,000 as burn-in, and is thinned by keeping every 250\textsuperscript{th} iteration, yielding 1,000 posterior samples per chain.

We initialise the network-level allocations, $\boldsymbol{C}$, using model-based clustering via the \verb|mclust| package with the flexible VVV model, yielding $G_0 = 3$. This replaces the previously used  $K$-means initialisation, which often led to poor posterior exploration.A sensitivity analysis varying $G_0 \in \{1,\dots,8\}$ and the $\mathcal{BNB}$ hyperparameters showed that $K$-means frequently produced singleton clusters for larger values of $G_0$. Based on these findings, the \verb|mclust| initialisation was used for final model fitting.

The mode of the optimal number of clusters, $\hat{G}_+ = 2$ across all 20 chains. Three chains failed the permutation test at the node level and were discarded, leaving 17 chains for analysis. The chain with the highest log-posterior value was selected for inference. Comparison with the remaining chains showed perfect agreement (ARI $= 1$), confirming the stability of the solution.

\begin{figure}[!t]
    \centering
    \includegraphics[width = \textwidth]{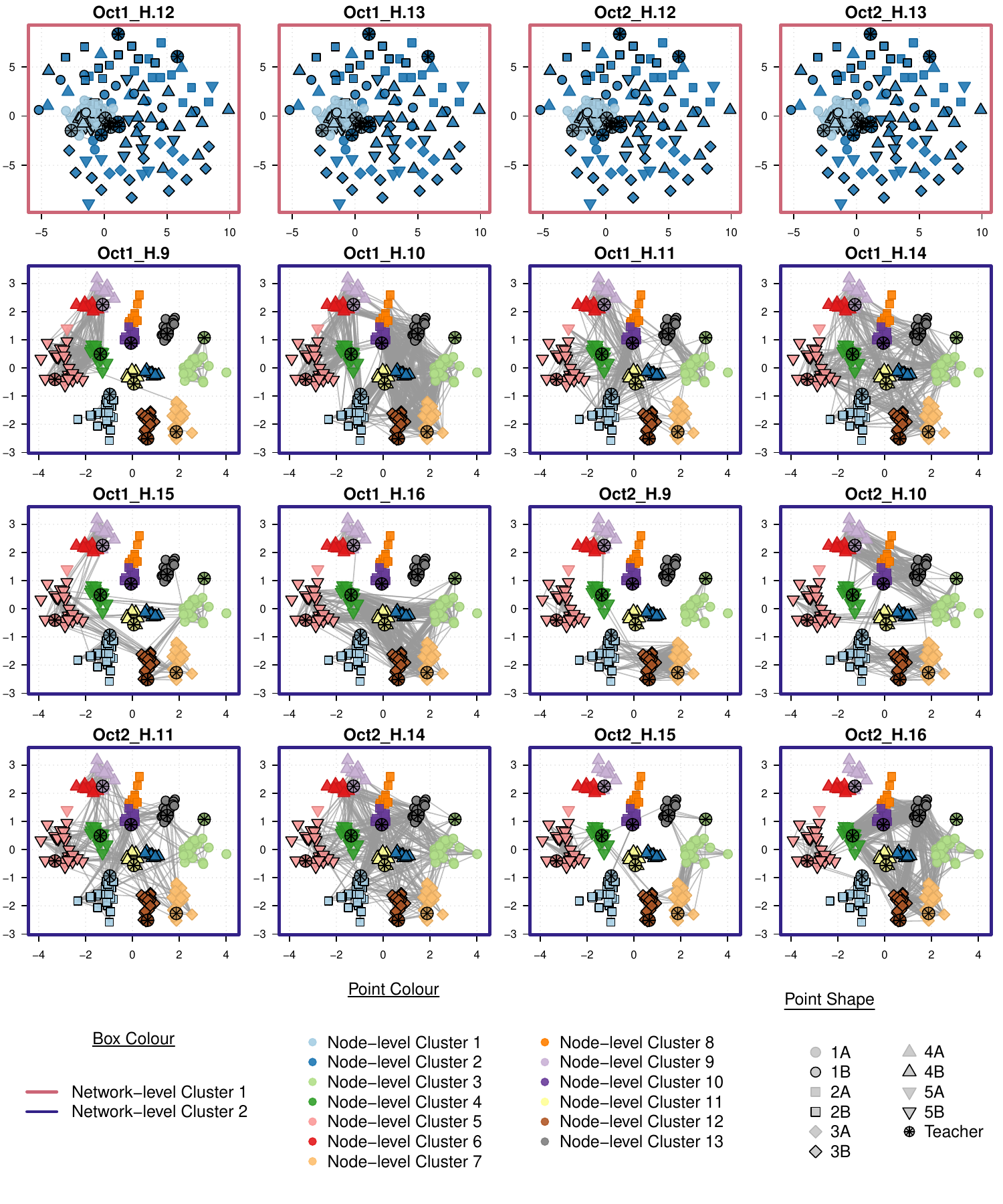}
    \caption{Latent space representations of the networks. Point colours indicate the class membership of each student or whether the individual is a teacher. The outline colour of each plotting panel denotes the network-level cluster assignment.}
    \label{fig:PSC_Clustered_Networks_LS_Class}
\end{figure}

Figure~\ref{fig:PSC_Clustered_Networks_LS_Class} shows the estimated latent spaces for the two inferred network-level clusters, with each network’s connections projected onto the shared space. Networks are indicated using the convention \textsf{OctD\_H.h}, where \textsf{D} is the day (1 or 2) and \textsf{h} the hour (9 to 16). As described previously, node-level metadata includes each individual’s class designation (e.g., 1A, 1B, $\dots$, 5A, 5B) or teacher status. The inferred network-level partition groups networks \textsf{Oct1\_H.12}, \textsf{Oct1\_H.13}, \textsf{Oct2\_H.12}, and \textsf{Oct2\_H.13} together, with the remaining 12 networks in a second cluster. In the latent space of the first network-level cluster, $\hat{\boldsymbol{Z}}_1$, the model found evidence of two node-level clusters: one comprising the large connected component of the network, and the other consisting of the remaining disconnected nodes. In the latent space of the second network-level cluster, $\hat{\boldsymbol{Z}}_2$, the model inferred 13 node-level clusters, which align strongly with class/status labels, yielding an ARI of 0.86.

The clustering and low-dimensional representation aligns well with key aspects of the social interactions in the data. During lessons, students interact mostly within their class, producing distinct class-based patterns. Around lunchtime (12:00–-14:00), cross-class interactions increase, reflecting freer movement in shared areas like the playground, a feature well-captured by the network-level clustering. Node-level clustering also reflects this structure: most links occur within classes, with some limited inter-class ties, consistent with occasional joint activities.

\subsubsection{Posterior Predictive Checks} \label{SubSubSec_PSC_PPCs}
As in Section~\ref{SubSec_Krackhardt} and Section~\ref{SubSec_Aarhus}, model fit is assessed using posterior predictive checks (PPCs) based on the final 500 samples from the posterior predictive distribution from the selected chain, in order to generate simulated multiplexes under the fitted model. Since this dataset comprises count-valued networks, unlike the binary-valued networks in the previous examples, we adopt alternative PPC metrics. Specifically, to compare observed and predicted networks, we evaluate: (i) the mean absolute difference in edge counts, (ii) the network distance proposed by \cite{Schieber:2017:Extra:Nature:Distance}, (iii) the true negative rate, and (iv) the log-empirical cumulative distribution function (ECDF) of counts greater than zero.

In previous applications, we used \verb|latentnet| for comparison. However, due to the size of the networks in this multiplex ($N = 242$), it was computationally infeasible to run \verb|latentnet| across all 16 networks and a range of cluster values. The required runtime for networks of this size was prohibitive, and thus \verb|latentnet| was not used in this analysis.

Table~\ref{tab:PSC_PPCs} presents the mean values (standard deviations, SD, in brackets) for three key PPC metrics across all networks: mean absolute difference in counts (MAD), \citet{Schieber:2017:Extra:Nature:Distance} network distance, and true negative rate (TNR). The overall average MAD is 0.4977 (SD = 0.0032), indicating that predicted edge counts deviate by less than one count from observed values on average, suggesting a good fit. The average network distance is 0.4423 (SD = 0.0033), which, given the metric is bounded between 0 and 1, reflects a reasonable structural similarity between the observed and predicted networks. The average TNR of 0.8746 (SD = 0.0013) demonstrates that the model is conservative in predicting edges, favouring the correct identification of non-edges and avoiding the introduction of spurious edges. 

Additional posterior predictive checks based on the ECDFs of positive counts (see Section~9 of the Supplementary Material) show that the model struggles to capture the distribution of non-zero edge counts, particularly underestimating large values. This suggests that the Poisson distribution may be too restrictive. Nonetheless, the model captures the overall network structure well and recovers clusters that align with students' class memberships and time-of-day patterns.

\begin{table*}[!bt]
\centering
\caption{Mean (standard deviation) values of the mean absolute difference (MAD) in counts, network distance, and true negative rate (TNR) across all networks.}
\label{tab:PSC_PPCs}
\begin{tabular}{@{}lccc@{}}
    \hline
    \multicolumn{1}{l}{Network} & \multicolumn{1}{c}{MAD} & \multicolumn{1}{c}{Distance} & \multicolumn{1}{c}{TNR} \\
    \hline

    \textsf{Oct1\_H.9} & 0.4438 (0.0032) & 0.4401 (0.0029) & 0.8744 (0.0013) \\ 
    \textsf{Oct1\_H.10} & 0.5888 (0.0032) & 0.3396 (0.0043) & 0.8749 (0.0014) \\ 
    \textsf{Oct1\_H.11} & 0.4654 (0.0031) & 0.4158 (0.0037) & 0.8749 (0.0014) \\ 
    \textsf{Oct1\_H.12} & 0.5990 (0.0032) & 0.4814 (0.0022) & 0.8741 (0.0014) \\ 
    \textsf{Oct1\_H.13} & 0.4938 (0.0032) & 0.4928 (0.0022) & 0.8743 (0.0013) \\ 
    \textsf{Oct1\_H.14} & 0.4059 (0.0033) & 0.4165 (0.0040) & 0.8749 (0.0013) \\ 
    \textsf{Oct1\_H.15} & 0.4510 (0.0033) & 0.4833 (0.0031) & 0.8748 (0.0013) \\ 
    \textsf{Oct1\_H.16} & 0.4732 (0.0032) & 0.4220 (0.0036) & 0.8744 (0.0014) \\ 
    \textsf{Oct2\_H.9} & 0.4685 (0.0033) & 0.4616 (0.0035) & 0.8744 (0.0014) \\ 
    \textsf{Oct2\_H.10} & 0.5660 (0.0031) & 0.4090 (0.0038) & 0.8750 (0.0014) \\ 
    \textsf{Oct2\_H.11} & 0.4913 (0.0033) & 0.4035 (0.0039) & 0.8749 (0.0013) \\ 
    \textsf{Oct2\_H.12} & 0.6079 (0.0033) & 0.4822 (0.0023) & 0.8739 (0.0012) \\ 
    \textsf{Oct2\_H.13} & 0.5465 (0.0034) & 0.4961 (0.0025) & 0.8742 (0.0012) \\ 
    \textsf{Oct2\_H.14} & 0.4204 (0.0033) & 0.4154 (0.0040) & 0.8744 (0.0013) \\ 
    \textsf{Oct2\_H.15} & 0.4497 (0.0033) & 0.4900 (0.0032) & 0.8748 (0.0014) \\ 
    \textsf{Oct2\_H.16} & 0.4926 (0.0032) & 0.4274 (0.0033) & 0.8746 (0.0013) \\ 
    \hline
\end{tabular}
\end{table*}

\section{Discussion} \label{Sec_Discussion}
The proposed latent position co-clustering model ({\em LaPCoM}) introduces a novel framework for jointly clustering networks within a multiplex alongside their constituent nodes. By leveraging a mixture-of-mixtures model, {\em LaPCoM} simultaneously achieves dimension reduction and two-level clustering of a multiplex network. The model is formulated under a Bayesian nonparametric framework using a mixture of finite mixtures (MFM), which places priors on the number of components at both the network and node levels. A sparse prior on the mixing proportions encourages the discarding of empty components and enables data-driven identification of the number of clusters. The performance of {\em LaPCoM} was evaluated through simulation studies and illustrative examples, where it successfully recovered interpretable clustering structures, demonstrating its practical utility across diverse multiplex datasets.

For interpretability and ease of visualisation, we fixed the dimension of the latent spaces to two in all analyses. However, this constraint may restrict the model's capacity to capture higher-dimensional latent structures in some data. A potential avenue of future work would be to introduce automatic latent dimension selection using BNP priors. Examples include the multiplicative gamma process shrinkage prior \citep{BhattacharyaDunson:2011:MGP}, which has shown success in similar contexts \cite{Gwee:2023:LSPCM}, or the cumulative shrinkage process prior \citep{Legramanti:2020:CUSP}, which has been extended to various factor analysis models \cite{KowalCanale:2023:CUSP_Ext, FS:2023:GCUSP}.

In Section~\ref{SubSec_PSC}, we applied {\em LaPCoM} to a time-varying multiplex. However, the model does not explicitly account for temporal dependencies between networks. This is a potential limitation, and a natural extension would involve the incorporation of a Markov process into the prior of the latent positions, an approach successfully employed in dynamic network modelling \citep{SewellChen:2016:DynamicMultiplex, SewellChen:2017:DynamicMultiplex}.

Future work could also explore more flexible modelling of edge formation in multiplex networks. The current model does not account for edge directionality. Incorporating sender-receiver effects into the latent space, as proposed by \citet{LiuLiu:2025:Directionality}, offers a valuable extension. Additionally, the application in Section~\ref{SubSec_PSC} suggests that the Poisson distribution may not adequately capture the observed edge count distribution. While clustering remains the primary goal of {\em LaPCoM}, adopting a zero-inflated Poisson model \citep{Lu:2025:ZeroInflatedPoissonLPCM}, or other flexible alternatives, could better accommodate under- or overdispersion and improve fit for count-valued multiplex data.

In summary, while {\em LaPCoM} offers a flexible and interpretable framework for the co-clustering of multiplex networks, these directions highlight potential methodological enhancements. Addressing these limitations would further improve the model's applicability across a broader range of multiplex network settings.

\section*{Acknowledgements}
This publication has emanated from research conducted with the financial support of Taighde Éireann (Research Ireland) under grant number 18/CRT/6049. For the purpose of Open Access, the author has applied a CC BY public copyright licence to any Author Accepted Manuscript version arising from this submission.


\clearpage
\appendix

\section*{Supplementary Material for ``A Latent Position Co-Clustering Model for Multiplex Networks"}

In this document, we provide supplementary material to the article 
\emph{``A Latent Position Co-Clustering Model for Multiplex Networks".} 

\begin{itemize}
    \item \textbf{Section~\ref{SuppMat:Notation}:} A comprehensive list of all notation used in the main article.
    \item \textbf{Section~\ref{SuppMat:HyperparameterChoices}:} The rationale behind the choice of hyperparameters in the prior distributions.
    \item \textbf{Section~\ref{SuppMat:FCs}:} Derivations of the full conditional distributions of the parameters.
    \item \textbf{Section~\ref{SuppMat:MCMC_Algorithm}:} Pseudocode for the MCMC algorithm discussed in Section 3 of the main article.
    \item \textbf{Section~\ref{SuppMat:Initialisation}:} Initialisation details for the MCMC algorithm discussed in Section 3 of the main article.
    \item \textbf{Section~\ref{SuppMat:SimStudies}:} Design details of the simulation studies discussed in Section 4 of the main article.
    \item \textbf{Section~\ref{SuppMat:PPCs:Krackhardt}:} Additional posterior predictive checks related to the Krackhardt \citep{Krackhardt:1987:Data} application discussed in Section 5.1 of the main article.
    \item \textbf{Section~\ref{SuppMat:PPCs:Aarhus}:} Additional posterior predictive checks related to the Aarhus \citep{Magnani:2013:Aarhus:Data} application discussed in Section 5.2 of the main article.
    \item \textbf{Section~\ref{SuppMat:PPCs:PSC}:} Additional posterior predictive checks related to the primary school \citep{Gemmetto:2014:App:PSC} application discussed in Section 5.3 of the main article.
    \item \textbf{Section~\ref{SuppMat:MFMvsOM}:} Comparison between a mixture of finite mixtures (MFM; \cite{FS:2021:Methods:MFM_TS}) model and an overfitted mixture model.
\end{itemize}

\clearpage
\section{Notation} \label{SuppMat:Notation}
$\mathcal{Y}$: The collection of $M$ networks.  \\
$\boldsymbol{Y}^{(m)}$: The $N \times N$ network adjacency matrix.   \\
$y_{ij}^{(m)}$: The value of the edge between nodes $i$ and $j$ in network $\boldsymbol{Y}^{(m)}$.  \\
$M$: Number of networks.    \\
$N$: Number of nodes.   \\
$\alpha$: The intercept parameter of the latent position model (LPM) that describes the overall level of connectivity in the network.   \\
$\delta_{\alpha}$: The scaling factor used in the standard deviation of the proposal distribution for the intercept parameter, $\alpha$.  \\
$G$: The number of network-level mixture components.    \\
$G_+$: The number of network-level clusters (active/non-empty mixture components).  \\
$\mathcal{D}$: Denotes the Dirichlet distribution.     \\
$e$: The Dirichlet concentration hyperparameter in the network-level mixing proportions prior distribution.    \\
$\boldsymbol{\tau}$: The network-level mixing proportions.    \\
$\tau_g$: The probability that a network belongs to network-level component $g$.   \\
$\boldsymbol{C}$: The $M \times G$ binary matrix indicating the membership of networks to network-level components.   \\
$\boldsymbol{C}_g$: The binary vector of length $M$ indicating which networks belong to network-level component $g$.  \\
$C_g^{(m)}$: The binary indicator of membership of network $\boldsymbol{Y}^{(m)}$ to network-level component $g$.    \\
$K_g$: The number of node-level mixture components within the $g$\textsuperscript{th} latent space.  \\
$K_{g+}$: The number of node-level clusters (active/non-empty mixture components) within the $g$\textsuperscript{th} latent space.   \\
$w_g$: The Dirichlet concentration hyperparameter in the node-level mixing proportions prior distribution.   \\
$\boldsymbol{\pi}_g$: The node-level mixing proportions within the $g$\textsuperscript{th} latent space.   \\
$\pi_{gk}$: The probability that a node belongs to node-level component $k$ within the $g$\textsuperscript{th} latent space.   \\
$\boldsymbol{S}_g$: The $N \times K_g$ binary matrix indicating the membership of nodes to node-level components within the $g$\textsuperscript{th} latent space.    \\
$\boldsymbol{S}_{gk}$: The binary vector of length $N$ indicating which nodes belong to node-level component $k$ within the $g$\textsuperscript{th} latent space.   \\
$S_{gk}^{(i)}$: The binary indicator of membership of node $i$ to network-level component $k$ within the $g$\textsuperscript{th} latent space. \\
$\mathcal{MVN}_p$: Denotes the multivariate Normal distribution of dimension $p$.      \\
$\boldsymbol{\mu}_{gk}$: The (2-dimensional) mean vector of the multivariate Normal distribution describing the $k$\textsuperscript{th} node-level component within the $g$\textsuperscript{th} latent space.    \\
$\boldsymbol{\Sigma}_{gk}$: The (diagonal) variance-covariance matrix of the multivariate Normal distribution describing the $k$\textsuperscript{th} node-level component within the $g$\textsuperscript{th} latent space.  \\
$\sigma_{gk,q}^2$: The covariance parameter of the multivariate Normal distribution in the $q$\textsuperscript{th} dimension, describing the $k$\textsuperscript{th} node-level component within the $g$\textsuperscript{th} latent space.  \\
$\boldsymbol{Z}_g$: The $N \times 2$ matrix of latent positions of the $g$\textsuperscript{th} network-level component. \\
$\boldsymbol{z}_{g, i}$: The (2-dimensional) latent position of node $i$ in the latent space of the $g$\textsuperscript{th} network-level component.    \\
$\delta_{Z}$: The scaling factor used in the standard deviation of the proposal distribution for the latent positions.  \\
$\boldsymbol{H}$: A collective term for the hyperparameters of the model. \\
$G_0$:  The initial number of network-level mixture components. \\
$G_{\text{max}}$:  The maximum number of network-level mixture components considered. \\
$\mathcal{N}$: Denotes the univariate Normal distribution.  \\
$m_{\alpha}$: The mean of the Normal prior on the intercept parameter, $\alpha$.   \\
$s_{\alpha}$: The standard deviation of the Normal prior on the intercept parameter, $\alpha$.  \\
$\mathcal{BNB}$: Denotes the Beta-Negative-Binomial distribution.   \\
$a_G$:  The ``number of successes until the experiment is stopped'' parameter of the $\mathcal{BNB}$ prior distribution on the number of network-level mixture components. \\
$b_G$: The first shape parameter of the $\mathcal{BNB}$ prior distribution on the number of network-level mixture components.  \\
$c_G$: The second shape parameter of the $\mathcal{BNB}$ prior distribution on the number of network-level mixture components. \\
$\mathcal{F}$: Denotes the Fisher-Snedecor distribution.    \\
$l_G$: The first degrees of freedom parameter of the $\mathcal{F}$ prior distribution on the Dirichlet concentration hyperparameter of the network-level mixing proportions prior distribution. \\
$r_G$: The second degrees of freedom parameter of the $\mathcal{F}$ prior distribution on the Dirichlet concentration hyperparameter of the network-level mixing proportions prior distribution. \\
$s_e$: The standard deviation of the proposal distribution for the Dirichlet concentration hyperparameter of the network-level mixing proportions prior distribution. \\
$K_0$: The initial number of node-level mixture components. \\
$K_{\text{max}}$: The maximum number of node-level mixture components considered. \\
$a_K$: The ``number of successes until the experiment is stopped'' parameter of the $\mathcal{BNB}$ prior distribution on the number of node-level mixture components. \\
$b_K$: The first shape parameter of the $\mathcal{BNB}$ prior distribution on the number of node-level mixture components. \\
$c_K$: The second shape parameter of the $\mathcal{BNB}$ prior distribution on the number of node-level mixture components. \\
$l_K$: The first degrees of freedom parameter of the $\mathcal{F}$ prior distribution on the Dirichlet concentration hyperparameter of the node-level mixing proportions prior distribution. \\
$r_K$: The first degrees of freedom parameter of the $\mathcal{F}$ prior distribution on the Dirichlet concentration hyperparameter of the node-level mixing proportions prior distribution. \\
$s_w$: The standard deviation of the proposal distribution for the Dirichlet concentration hyperparameter of the node-level mixing proportions prior distribution. \\
$\mathcal{G}$: Denotes the Gamma distribution.  \\
$\mathcal{IG}$: Denotes the Inverse-Gamma distribution.     \\
$u_{\sigma^2}$: The shape parameter of the $\mathcal{IG}$ prior distribution for any $\sigma_{gkq}^2, \ q = 1, 2$. \\
$v_{\sigma^2}$: The scale parameter of the $\mathcal{IG}$ prior distribution for any $\sigma_{gkq}^2, \ q = 1, 2$. \\
$T_S$: The total number of posterior samples after thinning and burn-in of the resulting MCMC chain.    \\
$T_B$:  The number of burn-in iterations to be discarded from the resulting MCMC chain. \\
$T_{\text{thin}}$: The number of iterations to thin the resulting MCMC chain by. \\
$T$: The total number of iterations the MCMC algorithm runs for, $T = (T_S * T_T) + T_B$.
$B$: Beta function.\\
$\Gamma$: Gamma function.

\clearpage
\section{Hyperparameter Choices} \label{SuppMat:HyperparameterChoices}
The choice of a $\mathcal{N}(0,1)$ prior for the intercept parameter $\alpha$ is motivated by the role of the intercept in latent position models (LPMs), where it represents the baseline propensity for edge formation across the entire network. Centring the prior around zero allows the data to primarily influence the intercept estimation, while a prior standard deviation of 1 maintains a relatively uninformative stance on $\alpha$ in the absence of strong prior knowledge.

The $\mathcal{MVN}(\boldsymbol{0},\boldsymbol{\mathbb{I}}_2)$ prior for the node-level cluster means $\boldsymbol{\mu}_{gk}$ was chosen to provide a non-informative basis, letting the data drive the estimation of cluster means in the latent spaces. Setting a mean of zero is appropriate, as the latent space is translation-invariant, while a diagonal covariance matrix respects the rotation and scaling invariance within the space.

The choice of hyperparameters $u_{\sigma^2}$ and $v_{\sigma^2}$ in the $\mathcal{IG}(u_{\sigma^2}, v_{\sigma^2})$ prior on the node-level cluster variances $\sigma_{gk,q}^2$ is guided by the need to control the spatial dispersion of nodes within each cluster in the latent space $\boldsymbol{Z}_g$. These parameters influence the expected within-cluster variance, ensuring that node positions remain sufficiently concentrated in a manner consistent with the assumptions of the latent position cluster model (LPCM), where clusters correspond to nodes located close together in the latent space with higher probabilities of connection. Given that latent positions drawn from $\mathcal{MVN}_2(\boldsymbol{0}, \boldsymbol{\mathbb{I}}_2)$ typically fall within the range $(-3, 3)$, the scale of the latent space implies that clusters should become increasingly tight as the network size $N$ grows to preserve interpretability. In networks with fewer than 60 nodes, it is reasonable to assume that a typical cluster contains at least $n_{\text{min}} = 5$ nodes, while for networks with $N \geq 60$, an average of at least $n_{\text{min}} = 10$ nodes per cluster is assumed. Accordingly, the expected variance $\mathbb{V}(N)$ within a cluster $k$ is defined as $\mathbb{V}(N) = \frac{1}{n_{\text{min}}\big(1 - \frac{n_{\text{min}}}{N}\big)}$, where $n_{\text{min}} = 5$ if $N < 60$, and $n_{\text{min}} = 5$ if $N \geq 60$. 

This function yields a variance around 0.2 for $N \in \{20, 30, 40, 50\}$ and approximately 0.1 for $N \in \{60, 70, 80, 90, 100\}$. To achieve these expected values, we place an $\mathcal{IG}(u_{\sigma^2},v_{\sigma^2})$ prior on $\sigma_{gk,q}^2$, where $\mathbb{E}\big[\sigma_{gk,q}^2\big] = \frac{v_{\sigma^2}}{u_{\sigma^2} - 1}$ (for $u_{\sigma^2} > 1$) and $\mathbb{V}\big[\sigma_{gk,q}^2\big] = \frac{v_{\sigma^2}^2}{\big(u_{\sigma^2} - 1\big)^2\big(u_{\sigma^2} - 2\big)}$ (for $u_{\sigma^2} > 2$). With $v_{\sigma^2} = 2$, the expected value of $\sigma_{gk,q}^2$ is 0.2 when $u_{\sigma^2} = 11$ (with variance 0.004) and 0.1 when $u_{\sigma^2} = 21$ (with variance 0.001). Thus, the priors $\mathcal{IG}(u_{\sigma^2} = 11, v_{\sigma^2} = 2)$ and $\mathcal{IG}(u_{\sigma^2} = 21, v_{\sigma^2} = 2)$ are reasonable choices for networks with $20 \leq N < 60$ and $60 \leq N \leq 100$, respectively, providing sufficient flexibility for variance while concentrating the probability mass close to smaller values.

\begin{figure}[!t]
    \centering
    \begin{subfigure}[t]{0.3\textwidth}
        \centering
        \includegraphics[width=\textwidth]{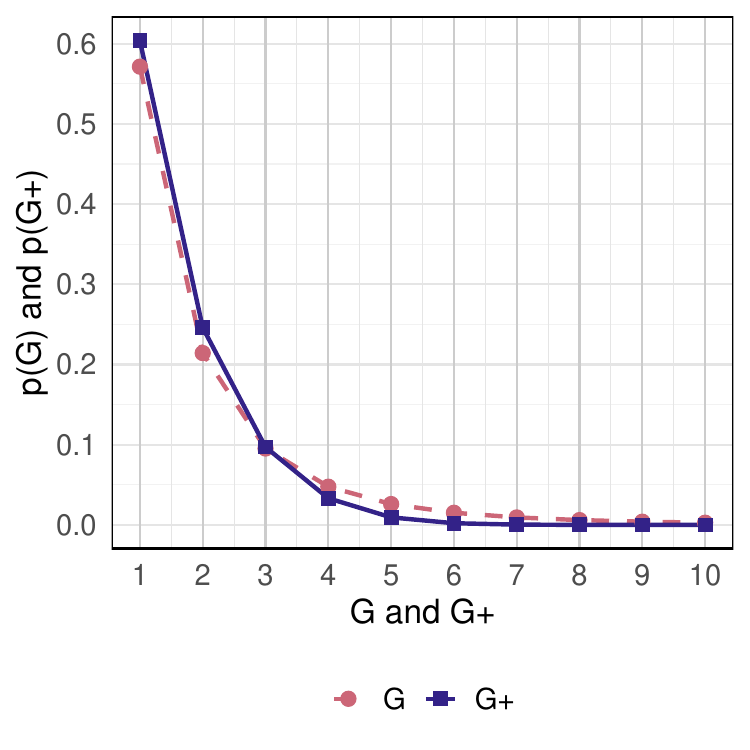}
        \caption{Comparison of the prior on $G$ (red) under a $\mathcal{BNB}(1, 4, 3)$ prior and the corresponding induced prior on $G_+$ (blue).}
        \label{fig:Hypers_G+_Induced_Prior_FS}
    \end{subfigure}
    \hspace{1em}
    \begin{subfigure}[t]{0.3\textwidth}
        \centering
        \includegraphics[width=\textwidth]{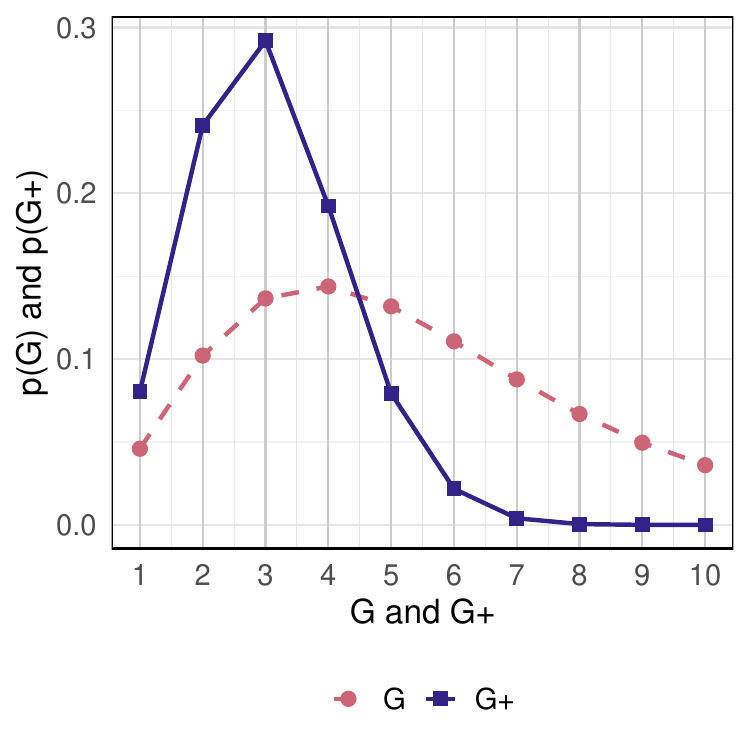}
        \caption{Comparison of the prior on $G$ (red) under a $\mathcal{BNB}(8,18,10)$ prior and the corresponding induced prior on $G_+$ (blue).}
        \label{fig:Hypers_G+_Induced_Prior}
    \end{subfigure}
    \hspace{1em}
    \begin{subfigure}[t]{0.3\textwidth}
        \centering
        \includegraphics[width=\textwidth]{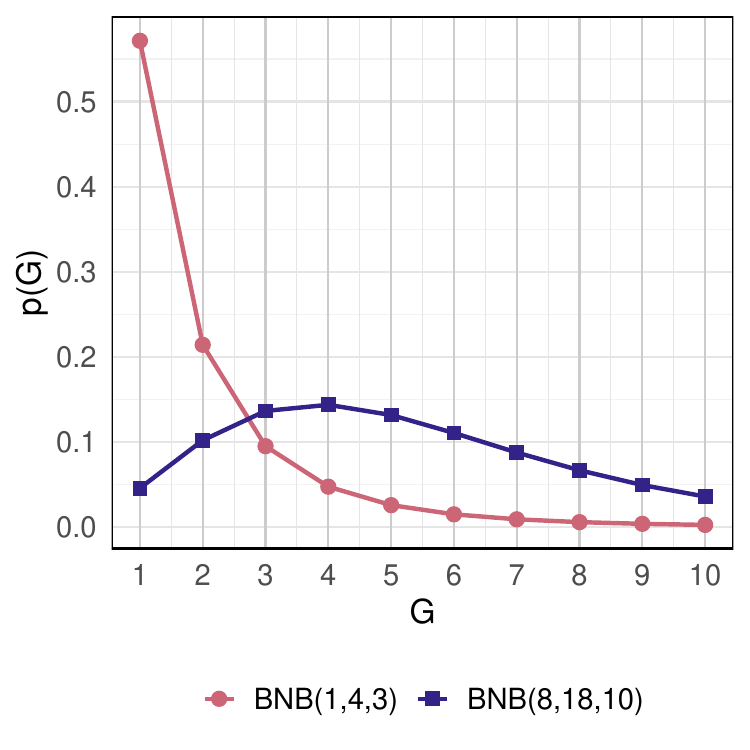}
        \caption{Comparison of the prior on $G$ under a $\mathcal{BNB}(1,4,3)$ prior (red) and under a $\mathcal{BNB}(8,18,10)$ prior (blue).}
        \label{fig:Hypers_BNB_Comparison}
    \end{subfigure}
    \caption{Illustration of how different $\mathcal{BNB}$ prior settings affect the prior distribution on the number of components $G$ and the corresponding induced prior on the number of clusters $G_+$.}
    \label{fig:enter-label}
\end{figure}

\citet{FS:2021:Methods:MFM_TS}, recommend to use a $\mathcal{BNB}(1, 4, 3) + 1$ translated prior distribution for the number of mixture components. This prior is weakly informative, concentrating on a small number of clusters while allowing for a broader range through its fat tail if the data support additional clusters. As shown in Figure~\ref{fig:Hypers_BNB_Comparison}, the $\mathcal{BNB}(1, 4, 3)$ prior places most of its probability mass on $G = 1$ or $G = 2$ components, naturally inducing the heavy shrinkage exploited in \citet{FS:2021:Methods:MFM_TS}. For comparison, a $\mathcal{BNB}(8,18,10)$ prior distribution is also displayed. 

The expected value and variance of $G$ under a $\mathcal{BNB}(a_G, b_G, c_G)$ prior distribution are given by $\mathbb{E}[G] = 1 + \frac{a_G c_G}{b_G - 1}$ and $\mathbb{V}[G] = 1 + \frac{a_G c_G (a_G + b_G - 1)(c_G + b_G - 1)}{(b_G - 2)(b_G - 1)^2}$. For a $\mathcal{BNB}(1, 4, 3)$ prior, this yields an expected value of 2 with a variance of 5; however, we found this expectation too small, as it could introduce unnecessary shrinkage of the mixture components, especially given that $G_+ \leq G$. Furthermore, a variance of 5 does not allow for sufficient prior mass on moderate to large values of $G$, potentially limiting their occurrence in the model and further promoting undue shrinkage. To address these concerns, we selected the $\mathcal{BNB}(8,18,10)$ prior, resulting in an expected $G$ of approximately 6 with a variance of about 12. This prior distribution encourages more moderate values of $G$ while still allowing for flexibility, making it a reasonable choice given our objectives and assumptions for $G-1$ and $K_g - 1 \ \forall g$. 

\citet{FS:2021:Methods:MFM_TS} discuss and derive the prior distribution induced on $G_+$ from the $\mathcal{BNB}$ prior distribution placed on $G$. In a dynamic mixture of finite mixtures (MFM) model, the priors $\mathbb{P}(G)$ and $\mathbb{P}(G_+)$ align closely only when the $\mathcal{BNB}$ prior has a small expected value. For priors $\mathbb{P}(G)$ with larger expected values, $\mathbb{P}(G_+)$ deviates notably, with its mass pulled toward smaller values of $G_+$. The $\mathcal{BNB}(1, 4, 3)$ prior, as suggested in \citet{FS:2021:Methods:MFM_TS}, leads to a weakly informative prior on $G_+$ that mirrors the prior on $G$, concentrating on a small number of clusters, predominantly $G_+ = 1$, but retaining a fat tail to allow estimation of a larger number of clusters, as shown in Figure~\ref{fig:Hypers_G+_Induced_Prior_FS}. By contrast, the $\mathcal{BNB}(8,18,10)$ prior we propose results in a more informative prior on $G_+$, as shown in Figure~\ref{fig:Hypers_G+_Induced_Prior}. This prior concentrates on a moderate number of clusters, primarily $G_+ \in \{2, 3, 4\}$, while still preserving the fat tail needed to account for potentially larger numbers of clusters. A direct comparison of the two priors on $G$ can be seen in Figure~\ref{fig:Hypers_BNB_Comparison}.

\citet{FS:2021:Methods:MFM_TS} suggest to set the initial number of clusters $G_0$ to overfit by a factor of two to three times the expected number of clusters. However, our choice of the $\mathcal{BNB}(8,18,10)$ prior over the $\mathcal{BNB}(1,4,3)$ for the number of mixture components $G$ was informed by a desire to avoid excessively low $G$ values, which might otherwise impose unwarranted shrinkage on the mixture components. By selecting a more moderate expected value for $G$ and allowing $G_+$ to adjust downward as needed, this prior enables the model to capture sufficient complexity without being overly restrictive. Additionally, we employ the same Dirichlet shrinkage prior on the mixing proportions as in \citet{FS:2021:Methods:MFM_TS}, which allows growth or shrinkage in accordance with the data. Unlike \citet{FS:2021:Methods:MFM_TS}, which recommends setting $G_0$ higher, we set $G_0 = 2$ (or $K_0 = 2$ for the node-level mixture) to leverage the adaptive growth permitted by our $\mathcal{BNB}(8,18,10)$ prior. Starting with a lower $G_0$ improves computational efficiency by reducing the number of components fitted in each iteration while remaining aligned with the structure of our multiplex data. Empirical results validate these hyperparameter choices, emphasising the need to tailor settings to dataset-specific characteristics, as \citet{FS:2021:Methods:MFM_TS} primarily addressed multivariate observations, in contrast to our focus on multiplex network data.

In the telescoping sampler (TS) procedure, the Markov chain Monte Carlo (MCMC) algorithm updates the value of $G$ by considering an upper bound $G_{\text{max}}$ and sampling from a multinomial distribution over the set $\{G_+, \ldots, G_{\text{max}}\}$ with probabilities proportional to the unnormalised posterior of $G$. While\citet{FS:2021:Methods:MFM_TS} set $G_{\text{max}} = 100$, this is unrealistic for our setting, where the maximum number of networks considered is $M = 100$. Since networks are clustered based on their structural similarity, it is plausible that a single network could form its own cluster. This stands in contrast to the node-level clustering, where we can reasonably assume a minimum average number of nodes per cluster, $n_{\text{min}}$, as above. At the network level, however, no such realistic lower bound exists, as assuming an expected minimum network-level cluster size would risk excluding structurally distinct networks that merit separation. However, permitting all $M$ networks to form their own clusters would defeat the purpose of clustering. To strike a balance, we define $G_{\text{max}}$ values that scale with the number of networks while still enabling meaningful clustering, as summarised in Table~\ref{tab:Hypers_Gmax}. 

Similarly, $K_g$ is updated by sampling from ${K_{g+}, \ldots, K_{\text{max}}}$, using weights based on the unnormalised posterior. Again, setting $K_{\text{max}} = 100$ is impractical in our setting. To enable flexible yet interpretable clustering, we define $K_{\text{max}}$ as a function of the number of nodes $N$ and a minimum average cluster size $n_{\text{min}}$, specifically, $K_{\text{max}} = \frac{N}{n_{\text{min}}} + 2$ (rounded to the nearest integer), where $n_{\text{min}} = 5$ for $N < 60$ and $n_{\text{min}} = 10$ for $N \geq 60$. Values  of $K_{\text{max}}$ for different values of $N$ are summarised in Table~\ref{tab:Hypers_Kmax}. In addition to interpretability and practicality, the specifications of $G_{\text{max}}$ and $K_{\text{max}}$ were also chosen to reduce computational burden while preserving model flexibility and avoiding the overhead of unnecessarily large upper bounds.

Lastly, the $\mathcal{F}(6,3)$ prior distribution on the Dirichlet concentration hyperparameters, as suggested in \citet{FS:2021:Methods:MFM_TS}, provides flexibility in the prior probability of homogeneity. We utilise this $\mathcal{F}(6,3)$ prior distribution on the Dirichlet concentration hyperparameters $e$ and $w_g, \ g = 1, \dots, G_+$ within the node-level mixture distribution, encoding an expectation of three for $e$ and $w_g$, $g = 1, \dots, G_+$, with infinite variance. 

\begin{table}[!bt]
\centering
    \begin{minipage}{0.45\textwidth}
        \caption{Value of $G_{\text{max}}$ for various $M$.}
        \label{tab:Hypers_Gmax}
        \centering
        \begin{tabular}{@{}cc@{}}
        \hline
        \multicolumn{1}{c}{$M$} & \multicolumn{1}{c}{$G_{\text{max}}$} \\
        \hline
        20   & 5 \\
        30   & 5 \\
        40   & 5 \\
        50   & 5 \\
        60   & 10 \\
        70   & 10 \\
        80   & 10 \\
        90   & 10 \\
        100  & 10 \\
        \hline
        \end{tabular}
    \end{minipage}%
    \hspace{0.05\textwidth} 
    \begin{minipage}{0.45\textwidth}
        \caption{Value of $K_{\text{max}}$ for various $N$.}
        \label{tab:Hypers_Kmax}
        \centering
        \begin{tabular}{@{}cc@{}}
        \hline
        \multicolumn{1}{c}{$N$} & \multicolumn{1}{c}{$K_{\text{max}}$} \\
        \hline
            20  & 6\\
            30  & 8\\
            40  & 10\\
            50  & 12\\
            60  & 8\\
            70  & 9\\
            80  & 10\\
            90  & 11\\
            100 & 12\\
        \hline
        \end{tabular}
    \end{minipage}
\end{table}

We conducted preliminary tests to compare our selected hyperparameters with those recommended by \citet{FS:2021:Methods:MFM_TS}, in order to validate our decision to diverge from their guidelines. The results, presented in Section~\ref{SuppMat:MFMvsOM} of this Supplementary Material, support the suitability of our choices for the specific network data and co-clustering model used.

\clearpage
\section{Derivation of Full Conditional Distributions} \label{SuppMat:FCs}
As discussed in Section 3 of the main article, the joint posterior distribution of {\em LaPCoM} is as follows:
\begin{eqnarray*}
    & & \mathbb{P} \bigg(G, e, \boldsymbol{\tau}, \boldsymbol{C}, \alpha, \Big\{K_g, w_g, \boldsymbol{\pi}_g, \boldsymbol{S}_g, \{\boldsymbol{\mu}_{gk}, \boldsymbol{\Sigma}_{gk}\}_{k=1}^{K_g}\Big\}_{g = 1}^G, \{\boldsymbol{Z}_g\}_{g = 1}^G \mid \boldsymbol{\mathcal{Y}} \bigg)\\
    & & \propto \mathbb{P}\big(\boldsymbol{\mathcal{Y}} \mid \boldsymbol{C}, \alpha, \{\boldsymbol{Z}_g\}_{g=1}^G\big) \times \mathbb{P}(\boldsymbol{C} \mid \boldsymbol{\tau}) \times \mathbb{P}(\boldsymbol{\tau} \mid G, e) \times \mathbb{P}(G \mid \boldsymbol{H}) \times \mathbb{P}(e \mid \boldsymbol{H}) \times \mathbb{P}(\alpha \mid \boldsymbol{H})\\
    & & \quad \times \prod_{g = 1}^G \mathbb{P}\Big(\boldsymbol{Z}_g \mid \boldsymbol{S}_g, \{\boldsymbol{\mu}_{gk}, \boldsymbol{\Sigma}_{gk}\}_{k=1}^{K_g}\Big) \times \prod_{g = 1}^G \prod_{k = 1}^{K_g} \mathbb{P}(\boldsymbol{\mu}_{gk} \mid \boldsymbol{H}) \times \prod_{g = 1}^G \prod_{k = 1}^{K_g} \mathbb{P}(\boldsymbol{\sigma}_{gk}^2 \mid \boldsymbol{H})\\
    & & \quad \times \prod_{g = 1}^G \mathbb{P}(\boldsymbol{S}_g \mid \boldsymbol{\pi}_{g}) \times \prod_{g = 1}^G \mathbb{P}(\boldsymbol{\pi}_{gk} \mid K_g, w_g) \times \prod_{g = 1}^G \mathbb{P}(K_g \mid \boldsymbol{H}) \times \prod_{g = 1}^G \mathbb{P}(w_g \mid \boldsymbol{H}).
\end{eqnarray*}

As an example, in the following derivations we consider a count-valued weighted multiplex, where each entry $y^{(m)}_{ij} \in \mathbb{N}$ represents the count shared between nodes $i$ and $j$ in view $m$. Accordingly, we model the edges using a Poisson distribution with rate parameter $\lambda_{g,ij}$ and apply the $\log$ function as the link function (see Section~2.2 of the main text). Therefore, each network view is modelled
$$
\boldsymbol{Y}^{(m)} \sim \sum_{g = 1}^G \tau_g \prod_{i \neq j} \frac{\lambda_{g,ij}^{y_{ij}^{(m)}} \exp(-\lambda_{g,ij})}{y_{ij}^{(m)}!}, \quad m = 1, \dots, M.
$$
Throughout the derivation of the full conditional distributions, we adopt the standard convention that the conditioning set, denoted by $\ldots$, includes the observed data and all other parameters not currently being updated. Analogous derivations apply in the case of a binary multiplex, where edges are modelled using a Bernoulli distribution and the logit function is used as the link function.

The full conditional distribution of the probability of network $\boldsymbol{Y}^{(m)}$ belonging to network-level cluster $g$ is derived as follows:
\begin{eqnarray*}
    Pr \Big(C^{(m)} = g \mid  \ldots \Big) & \propto & \mathbb{P}(\boldsymbol{Y}^{(m)} \mid C_g^{(m)}, \alpha, \boldsymbol{Z}_g) \times \mathbb{P}(C_g^{(m)} \mid \tau_g) \\
    & \propto & \tau_g \prod_{i \neq j} \frac{\lambda_{g,ij}^{y_{ij}^{(m)}} \exp(-\lambda_{g,ij})}{y_{ij}^{(m)}!}\\
    & = & \frac{\tau_g \prod_{i \neq j} \frac{\lambda_{g,ij}^{y_{ij}^{(m)}} \exp(-\lambda_{g,ij})}{y_{ij}^{(m)}!}}{\sum_{h=1}^G \tau_h \prod_{i \neq j} \frac{\lambda_{h,ij}^{y_{ij}^{(m)}} \exp(-\lambda_{h,ij})}{y_{ij}^{(m)}!}}.
\end{eqnarray*}

The full conditional distribution of the network-level mixing proportions $\boldsymbol{\tau}$ is derived as follows:
\begin{eqnarray*}
    \mathbb{P}(\boldsymbol{\tau} \mid \ldots) & \propto & \mathbb{P}(\boldsymbol{C} \mid \boldsymbol{\tau}) \times \mathbb{P}(\boldsymbol{\tau} \mid G, e)\\
    & \propto & \prod_{g = 1}^G \prod_{m = 1}^M \tau_g^{C^{(m)}} \times \prod_{g = 1}^G \tau_g^{\frac{e}{G} - 1}\\
    & = & \prod_{g = 1}^G \tau_g^{M_g + \frac{e}{G} - 1}\\
    & \sim & \mathcal{D}(\zeta_1, \dots, \zeta_G),
\end{eqnarray*}
where $\zeta_g = M_g + \frac{e}{G}$ and $M_g = \sum_{m = 1}^M \mathbb{I} \{C_g^{(m)} = 1\}$.

The full conditional distribution of the intercept $\alpha$ is derived as follows:
\begin{eqnarray*}
    \mathbb{P}(\alpha \mid \ldots) & \propto & \mathbb{P}(\boldsymbol{\mathcal{Y}} \mid \boldsymbol{C}, \alpha, \{\boldsymbol{Z}_g\}_{g=1}^G) \times \mathbb{P}(\alpha \mid \boldsymbol{H})\\
    & \propto & \prod_{m = 1}^M \prod_{g = 1}^G \Bigg[ \prod_{i \neq j} \frac{\lambda_{g,ij}^{y_{ij}^{(m)}} \exp(-\lambda_{g,ij})}{y_{ij}^{(m)}!} \Bigg]^{C_g^{(m)}} \\
    & & \times \frac{1}{\sqrt{2\pi}s_{\alpha}} \exp \Bigg(-\frac{1}{2}\frac{(\alpha - m_{\alpha})^2}{s_{\alpha}^2}\Bigg),
\end{eqnarray*}
\noindent which is sampled using a Metropolis-Hastings (MH) update, as it is not of recognisable/closed form.

The full conditional distribution of the latent space corresponding to network-level cluster $g$, $\boldsymbol{Z}_g$, is derived as follows: 
\begin{eqnarray*}
    \mathbb{P}(\boldsymbol{Z}_g \mid \ldots) & \propto & \mathbb{P}(\boldsymbol{\mathcal{Y}} \mid \boldsymbol{C}, \alpha, \boldsymbol{Z}_g) \times \mathbb{P}(\boldsymbol{Z}_g \mid \boldsymbol{S}_g, \{\boldsymbol{\mu}_{gk}, \boldsymbol{\Sigma}_{gk}\}_{k = 1}^{K_g})\\
    & \propto & \prod_{m: C_g^{(m)} = 1} \prod_{i \neq j} \frac{\lambda_{g,ij}^{y_{ij}^{(m)}} \exp(-\lambda_{g,ij})}{y_{ij}^{(m)}!} \\
    & & \times \prod_{i = 1}^N \prod_{k = 1}^{K_g} \bigg\{  \mid \boldsymbol{\Sigma}_{gk} \mid ^{-\frac{1}{2}} \exp\bigg[-\frac{1}{2} (\boldsymbol{z}_{g,i} - \boldsymbol{\mu}_{gk})^T \boldsymbol{\Sigma}_{gk}^{-1}(\boldsymbol{z}_{g,i} - \boldsymbol{\mu}_{gk})\bigg]\bigg\}^{S_{gk}^i},
\end{eqnarray*}
which is sampled using a Metropolis-Hastings update, as it is not of recognisable/closed form.

The number of network-level components $G$ is sampled from a multinomial distribution, considering options from $G_+, \dots, G_{\text{max}}$, with the following probabilities:
\begin{eqnarray*}
    Pr(G = g \mid \ldots) \propto \mathbb{P}(g) \frac{e^{G_+} g!}{g^{G_+} (g - G_+)!} \prod_{h = 1}^{G_+} \frac{\Gamma(M_h + \frac{e}{g})}{\Gamma(1 + \frac{e}{g})},
\end{eqnarray*}
where $\mathbb{P}(g) = \frac{\Gamma(a_G + g - 1) B(a_G + b_G, g - 1 + c_G)}{\Gamma(a_G)\Gamma(g)B(b_G, c_G)}$, as $\mathbb{P}(g - 1) \sim \mathcal{BNB}(a_G, b_G, c_G)$.

The full conditional distribution of the network-level Dirichlet concentration hyperparameter, $e$, is derived as follows:
\begin{eqnarray*}
    \mathbb{P}(e \mid \ldots) & \propto & \mathbb{P}(e) \frac{e^{G_+} \Gamma(e)}{\Gamma(M + e)} \prod_{h = 1}^{G_+} \frac{\Gamma(M_h + \frac{e}{G})}{\Gamma(1 + \frac{e}{G})}\\
    & \propto & \frac{\sqrt{\frac{(l_G e)^{l_G} r_G^{r_G}}{(l_G e + r_G)^{l_G + r_G}}}}{e B(\frac{l_G}{2}, \frac{r_G}{2})} \frac{e^{G_+} \Gamma(e)}{\Gamma(M + e)} \prod_{h = 1}^{G_+} \frac{\Gamma(M_h + \frac{e}{G})}{\Gamma(1 + \frac{e}{G})},
\end{eqnarray*}
which is sampled using a Metropolis-Hastings update, as it is not of recognisable/closed form.

The full conditional distribution of the assignment of node $i$ to node-level cluster $k$ within network-level cluster $g$ is a multinomial with probability derived as follows:
\begin{eqnarray*}
    Pr(S_g^{(i)} = k \mid \ldots) & \propto & \mathbb{P}(\boldsymbol{Z}_g \mid \boldsymbol{S}_g, \boldsymbol{\mu}_{gk}, \boldsymbol{\Sigma}_{gk}, \pi_{gk}) \times \mathbb{P}(\boldsymbol{S}_g \mid \pi_{gk})\\
    & \propto & \pi_{gk}  \mid \boldsymbol{\Sigma}_{gk} \mid ^{-\frac{1}{2}} \exp\bigg[-\frac{1}{2} (\boldsymbol{z}_{g,i} - \boldsymbol{\mu}_{gk})^T \boldsymbol{\Sigma}_{gk}^{-1}(\boldsymbol{z}_{g,i} - \boldsymbol{\mu}_{gk})\bigg]\\
    & = & \frac{\pi_{gk}  \mid \boldsymbol{\Sigma}_{gk} \mid ^{-\frac{1}{2}} \exp\bigg[-\frac{1}{2} (\boldsymbol{z}_{g,i} - \boldsymbol{\mu}_{gk})^T \boldsymbol{\Sigma}_{gk}^{-1}(\boldsymbol{z}_{g,i} - \boldsymbol{\mu}_{gk})\bigg]}{\sum_{h = 1}^K \pi_{gh}  \mid \boldsymbol{\Sigma}_{gh} \mid ^{-\frac{1}{2}} \exp\bigg[-\frac{1}{2} (\boldsymbol{z}_{g,i} - \boldsymbol{\mu}_{gh})^T \boldsymbol{\Sigma}_{gh}^{-1}(\boldsymbol{z}_{g,i} - \boldsymbol{\mu}_{gh})\bigg]}.
\end{eqnarray*}

The full conditional distribution of the mean of node-level cluster $k$ within network-level cluster $g$, $\boldsymbol{\mu}_{gk}$ is derived as follows: 
\begin{eqnarray*}
    \mathbb{P}(\boldsymbol{\mu}_{gk} \mid \ldots) & \propto & \mathbb{P}(\boldsymbol{Z}_g \mid \boldsymbol{S}_g, \boldsymbol{\mu}_{gk}, \boldsymbol{\Sigma}_{gk}, \pi_{gk}) \times \mathbb{P}(\boldsymbol{\mu}_{gk} \mid \boldsymbol{H})\\
    & \propto & \prod_{i: S_g^{(i)} = k}  \mid \boldsymbol{\Sigma}_{gk} \mid ^{-\frac{1}{2}} \exp \bigg[ -\frac{1}{2} (\boldsymbol{z}_{g,i} - \boldsymbol{\mu}_{gk})^T \boldsymbol{\Sigma}_{gk}^{-1} (\boldsymbol{z}_{g,i} - \boldsymbol{\mu}_{gk}) \bigg] \\
    & & \times \mid \boldsymbol{\mathbb{I}}_2 \mid ^{-\frac{1}{2}} \exp\Big[ (\boldsymbol{\mu}_{gk} - \boldsymbol{0})^T \boldsymbol{\mathbb{I}}_2^{-1} (\boldsymbol{\mu}_{gk} - \boldsymbol{0}) \Big]\\
    & \sim & \mathcal{MVN}_2 \Bigg(\boldsymbol{\Sigma}_{gk}^* \Bigg[ \boldsymbol{\Sigma}_{gk}^{-1} \Bigg(\sum_{i:S_g^{(i)} = k} \boldsymbol{z}_{g,i}\Bigg) + \boldsymbol{\mathbb{I}}_2^{-1}\boldsymbol{0}\Bigg], \big[N_{gk} \boldsymbol{\Sigma}_{gk}^{-1} + \boldsymbol{\mathbb{I}}_2^{-1}\big]^{-1}\Bigg),
\end{eqnarray*}
where $N_{gk} = \sum_{i = 1}^N \mathbb{I}\{S_g^{(i)} = k\}$.

The full conditional distribution of the variance of dimension $q$, $q = 1, 2$, of node-level cluster $k$ within network-level cluster $g$, $\sigma_{gkq}^2$ is derived as follows:
\begin{eqnarray*}
    \mathbb{P}(\sigma_{gkq}^2 \mid \ldots) & \propto & \mathbb{P}(\boldsymbol{Z}_g \mid \boldsymbol{S}_g, \boldsymbol{\mu}_{gk}, \sigma_{gkq}^2) \times \mathbb{P}(\sigma_{gkq}^2 \mid \boldsymbol{H})\\
    & \propto & \prod_{i:S_g^{(i)} = k} ({\sigma_{gkq}^2})^{-\frac{1}{2}} \exp\bigg[-\frac{1}{2} (\boldsymbol{z}_{g,i} - \boldsymbol{\mu}_{gk})^T \frac{1}{\sigma_{gk}^2} (\boldsymbol{z}_{g,i} - \boldsymbol{\mu}_{gk})\bigg] \\
    & & \times ({\sigma_{gkq}^2})^{-u_{\sigma^2} -1} \exp\bigg[-\frac{v_{\sigma^2}}{\sigma_{gkq}^2}\bigg]\\
    & \sim & \mathcal{IG}(u_{\sigma^2}^*, v_{\sigma^2}^*),
\end{eqnarray*}
where $u_{\sigma^2}^* = \frac{N_{gk}}{2} + u_{\sigma^2}$ and $v_{\sigma^2}^* = v_{\sigma^2} + \frac{1}{2} \sum_{i:S_g^{(i)} = k} (\boldsymbol{z}_{g,i} - \boldsymbol{\mu}_{gk})^T(\boldsymbol{z}_{g,i} - \boldsymbol{\mu}_{gk})$.

The full conditional distribution of the node-level mixing proportions $\boldsymbol{\pi}_g$ within network-level cluster $g$ is derived as follows:
\begin{eqnarray*}
    \mathbb{P}(\boldsymbol{\pi}_g \mid \ldots) & \propto & \mathbb{P}(\boldsymbol{S}_g \mid \boldsymbol{\pi}_g) \times \mathbb{P}(\boldsymbol{\pi}_g \mid K_g, w_g)\\
    & \propto & \prod_{k = 1}^{K_g} \prod_{i = 1}^N \pi_{gk}^{S_{gk}^{(i)}} \times \prod_{k = 1}^{K_g} \pi_{gk}^{\frac{w_g}{K_g} - 1}\\
    & \propto & \prod_{k = 1}^{K_g} \pi_{gk}^{N_{gk} + \frac{w_g}{K_g} - 1}\\
    & \sim & \mathcal{D}(\psi_1, \dots, \psi_{K_g}),
\end{eqnarray*}
where $\psi = N_{gk} + \frac{w_g}{K_g}$ and $N_{gk}$ is defined as before.

The number of node-level components $K_g$, in network-level component $g$, is sampled from a multinomial distribution, considering options from $K_{g+}, \dots, K_{\text{max}}$, with the following probabilities:
\begin{eqnarray*}
    Pr(K_g = k \mid \ldots) \propto \mathbb{P}(k) \frac{w_g^{K_{g+}} k!}{k^{K_{g+}} (k - K_{g+})!} \prod_{k^* = 1}^{K_{g+}} \frac{\Gamma(N_{gk^*} + \frac{w_g}{k})}{\Gamma(1 + \frac{w_g}{k})},
\end{eqnarray*}
where $\mathbb{P}(k) = \frac{\Gamma(a_K + k - 1) B(a_K + b_K, k - 1 + c_K)}{\Gamma(a_K)\Gamma(k)B(b_K, c_K)}$, as $\mathbb{P}(k - 1) \sim \mathcal{BNB}(a_K, b_K, c_K)$.

The full conditional distribution of the node-level Dirichlet concentration hyperparameter, $w_g$, in network-level component $g$, is derived as follows:
\begin{eqnarray*}
    \mathbb{P}(w_g \mid \ldots) & \propto & \mathbb{P}(w_g) \frac{w_g^{K_{g+}} \Gamma(w_g)}{\Gamma(N + w_g)} \prod_{k^* = 1}^{K_{g+}} \frac{\Gamma(N_{gk^*} + \frac{w_g}{K_g})}{\Gamma(1 + \frac{w_g}{K_g})}\\
    & \propto & \frac{\sqrt{\frac{(l_K w_g)^{l_K} r_K^{r_K}}{(l_K w_g + r_K)^{l_K + r_K}}}}{w_g B(\frac{l_K}{2}, \frac{r_K}{2})} \frac{w_g^{K_{g+}} \Gamma(w_g)}{\Gamma(N + w_g)} \prod_{k^* = 1}^{K_{g+}} \frac{\Gamma(N_{gk^*} + \frac{w_g}{K_g})}{\Gamma(1 + \frac{w_g}{K_g})},
\end{eqnarray*}
which is sampled using a Metropolis-Hastings update, as it is not of recognisable/closed form.

\clearpage
\section{Pseudocode of the MCMC algorithm} \label{SuppMat:MCMC_Algorithm}
\begin{algorithm}[H]\footnotesize
    \caption{MCMC Algorithm}\label{alg:MCMC_Inference}
        \begin{algorithmic}
            \State Initialise $G^{[1]}, \boldsymbol{\tau}^{[1]}, e^{[1]}, \boldsymbol{C}^{[1]}, \big\{\boldsymbol{Z}_g^{[1]}\big\}_{g=1}^{G^{[1]}},\alpha^{[1]}, \Big\{K_g^{[1]}, \boldsymbol{\pi}^{[1]}, w_g^{[1]}, \boldsymbol{S}_g^{[1]}, \big\{\boldsymbol{\mu}_{gk}^{[1]}, \boldsymbol{\Sigma}_{gk}^{[1]}\big\}_{k=1}^{K_g^{[1]}}\Big\}_{g=1}^{G^[1]}$
            
            \For{$t = 1, \dots, T$}

                \For{$m = 1,\dots,M$}
                    \State Sample $C_g^{(m)[t + 1]}$ from a multinomial distribution with  $Pr(C^{(m)[t + 1]} = g \mid \dots)$ as 
                    \State derived in Section~\ref{SuppMat:FCs}
                \EndFor 

                \State Compute $M_g^{[t+1]}$ for $g=1,\dots,G^{[t]}$, determine $G_+^{[t+1]}$, and relabel to ensure the 
                \State first $G_+^{[t+1]}$ (network) mixture components are non-empty

                \For{$g = 1,\dots,G_+^{[t+1]}$}
                    \State Sample $\boldsymbol{Z}_g^{[t+1]}$ using a block MH step, with $\hat{\boldsymbol{z}}_{g,i} \sim \mathcal{MVN}_2\Big(\boldsymbol{z}_{g,i}^{[t]}, \delta_{\boldsymbol{Z}}^2 \boldsymbol{\Sigma}_{gk}^{[t]}\Big)$

                    \For{$i = 1,\dots,N$}
                        \State Sample $S_{gk}^{(i)[t + 1]}$ from a multinomial distribution with $Pr\big(S_g^{(i)[t + 1]} \mid \dots\big)$ as 
                        \State derived in Section~\ref{SuppMat:FCs}
                    \EndFor 
    
                    \State Compute $N_{gk}^{[t+1]}$ for $k=1,\dots,K_{g}^{[t]}$, determine $K_{g+}^{[t+1]}$, and relabel to ensure 
                    \State the first $K_{g+}^{[t+1]}$ (network) mixture components are non-empty

                    \For{$k=1,\dots,K_{g+}^{[t+1]}$}
                        \State Sample $\boldsymbol{\mu}_{gk}^{[t+1]} \sim \mathcal{MVN}_2(\boldsymbol{\mu}_{gk}^*, \boldsymbol{\Sigma}_{gk}^*)$ and $\sigma_{gk,q}^{2[t+1]} \sim \mathcal{IG}(u_{\sigma^2}^*, v_{\sigma^2}^*), \ q = 1, 2$
                    \EndFor

                    \State Sample $K_g^{[t+1]}$ from a multinomial distribution with $\mathbb{P}\big(k^* \mid \ldots)$ as derived in \State Section~\ref{SuppMat:FCs}, considering $k^* = K_{g+}^{[t]}, \dots, K_{\text{max}}$
                    
                    \State Sample $w_g^{[t+1]}$ from a MH step, with proposal $\log(\hat{w}_g) \sim \mathcal{N}\big(\log\big(w_g^{[t]}\big), s_w^2\big)$

                    \If{$K_g^{[t+1]} > K_{g+}^{[t+1]}$}
                        \For{$k = K_{g+}^{[t+1]} + 1, \dots, K_g^{[t+1]}$}
                            \State Add an empty node-level mixture component, sampling $\boldsymbol{\mu}_{gk}^{[t+1]}$ and
                            \State $\sigma_{gk,q}^{2[t+1]}$ from the appropriate prior distributions
                        \EndFor
                    \EndIf
                    
                    \State Sample $\boldsymbol{\pi}^{[t+1]} \sim \mathcal{D}\Big(\psi_{g1}^{[t+1]},\dots,\psi_{g{K_g^{[t+1]}}}^{[t+1]}\Big)$, where $\psi_{gk}^{[t+1]} = \frac{w_g^{[t+1]}}{K_g^{[t+1]}} + N_{gk}^{[t+1]}$

                \EndFor 

                \State Sample $\alpha^{[t+1]}$ from a Metropolis-Hastings step, with proposal $\hat{\alpha} \sim \mathcal{N}(\alpha^{[t]}, \delta_{\alpha}^2 s_{\alpha}^2)$
                \State Sample $G^{[t+1]}$ from a multinomial distribution with $\mathbb{P}\big(g^* \mid \ldots)$ as derived in \State Section~\ref{SuppMat:FCs}, considering $g^* = G_+^{[t]}, \dots, G_{\text{max}}$
                \State Sample $e^{[t+1]}$ from a MH step, with proposal $\log(\hat{e}) \sim \mathcal{N}\big(\log(e^{[t]}), s_e^2\big)$

                \If{$G^{[t+1]} > G_+^{[t+1]}$}
                    \For{$g = G_+^{[t+1]} + 1, \dots, G^{[t+1]}$}
                        \State Add an empty node-level mixture component, sampling all related 
                        \State parameters from the appropriate prior distribution, assuming no clustering 
                        \State structure in the additional latent spaces
                    \EndFor
                \EndIf

                \State Sample $\boldsymbol{\tau}^{[t+1]} \sim \mathcal{D}\Big(\zeta_1^{[t+1]}, \dots, \zeta_{G^{[t+1]}}^{[t+1]}\Big)$, where $\zeta_g = \frac{e^{[t+1]}}{G^{[t+1]}} + M_g^{[t+1]}$
                
            \EndFor 
            
        \end{algorithmic}
\end{algorithm}

\clearpage
\section{Initialisation of the Model Parameters} \label{SuppMat:Initialisation}
We initialise the model parameters in the following manner, for $t = 1$:

\begin{enumerate}

    \item Initialise the number of network-level mixture components as $G^{[1]} = G_0$  (Section~\ref{SuppMat:HyperparameterChoices}), with all components active.
    
    \item Set network-level mixing proportions uniformly as $\boldsymbol{\tau}^{[1]} = \big(\frac{1}{G^{[1]}}, \dots, \frac{1}{G^{[1]}}\big)$.

    \item Set the network-level Dirichlet hyperparameter to a small value $e^{[1]} = 10^{-5}$ to encourage shrinkage \citep{MW:2016:Methods:SFM:Original,FSMW:2019:Methods:SFMvsDPM}.

    \item Initialise network-level allocations $\boldsymbol{C}^{[1]}$ by computing pairwise network distances \citep{Schieber:2017:Extra:Nature:Distance}, applying multidimensional scaling (MDS) to obtain a 2-dimensional representation \citep{Cox:2008:Extra:MDS}, then clustering (using $K$-means \citep{Wu:2012:Extra:Kmeans} or \texttt{mclust} \citep{Scrucca:2016:Extra:mclust}) into $G^{[1]}$ groups. The initial $\boldsymbol{C}^{[1]}$ is set to the partition obtained.

    \item Initialise network-level latent spaces $\{\boldsymbol{Z}_g^{[1]}\}_{g=1}^{G^{[1]}}$ by averaging geodesic distances within each cluster $g$ and applying MDS to obtain 2D representations.

    \item Initialise the intercept $\alpha^{[1]}$ by calculating $\hat{\alpha}_g^{(m)[1]} = \log \left(\frac{1}{N} \sum_{i,j} y_{ij}^{(m)} \right) + \frac{1}{N} \sum_{i \leq j} \|\boldsymbol{z}^{[1]}_{g,i} - \boldsymbol{z}^{[1]}_{g,j}\|_2^2$  \citep{Hoff:2002:Methods:LPM} and averaging across networks $m$ and clusters $g$.

    \item For each $g = 1, \dots, G^{[1]}$, initialise node-level parameters as follows:

    \begin{enumerate}

        \item Set the number of node-level components $K_g^{[1]} = K_0$ (Section~\ref{SuppMat:HyperparameterChoices}).

        \item Initialise mixing proportions uniformly: $\boldsymbol{\pi}_g^{[1]} = \big(\frac{1}{K_g^{[1]}}, \dots, \frac{1}{K_g^{[1]}}\big)$.

        \item Draw the Dirichlet hyperparameter $w_g^{[1]}$ from the $\mathcal{F}(l_K, r_K)$ prior.

        \item Initialise node-level allocations $\boldsymbol{S}_g^{[1]}$ and means $\boldsymbol{\mu}_g^{[1]}$ via $K$-means clustering of $\boldsymbol{Z}_g^{[1]}$ into $K_g^{[1]}$ clusters; set $\boldsymbol{\mu}_g^{[1]}$ to cluster centres.

        \item For each $k=1,\dots,K_g^{[1]}$, sample diagonal covariance elements of $\boldsymbol{\Sigma}_{gk}^{[1]}$ independently from $\mathcal{IG}(u_{\sigma^2}, v_{\sigma^2})$ with zero off-diagonal entries.

    \end{enumerate}
    
\end{enumerate}

\clearpage
\section{Simulation Study Designs} \label{SuppMat:SimStudies}
This section provides the details of the data-generating mechanisms used in Simulation Studies 1 and 2, discussed in Section~4 of the main article. Each study investigates clustering performance under varying structural properties of the generated multiplex. 

The node-level cluster means $\boldsymbol{\mu}_{gk}$ and variances $\boldsymbol{\Sigma}_{gk}$ were kept consistent across both simulation studies. Specifically, when the number of node-level clusters within the $g$\textsuperscript{th} network-level cluster was one, the cluster mean was set to $\boldsymbol{\mu}_{g1} = (0,0)$. For two node-level clusters, the means were $\boldsymbol{\mu}_{g1} = (-0.8, 0.8)$ and $\boldsymbol{\mu}_{g2} = (0.8, -0.8)$. For three node-level clusters, the means were $\boldsymbol{\mu}_{g1} = (-0.9, -0.9)$, $\boldsymbol{\mu}_{g2} = (1.4, 0.4)$ and $\boldsymbol{\mu}_{g3} = (-0.9, 1.4)$. The node-level cluster variance was fixed to $\sigma_{gk, q}^2 = 0.25$ for all dimensions $q = 1,2$, for all node-level clusters $k$ and for all network-level clusters $g$.

In each scenario, simulated network multiplexes comprise $M$ networks, each containing $N$ nodes. The networks are generated from $G^*$ network-level clusters with proportions defined by $\boldsymbol{\tau}$, and each network-level cluster relates to a corresponding latent space generated with $K_g$ node-level clusters, characterised by proportions $\boldsymbol{\pi}_g$.

Table~\ref{tab:SS1_Setup} summarises the parameter configurations for Simulation Study 1. In this study, we generated undirected, count-valued networks where edges follow a Poisson distribution with a log link function. For smaller networks ($N = 30$), $\alpha = 0.6$ was used, whereas larger networks ($N > 30$) used $\alpha = -0.4$.

\begin{table*}[!b]
\centering
    \caption{Settings for simulation study 1.}
    \label{tab:SS1_Setup}
    \begin{tabular}{@{}ccccccc@{}}
        \hline
        \multicolumn{1}{c}{Scenario} & \multicolumn{1}{c}{$M$} & \multicolumn{1}{c}{$N$} & \multicolumn{1}{c}{$G^*$} & \multicolumn{1}{c}{$\boldsymbol{\tau}$} & \multicolumn{1}{c}{$K_g$} & \multicolumn{1}{c}{$\boldsymbol{\pi}$} \\
        \hline
        A          & 20    & 30    & 2     & \{0.6, 0.4\}      & \{1, 1\}      & \{\{1\}, \{1\}\}\\
        B          & 20    & 50    & 2     & \{0.6, 0.4\}      & \{1, 1\}      & \{\{1\}, \{1\}\}\\
        C          & 20    & 30    & 2     & \{0.6, 0.4\}      & \{1, 2\}      & \{\{1\}, \{0.5, 0.5\}\}\\
        D          & 20    & 50    & 2     & \{0.6, 0.4\}      & \{1, 2\}      & \{\{1\}, \{0.5, 0.5\}\}\\
        E          & 20    & 30    & 2     & \{0.6, 0.4\}      & \{2, 3\}      & \{\{0.7, 0.3\}, \{0.4, 0.3, 0.3\}\}\\
        F          & 20    & 60    & 2     & \{0.6, 0.4\}      & \{2, 3\}      & \{\{0.7, 0.3\}, \{0.4, 0.3, 0.3\}\}\\
        G          & 50    & 30    & 2     & \{0.6, 0.4\}      & \{2, 3\}      & \{\{0.7, 0.3\}, \{0.4, 0.3, 0.3\}\}\\
        H          & 50    & 60    & 2     & \{0.6, 0.4\}      & \{2, 3\}      & \{\{0.7, 0.3\}, \{0.4, 0.3, 0.3\}\}\\
        \hline
    \end{tabular}
\end{table*}

\begin{table*}[!b]
\centering
\caption{Settings for simulation study 2.}
\label{tab:SS2_Setup}
\centering
\resizebox{\linewidth}{!}{ 
\begin{tabular}{@{}ccccccc@{}}
    \hline
    Scenario & $M$ & $N$ & $G^*$ & $\boldsymbol{\tau}$ & $K_g$ & $\boldsymbol{\pi}_g$ \\
    \hline
    I          & 20    & 50    & 2     & $\{\nicefrac{3}{5}, \nicefrac{2}{5}\}$              & $\{1, 1\}$          & $\{\{1\}, \{1\}\}$ \\
    II         & 50    & 30    & 2     & $\{\nicefrac{3}{5}, \nicefrac{2}{5}\}$              & $\{2, 3\}$          & $\{\{\nicefrac{7}{10}, \nicefrac{3}{10}\}, \{\nicefrac{2}{5}, \nicefrac{3}{10}, \nicefrac{3}{10}\}\}$ \\
    III        & 50    & 30    & 3     & $\{\nicefrac{2}{5}, \nicefrac{3}{10}, \nicefrac{3}{10}\}$         & $\{1, 2, 3\}$       & $\{\{1\}, \{\nicefrac{7}{10}, \nicefrac{3}{10}\}, \{\nicefrac{2}{5}, \nicefrac{3}{10}, \nicefrac{3}{10}\}\}$ \\
    IV         & 50    & 60    & 4     & $\{\nicefrac{3}{10}, \nicefrac{3}{10}, \nicefrac{1}{5}, \nicefrac{1}{5}\}$    & $\{1, 2, 2, 3\}$    & $\{\{1\}, \{\nicefrac{1}{2}, \nicefrac{1}{2}\}, \{\nicefrac{7}{10}, \nicefrac{3}{10}\}, \{\nicefrac{2}{5}, \nicefrac{3}{10}, \nicefrac{3}{10}\}\}$ \\
    V          & 100   & 60    & 4     & $\{\nicefrac{3}{10}, \nicefrac{3}{10}, \nicefrac{1}{5}, \nicefrac{1}{5}\}$    & $\{1, 2, 2, 3\}$    & $\{\{1\}, \{\nicefrac{1}{2}, \nicefrac{1}{2}\}, \{\nicefrac{7}{10}, \nicefrac{3}{10}\}, \{\nicefrac{2}{5}, \nicefrac{3}{10}, \nicefrac{3}{10}\}\}$ \\
    \hline
\end{tabular}
}
\end{table*}

Table~\ref{tab:SS2_Setup} summarises the parameter configurations for Simulation Study 2. In this study, we generated undirected, binary-valued networks where edges follow a Bernoulli distribution with a logit link function. For smaller networks ($N = 30$), $\alpha = 0.6$ was used, whereas larger networks ($N > 30$) used $\alpha = -0.4$.

\clearpage
\section{Posterior Predictive Checks for the Krackhardt Multiplex} \label{SuppMat:PPCs:Krackhardt}
In this section, we present the results for the remaining four posterior predictive check (PPC) metrics not included in Section 5.1 of the main article. These metrics further assess model fit by comparing the observed data to the data simulated from the posterior predictive distributions. While the primary metric highlighted in the main article showed a clear advantage for {\em LaPCoM}, the results for these additional metrics are either comparable across both {\em LaPCoM} and \verb|latentnet| or show a slight preference for {\em LaPCoM}. The details of these metrics and their interpretations are provided below.

\begin{figure}[!t]
    \centering
    \includegraphics[width = \textwidth]{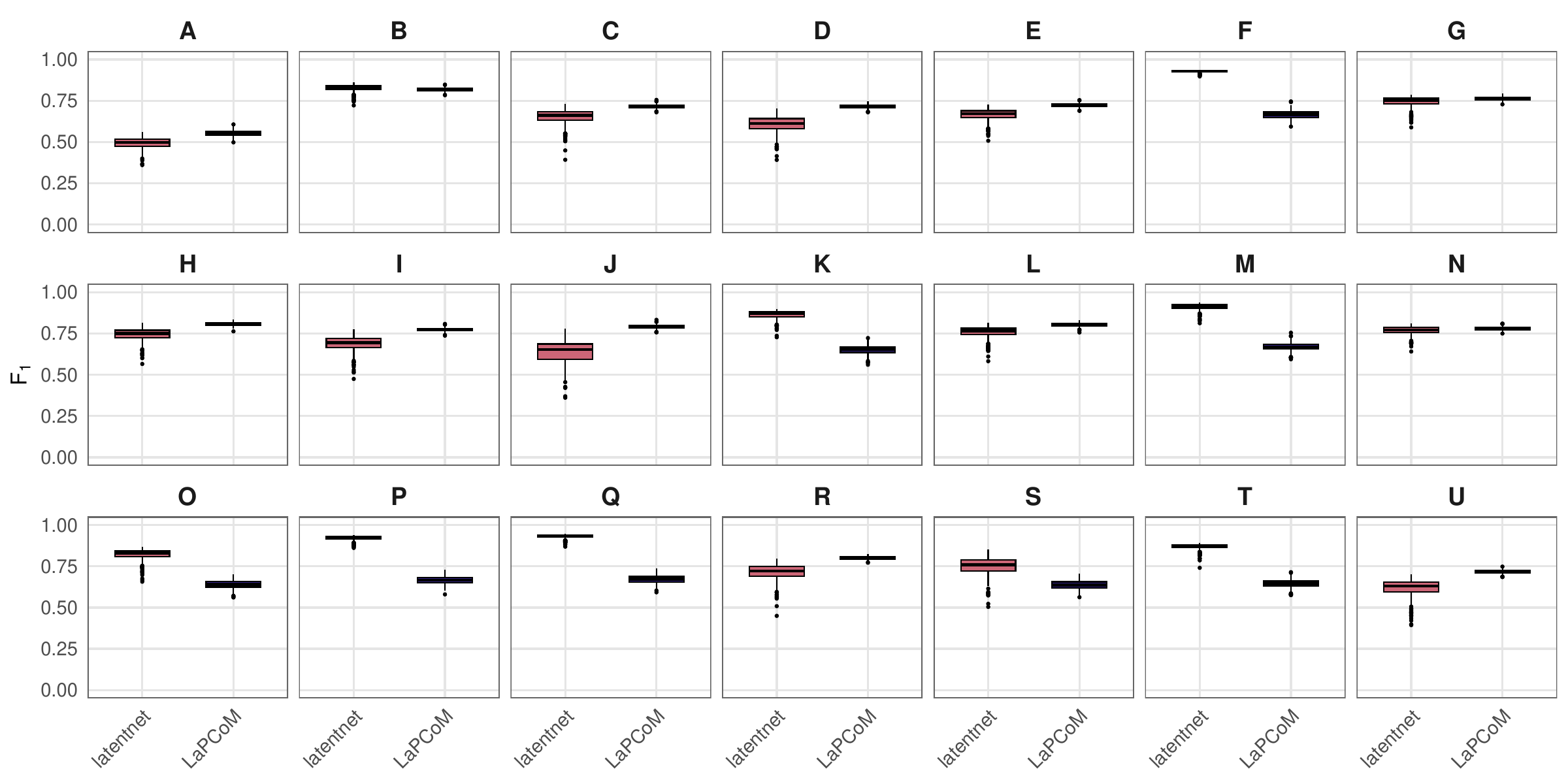}
    \caption{Boxplots showing the distribution of $F_1$-scores across 500 multiplexes generated from the posterior predictive distribution, using both \texttt{latentnet} and {\em LaPCoM}, for the Krackhardt application.}
    \label{fig:Krack_PPC_F1}
\end{figure}

The $F_1$-score, defined as the harmonic mean of precision and recall, is calculated as follows: $F_1 = 2 \times \frac{\mathrm{precision} \times \mathrm{recall}}{\mathrm{precision} + \mathrm{recall}}$. Precision is the proportion of true positive ties among all predicted ties, while recall is the proportion of true positive ties among all actual ties. For any network, we calculate the $F_1$-score using the simulated network and the observed network across all $M$ networks in the multiplex and the 500 simulated multiplexes. The $F_1$-score ranges from 0 to 1 with higher values representing a better fit. Figure~\ref{fig:Krack_PPC_F1} presents a boxplot of the 500 $F_1$-scores for each network, using both \verb|latentnet| and {\em LaPCoM}. For \verb|latentnet|, the average median $F_1$-score was 0.76. In comparison, {\em LaPCoM} achieved a comparable but slightly lower average median of 0.72.

\begin{figure}[!t]
    \centering
    \includegraphics[width = \textwidth]{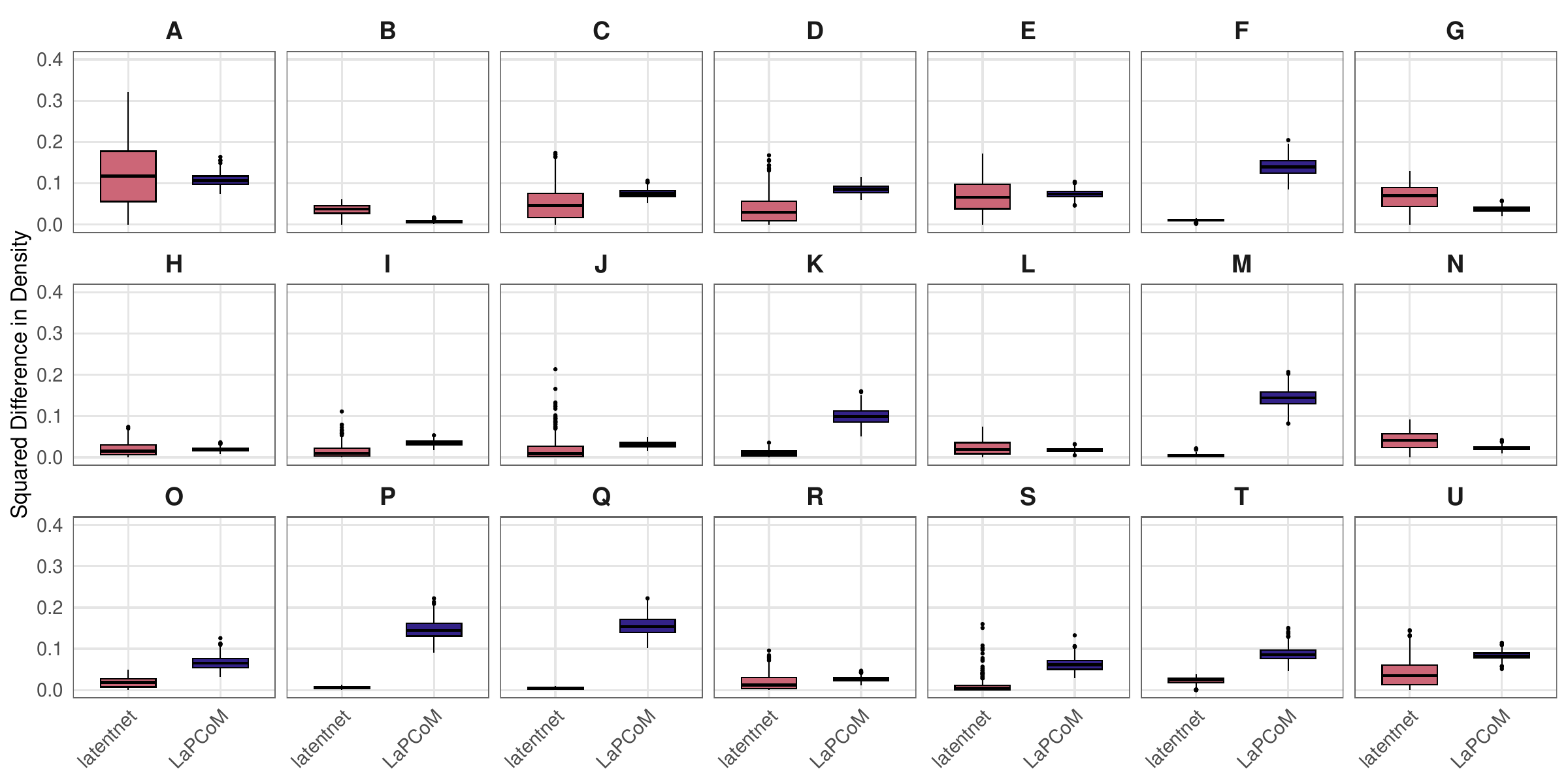}
    \caption{Boxplots showing the distribution of squared differences (in density) across 500 multiplexes generated from the posterior predictive distribution, using both \texttt{latentnet} and {\em LaPCoM}, for the Krackhardt application.}
    \label{fig:Krack_PPC_Density}
\end{figure}

Network properties such as density are also considered in our analysis. A network’s density is the ratio of the number of observed dyadic connections to the number of possible dyadic connections. For any network, we calculate the squared difference between the density of the simulated network and the density of the observed network across all $M$ networks in the multiplex and the 500 simulated multiplexes. A lower squared difference indicates a better model fit. Figure~\ref{fig:Krack_PPC_Density} presents a boxplot of the 500 squared density differences for each network, using both \verb|latentnet| and {\em LaPCoM}. For \verb|latentnet|, the average median squared difference was 0.028. {\em LaPCoM} showed a marginally higher median squared difference of 0.072, still indicating a good fit overall. 

\begin{figure}[!t]
    \centering
    \includegraphics[width = \textwidth]{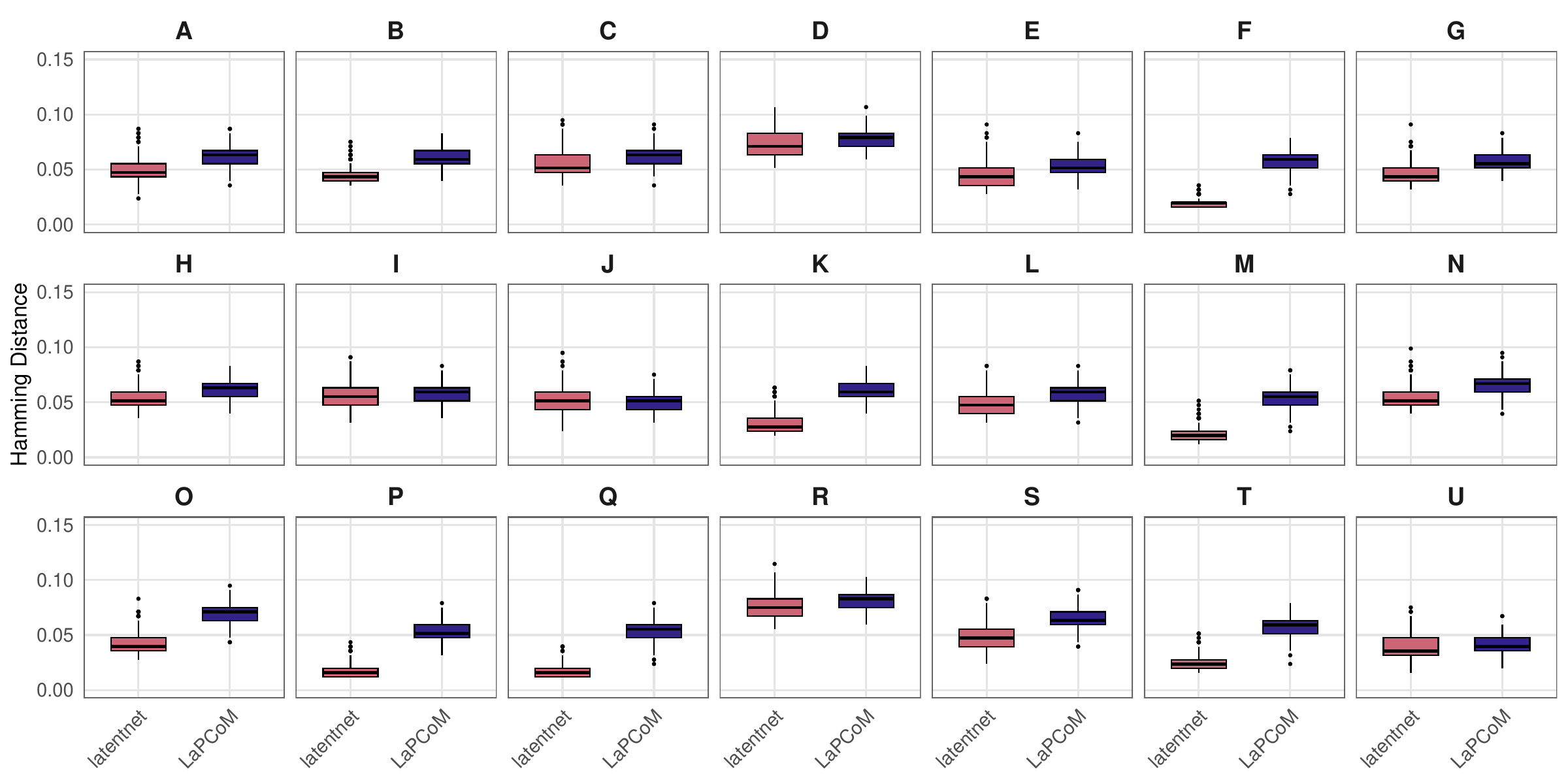}
    \caption{Boxplots showing the distribution of Hamming distances across 500 multiplexes generated from the posterior predictive distribution, using both \texttt{latentnet} and {\em LaPCoM}, for the Krackhardt application.}
    \label{fig:Krack_PPC_Hamming_Distances}
\end{figure}

Figure~\ref{fig:Krack_PPC_Hamming_Distances} presents a boxplot of the 500 Hamming distances for each network, using both \verb|latentnet| and {\em LaPCoM}. The \verb|latentnet| results had an average median of 0.04, compared to 0.06 for {\em LaPCoM}. 

\begin{figure}[!t]
    \centering
    \includegraphics[width = \textwidth]{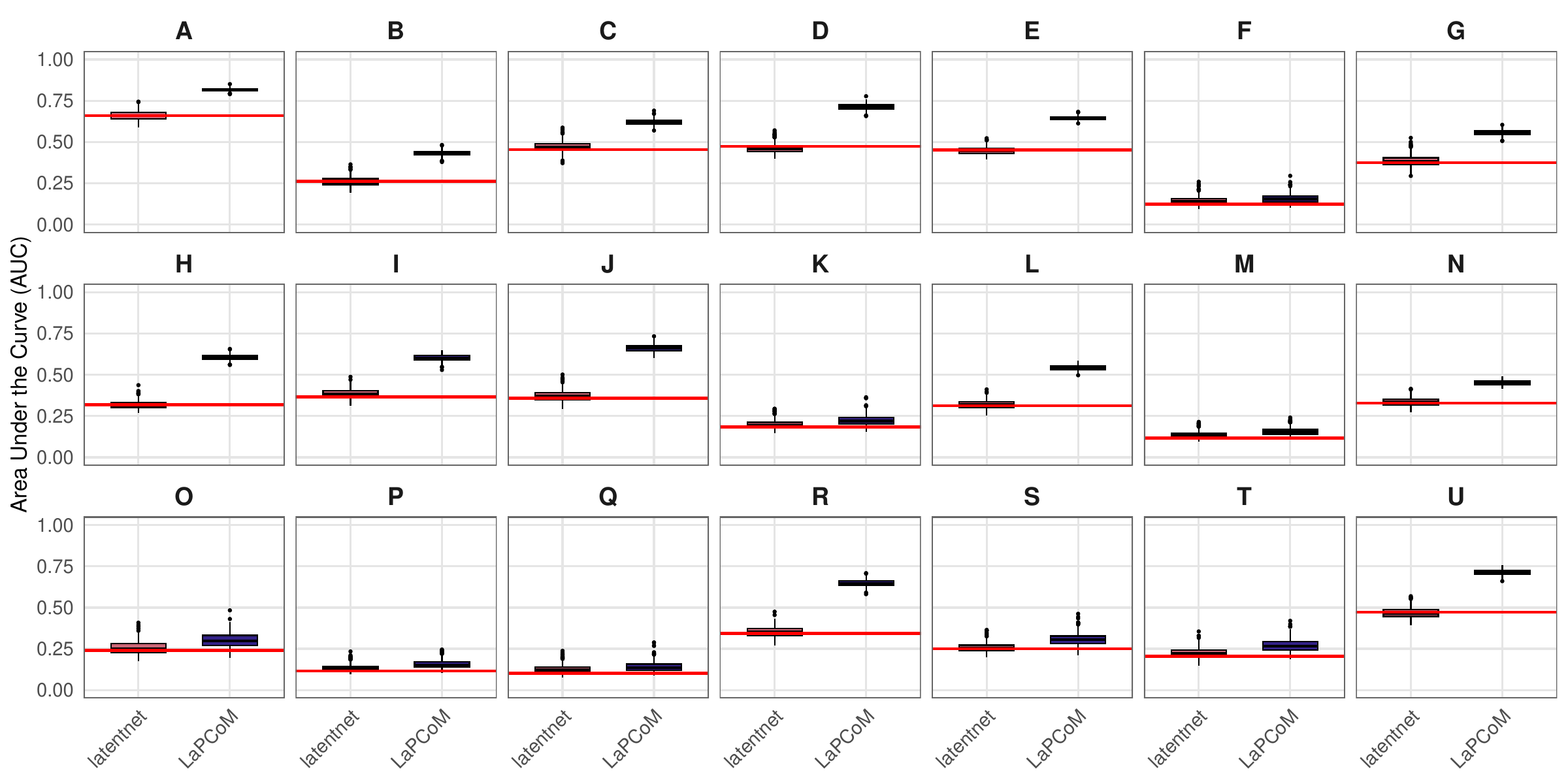}
    \caption{Boxplots showing the distribution of area under the curve (AUC) values across 500 multiplexes generated from the posterior predictive distribution, using both \texttt{latentnet} and {\em LaPCoM}.}
    \label{fig:Krack_PPC_AUC}
\end{figure}

For each network, we compute the precision-recall curve using the simulated tie probabilities and the vectorised observed adjacency matrices across all $M$ networks and 500 simulated multiplexes. This curve evaluates the trade-off between precision (true positives among predicted ties) and recall (true positives among actual ties) without relying on a specific threshold. We summarise performance by the area under the curve (AUC), a scalar measure of model performance. Figure~\ref{fig:Krack_PPC_AUC} shows boxplots of AUC distributions across the 500 posterior predictive multiplexes, comparing \verb|latentnet| and {\em LaPCoM}. The observed network density is marked by a red horizontal line in each panel. For \verb|latentnet|, the median AUC aligns with the observed density, indicating an average fit, whereas for {\em LaPCoM}, the entire distribution lies above the observed density, demonstrating a markedly better fit.

\clearpage
\section{Posterior Predictive Checks for the Aarhus Multiplex} \label{SuppMat:PPCs:Aarhus}
In this section, we present additional PPC metric results for the application presented in Section~5.2 of the main article. As in the previous section, {\em LaPCoM} generally demonstrates better or comparable performance relative to \verb|latentnet|. 

\begin{figure}[!t]
    \centering
    \includegraphics[width = \textwidth]{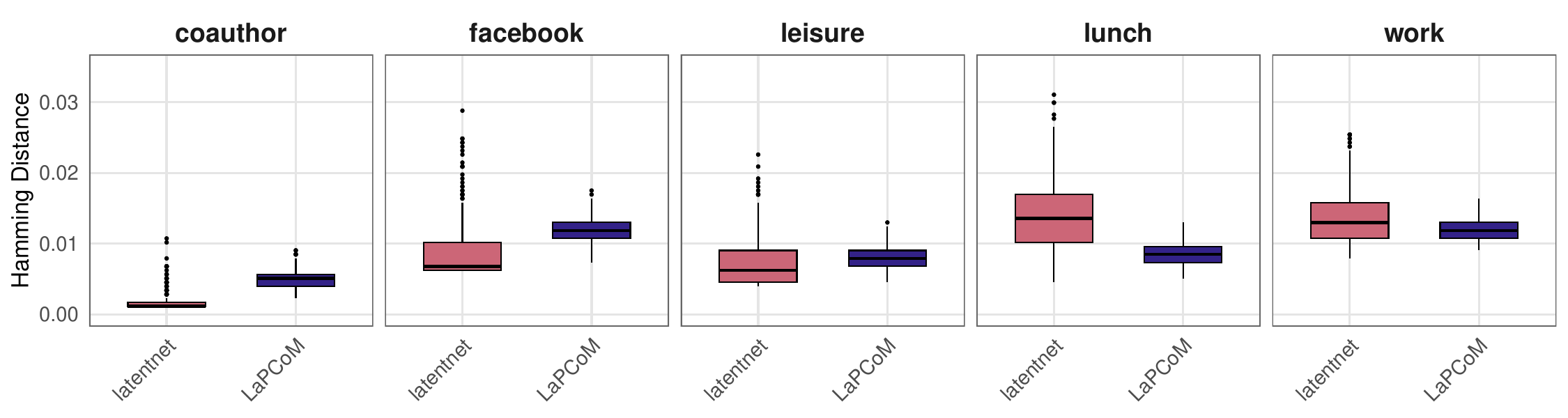}
    \caption{Boxplots showing the distribution of the Hamming distances across 500 multiplexes generated from the posterior predictive distribution, using both \texttt{latentnet} and {\em LaPCoM}, for the Aarhus application.}
    \label{fig:Aarhus_PPC_Ham_Dist}
\end{figure}

Figure~\ref{fig:Aarhus_PPC_Ham_Dist} presents a boxplot of the 500 Hamming distances for each network, using both \verb|latentnet| and {\em LaPCoM}. {\em LaPCoM} has an average median Hamming distance of 0.009 with an interquartile range (IQR) of 0.002, while \verb|latentnet| has an average median of 0.008 with a median IQR of 0.004. Although the median Hamming distances for \verb|latentnet| tend to be, on average, slightly smaller, {\em LaPCoM} is generally more precise, with less deviation. This indicates that both models perform at a comparable level with respect to Hamming distance.

\begin{figure}[!t]
    \centering
    \includegraphics[width = \textwidth]{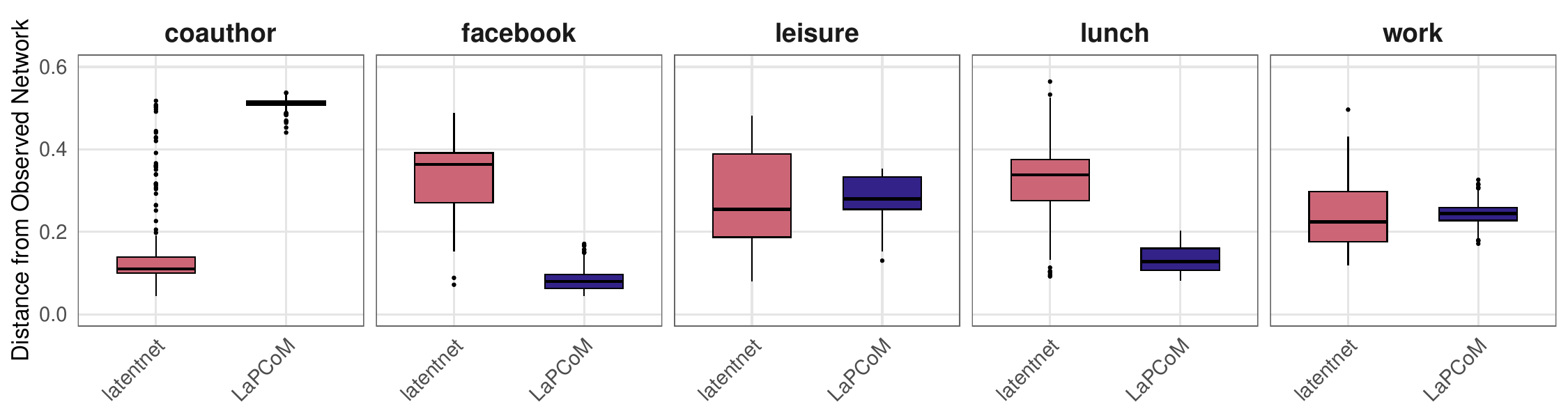}
    \caption{Boxplots showing the distribution of the network distances across 500 multiplexes generated from the posterior predictive distribution, using both \texttt{latentnet} and {\em LaPCoM}, for the Aarhus application.}
    \label{fig:Aarhus_PPC_Net_Dist}
\end{figure}

Figure~\ref{fig:Aarhus_PPC_Net_Dist} presents a boxplot of the 500 \citet{Schieber:2017:Extra:Nature:Distance} network distances for each network, using both \verb|latentnet| and {\em LaPCoM} (lower values are better). The average median network distance for {\em LaPCoM} was 0.25, slightly lower than the \verb|latentnet| average median of 0.27. Additionally, the average IQR for \verb|latentnet| was 0.12, compared to 0.04 for {\em LaPCoM}. These results suggest that {\em LaPCoM} provides a better fit, on average, to the data based on this metric.

\begin{figure}[!t]
    \centering
    \includegraphics[width = \textwidth]{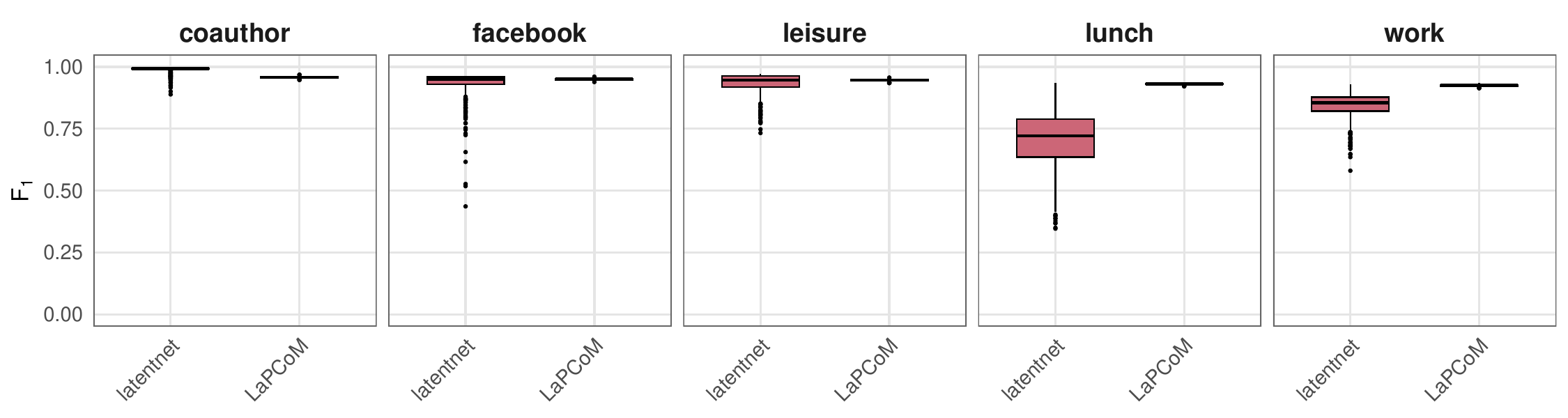}
    \caption{Boxplots showing the distribution of $F_1$-scores across 500 multiplexes generated from the posterior predictive distribution, using both \texttt{latentnet} and {\em LaPCoM}.}
    \label{fig:Aarhus_PPC_F1}
\end{figure}

Figure~\ref{fig:Aarhus_PPC_F1} shows boxplots of the 500 $F_1$-scores for each network under both \verb|latentnet| and {\em LaPCoM}. The models perform similarly for three of the five networks: {\em LaPCoM} yields an average median $F_1$-score of 0.95 with an average IQR of 0.005, while \verb|latentnet| gives a slightly higher median of 0.96 but with a wider IQR of 0.03, indicating lower precision. More pronounced differences appear in the ``lunch'' and ``work'' networks. For ``work'', {\em LaPCoM} achieves a median $F_1$-score of 0.93 (IQR 0.004), outperforming \verb|latentnet| (median 0.86, IQR 0.06). For ``lunch'', {\em LaPCoM} again excels with a median of 0.93 and IQR of 0.005, compared to \verb|latentnet|'s median of 0.72 and IQR of 0.15, indicating both higher accuracy and lower variability.

\begin{figure}[!t]
    \centering
    \includegraphics[width = \textwidth]{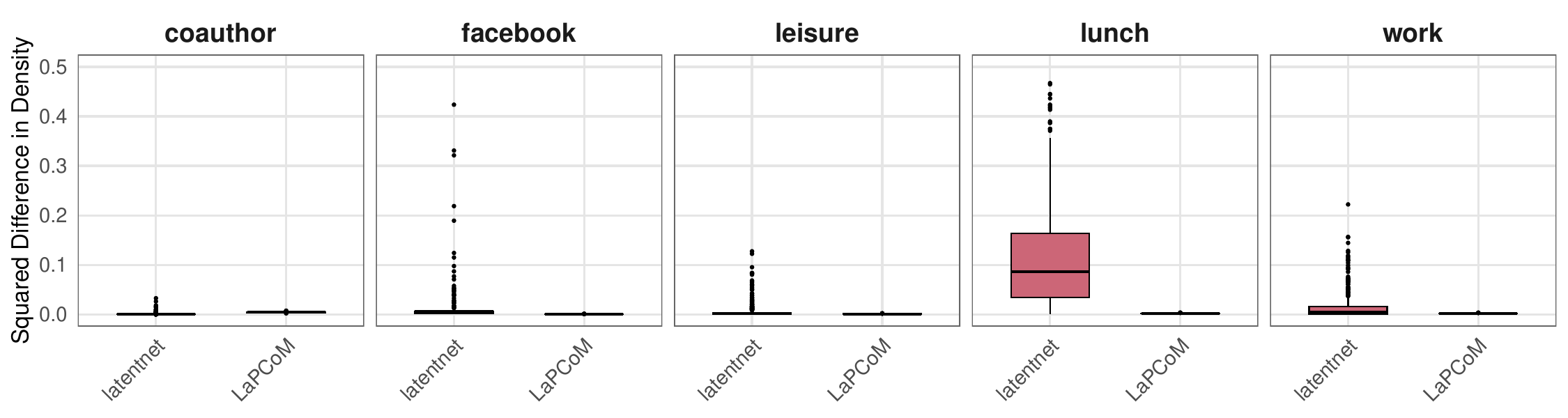}
    \caption{Boxplots showing the distribution of squared differences (in density) across 500 multiplexes generated from the posterior predictive distribution, using both \texttt{latentnet} and {\em LaPCoM}.}
    \label{fig:Aarhus_PPC_Density}
\end{figure}

Figure~\ref{fig:Aarhus_PPC_Density} presents boxplots of the 500 squared differences in density for each network under both \verb|latentnet| and {\em LaPCoM}. Across four networks, the models are broadly comparable: {\em LaPCoM} yields an average median of 0.002 with an average IQR of 0.0006, while \verb|latentnet| has a slightly higher average median of 0.003 and a wider IQR of 0.005. Although the differences are modest, {\em LaPCoM} demonstrates greater precision, with notably fewer outliers. For the ``lunch'' network, {\em LaPCoM} performs substantially better, with a median of 0.002 and IQR of 0.0007, in contrast to \verb|latentnet|, which shows a median of 0.09 and IQR of 0.13, indicating poorer fit.

\clearpage
\section{Posterior Predictive Checks for the Primary School Interactions Multiplex} \label{SuppMat:PPCs:PSC}
In this section, we present results for an additional PPC metric in relation to the primary school network data described in Section~5.3 of the main article. This metric further evaluates model fit by comparing the observed data to data simulated from the posterior predictive distribution. Specifically, we focus on the empirical cumulative distribution function (ECDF) of the positive edge counts (on the log scale), which provides insight into how well the model captures the distributional structure of non-zero edge weights across the network.

\begin{figure}[!t]
    \centering
    \includegraphics[width=\linewidth]{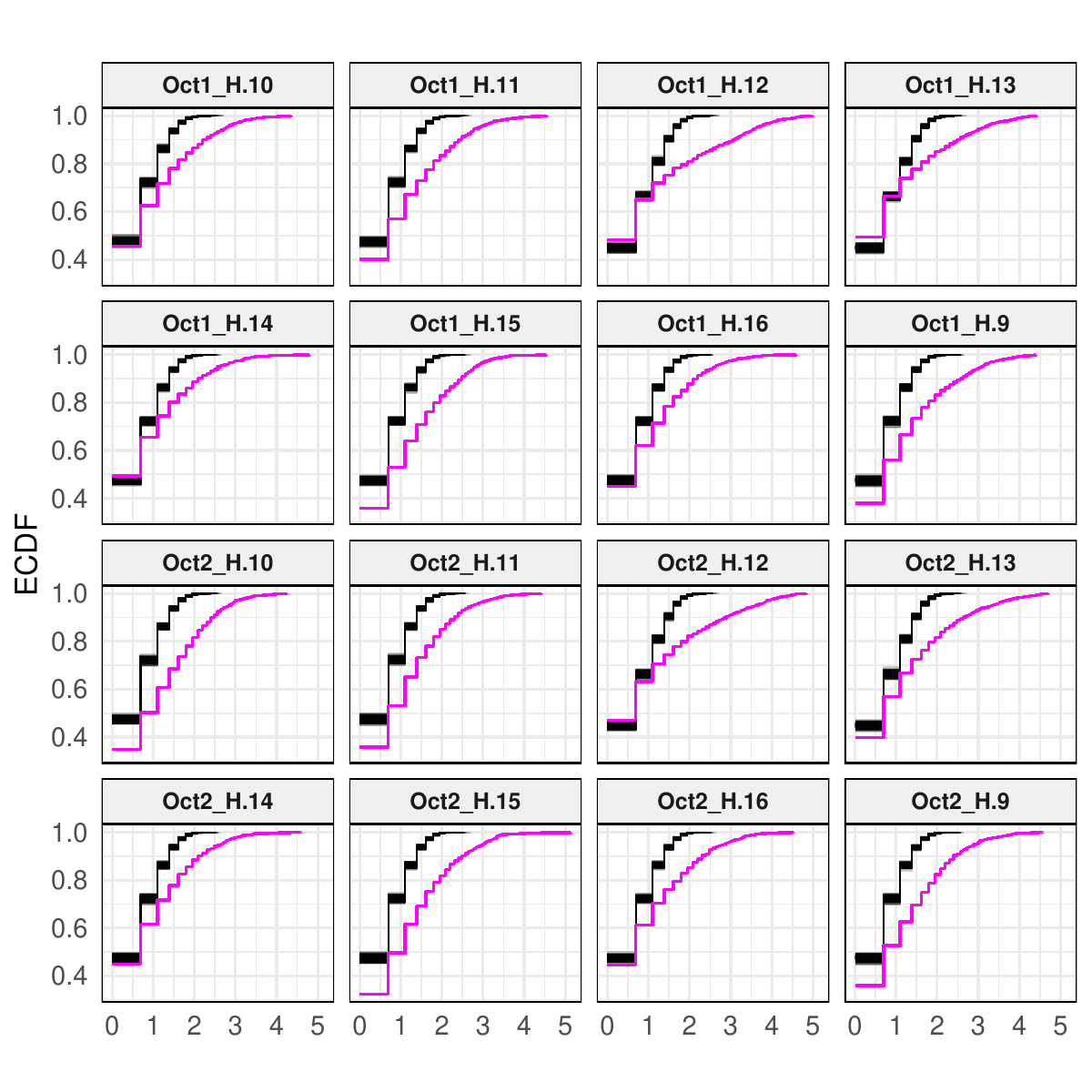}
    \caption{Empirical cumulative distribution functions of log positive counts for each network. The pink line shows the observed ECDF, and black lines show 500 posterior predictive replicates.}
    \label{fig:PSC_PPC_ECDF}
\end{figure}

Figure~\ref{fig:PSC_PPC_ECDF} shows the ECDF of the positive counts (on a log scale) for each network. The true ECDF is indicated by the pink line, while the black lines represent 500 ECDFs from posterior predictive samples overlaid for each network. The deviation from the observed ECDF indicates that the model tends to underestimate the occurrence of large edge counts. The edge count distribution in the multiplex data is characterised by overdispersion. Therefore, the Poisson distribution may be too restrictive to capture this aspect of the data. Nonetheless, the model still captures the overall structure of the network reasonably well, and it provides a clustering which aligns with the membership of the students to the different classes and to specific times of the day.

\clearpage
\section{A Mixture of Finite Mixtures Model versus an Overfitted Mixture Model} \label{SuppMat:MFMvsOM}
As discussed in Section~2.4 of the main article, we chose to implement a mixture of finite mixtures (MFM) model with TS \citep{MillerHarrison:2015:MFM,FS:2021:Methods:MFM_TS}, rather than an overfitted mixture (OM) approach \citep{MW:2016:Methods:SFM:Original,FSMW:2019:Methods:SFMvsDPM}. The OM model fixes the number of components to a large value and assumes a sparse prior on the mixing proportions to recover the underlying number of clusters. In contrast, the MFM places a prior on the number of components (specifically, a translated beta-negative-binomial distribution) and differentiates between total and active (non-empty) components, which are dynamically updated through the TS procedure. This flexibility made the MFM more suitable for our setting, which involves simultaneous inference of network- and node-level clusters.

In the following, we compare three different LaPCoM formulations: the MFM formulation proposed in the main text (MFMTS1), an MFM variant with hyperparameters set following \citet{FS:2021:Methods:MFM_TS} (MFMTS2), and a formulation employing the OM approach (OM). The aim is to determine the effect of the prior specification for the mixture structure of the model at both network and node level, especially in regards to clustering performance at the network level. 

In MFMTS1, we employed our specified hyperparameters, where at the network-level we specify $G - 1\sim \mathcal{BNB}(8,18,10)$ with $G_0 = 2$ and $G_{\text{max}}$ is as in Table~\ref{tab:Hypers_Gmax}; at the node-level we specify $K_g - 1\sim \mathcal{BNB}(8,18,10)$ with $K_0 = 2$ and $K_{\text{max}}$ is as in Table~\ref{tab:Hypers_Kmax}. Formulation MFMTS2, followed the hyperparameter choices recommended in \citet{FS:2021:Methods:MFM_TS}, where at the network-level we specify $G - 1 \sim \mathcal{BNB}(1,4,3)$ with $G_0=6$ and $G_{\text{max}} = 10$, while at the node-level we specify $K_g - 1 \sim \mathcal{BNB}(1,4,3)$ with $K_0=3$ and $K_{\text{max}} = 15$. Lastly, we used an OM approach with $G_{\text{max}} = 10$ and $K_{\text{max}} = 10$. The MFMTS2 configuration was based on recommendations in \citet{FS:2021:Methods:MFM_TS}, where we selected $G_0 = 6$, approximately 2-3 times the expected true value $G^{*} = 2$ used in our simulated data experiments. For $K_0 = 3$, we aimed to reflect the range of values observed in the simulations, $K_g^{*} = \{1,2,3\}$. For the OM configuration, we chose $G_{\text{max}} = 10$, a large yet practical choice inspired by \citet{FS:2021:Methods:MFM_TS}, where they empirically tested values as high as $G_{\text{max}} = 100$. However, such a large value was not feasible in our setting. Similarly, we set $K_{\text{max}} = 10$. 

We implemented simulation studies using the same setup followed in both simulation study 1 (SS1) and simulation study 2 (SS2) of the main text (details are in Section~4 of the main article and in Section~\ref{SuppMat:SimStudies} here). In both cases, MCMC was run for 10 replications of 13,000 iterations, including a burn-in period of 3,000 iterations, which was discarded. Samples were thinned at an interval of 10, resulting in a total of 1,000 samples.

The following results focus solely on network-level clustering solutions to assess each method’s efficiency in discerning network-level structure.Table~\ref{tab:SuppMat:MFMvsOM_Results_G+_SS1} summarises the posterior estimates of the optimal number of network-level clusters, $\hat{G}$, as determined by each of the three approaches tested over 10 replications for each of the eight scenarios considered in SS1. Our configuration, MFMTS1, consistently identified the correct $\hat{G} = 2$ across all scenarios, supporting the suitability of our chosen hyperparameters. In contrast, MFMTS2 underperformed, identifying the correct number of clusters in none of the scenarios. The OM model showed competitive performance to MFMTS1, estimating the correct $\hat{G}$ in seven out of eight scenarios. Table~\ref{tab:SuppMat:MFMvsOM_Results_Network_Clustering_SS1} presents the adjusted rand index (ARI, \citet{Hubert:1985:Extra:ARI}) values for the network-level clustering of each approach. The proposed MFMTS1 approach achieved an ARI of one in all eight scenarios. The MFMTS2 configuration displayed lower ARI values. The OM approach performed comparably to MFMTS1, achieving an ARI of one in four out of eight scenarios. These findings substantiate our choice to use a MFM model with the TS procedure over the OM approach, and further validate our specific hyperparameter settings over those recommended in \citet{FS:2021:Methods:MFM_TS}. The MFM configuration proposed in the main text consistently achieved greater accuracy in identifying network-level clusters across diverse scenarios.

\begin{table}[!b]
\centering
\begin{minipage}{0.45\textwidth}
    \caption{Posterior mode of network-level clusters $\hat{G}_+$.}
    \label{tab:SuppMat:MFMvsOM_Results_G+_SS1}
    \resizebox{\textwidth}{!}{%
    \begin{tabular}{@{}cccc@{}}
    \hline
    Scenario & OM & MFMTS1 & MFMTS2 \\
    \hline
    A & 8 & 2 & 4 \\
    B & 2 & 2 & 4 \\
    C & 2 & 2 & 3 \\
    D & 2 & 2 & 6 \\
    E & 2 & 2 & 3 \\
    F & 2 & 2 & 4 \\
    G & 2 & 2 & 3 \\
    H & 2 & 2 & 4 \\
    \hline
    \end{tabular}%
    }
\end{minipage}%
\hspace{0.05\textwidth}
\begin{minipage}{0.45\textwidth}
    \caption{Adjusted Rand index (ARI) of network-level clustering.}
    \label{tab:SuppMat:MFMvsOM_Results_Network_Clustering_SS1}
    \resizebox{\textwidth}{!}{%
    \begin{tabular}{@{}cccc@{}}
    \hline
    Scenario & OM & MFMTS1 & MFMTS2 \\
    \hline
    A & 0.56 & 1.00 & 0.66 \\
    B & 0.63 & 1.00 & 0.69 \\
    C & 1.00 & 1.00 & 0.70 \\
    D & 1.00 & 1.00 & 0.42 \\
    E & 0.92 & 1.00 & 0.77 \\
    F & 0.93 & 1.00 & 0.72 \\
    G & 1.00 & 1.00 & 0.95 \\
    H & 1.00 & 1.00 & 0.82 \\
    \hline
    \end{tabular}%
    }
\end{minipage}
\end{table}

Table~\ref{tab:SuppMat:MFMvsOM_Results_G+_SS2} summarises the posterior estimates of the optimal number of network-level clusters, $\hat{G}$, as determined by each of the three model specifications tested over 10 replications for each of the five scenarios considered in SS2. The proposed MFMTS1, consistently identified the correct $\hat{G}$ across all scenarios. In contrast, MFMTS2 slightly underperformed, accurately capturing $\hat{G}$ in three of the five scenarios. The model based on the OM approach showed poor performance in this setting, underestimating the number of network-level clusters in four scenarios. Table~\ref{tab:SuppMat:MFMvsOM_Results_Network_Clustering_SS2} presents the ARI values for network-level clustering under each approach. Our proposed MFMTS1 approach achieved a perfect or near to perfect ARI. The MFMTS2 configuration displayed slightly lower ARI values, with an ARI of 0.53 in one scenario. The OM approach showed the weakest performance, achieving an ARI of one in only one scenario, and an average ARI of 0.45 across the remaining four scenarios. These results further support our choice to use a MFM model with TS, particularly with our chosen hyperparameters, which outperformed both the OM method and the parameter settings recommended in \citet{FS:2021:Methods:MFM_TS}. 

\begin{table}[!bt]
\centering
\begin{minipage}{0.45\textwidth}
    \caption{Posterior mode of network-level clusters $\hat{G}_+$.}
    \label{tab:SuppMat:MFMvsOM_Results_G+_SS2}
    \resizebox{\textwidth}{!}{%
    \begin{tabular}{@{}ccccc@{}}
    \hline
    Scenario & $G^*$ & OM & MFMTS1 & MFMTS2 \\
    \hline
    I   &   2   & 9 & 2 & 4 \\
    II  &   2   & 2 & 2 & 2 \\
    III &   3   & 2 & 3 & 3 \\
    IV  &   4   & 2 & 4 & 5 \\
    V   &   4   & 2 & 4 & 4 \\
    \hline
    \end{tabular}%
    }
\end{minipage}%
\hspace{0.05\textwidth}
\begin{minipage}{0.45\textwidth}
    \caption{Adjusted Rand index (ARI) of network-level clustering.}
    \label{tab:SuppMat:MFMvsOM_Results_Network_Clustering_SS2}
    \resizebox{\textwidth}{!}{%
    \begin{tabular}{@{}cccc@{}}
    \hline
    Scenario & OM & MFMTS1 & MFMTS2 \\
    \hline
    A & 0.32 & 0.98 & 0.53 \\
    B & 1.00 & 1.00 & 0.99 \\
    C & 0.56 & 1.00 & 1.00 \\
    D & 0.47 & 1.00 & 0.97 \\
    E & 0.43 & 0.99 & 1.00 \\
    \hline
    \end{tabular}%
    }
\end{minipage}
\end{table}

Because the MCMC algorithms in this comparison were run for fewer iterations than in the main simulations in the main text, in some instances the permutation test failed to identify valid classification sequences of $1, \dots, \hat{G}_+$ (see Section~3.3 of the main article). This occurred in the SS1 set-up for MFMTS1 (one simulation, Scenario F), MFMTS2 (two simulations, Scenario F), and OM (one simulation each in Scenarios A and F). These instances reflected random variation or insufficient number of iterations and were discarded.

\clearpage
\bibliographystyle{apalike}
\bibliography{references}

\end{document}